\documentclass[11pt]{article}

\usepackage{graphicx}
\usepackage{color}
\usepackage{rotating}
\usepackage{amsmath}
\usepackage{amssymb}
\usepackage{amsthm}
\usepackage{float}
\usepackage{ulem}
 \usepackage{url}
\usepackage{subfigure}

\usepackage{enumerate}
\usepackage{enumitem}
 
 \usepackage{algorithm}
\usepackage{algorithmic}

\usepackage{array,arydshln}

\newcommand\ba {\mathbf a}
\newcommand\bb {\mathbf b}
\newcommand\bc {\mathbf c}

\newcommand\bw {\mathbf w}
\newcommand\bx {\mathbf x}

\newcommand\bz {\mathbf z}

\newcommand\bB {\mathbf B}

\newcommand\indica {\mathbb{I}}

\newcommand\wa {\widehat{{a}}}
\newcommand\wba {\widehat{\ba}} 
\newcommand\wb {\widehat{{b}}}
\newcommand\wbb {\widehat{{\bb}}}

\newcommand\wy {\widehat{y}}

\newcommand\itA {{\mathcal{A}}}
\newcommand\itB {{\mathcal{B}}}
\newcommand\itC {{\mathcal{C}}}

\newcommand\itF {{\mathcal{F}}}
\newcommand\itG {{\mathcal{G}}}
\newcommand\itH {{\mathcal{H}}}
\newcommand\itI {{\mathcal{I}}}
\newcommand\itJ {{\mathcal{J}}}

\newcommand\itL {{\mathcal{L}}}
\newcommand\itM {{\mathcal{M}}}
\newcommand\itN {{\mathcal{N}}}

\newcommand\itT {{\mathcal{T}}}
\newcommand\itU {{\mathcal{U}}}
\newcommand\itV {{\mathcal{V}}}

\newcommand\itX {{\mathcal{X}}}

\newcommand{\itZ}{\mathcal{Z}}

\newcommand\bgama {\mbox{\boldmath $\gamma$}}

\newcommand\wbeta {\widehat{\beta}}

\newcommand\weta {\widehat{\eta}}

\newcommand\wgamma {\widehat{\gamma}}

\newcommand\wsigma {\widehat{\sigma}}

\newcommand\wtheta {\widehat{\theta}}

\newcommand\wxi {\widehat{\xi}}

\newcommand\wtbeta {\widetilde{\beta}}

\newcommand\wteta {\widetilde{\eta}}

\newcommand\wttheta {\widetilde{\theta}}

\def\real{\mathbb{R}}
\def\natu{\mathbb{N}}
\def\qu{\mathbb{Q}}

\newcommand{\esp}{\mathbb{E}}
\newcommand{\prob}{\mathbb{P}}

\newcommand{\var}{\mbox{\sc Var}}

\newcommand{\convpp}{ \buildrel{a.s.}\over\longrightarrow}

\newcommand{\convprob  }{ \buildrel{p}\over\longrightarrow}

\newcommand{\trasp}{^{\mbox{\footnotesize \sc t}}}

\newcommand\bcero {{\bf{0}}}

\def\dst{\displaystyle}

\def\mad{\mathop{\mbox{\sc mad}}}

\def\argmin{\mathop{\mbox{argmin}}}

\def\MISE{\mathop{\rm MISE}}

\newcommand\noi{\noindent}
\def\dst{\displaystyle}

\def\square{\ifmmode\sqr\else{$\sqr$}\fi}
\def\sqr{\vcenter{
         \hrule height.1mm
         \hbox{\vrule width.1mm height2.2mm\kern2.18mm
\vrule width.1mm}
         \hrule height.1mm}}

\newcommand{\tuk}{\mbox{\scriptsize \sc tuk}}

\newcommand{\ini}{\mbox{\footnotesize \sc ini}}

\newcommand{\eme}{\mbox{\footnotesize \sc m}} 
\newcommand\ls  {\mbox{\footnotesize\sc ls}}

\newcommand\trim {\mbox{\footnotesize  \sc tr}}

\newcommand{\clean}{\mbox{\scriptsize \sc clean}}

\newcommand{\monmod}{\mbox{\scriptsize \sc mod}}

\begin{document}


\title{Robust estimation for semi-functional linear regression models}
\author{Graciela Boente$^a$, Mat\'\i as Salibian-Barrera$^b$ and Pablo Vena$^a$ \\ 
  $^a$ Universidad de Buenos Aires and CONICET, Argentina \\
$^b$ University of British Columbia, Canada
}
\date{}

\maketitle

\begin{abstract}
Semi-functional linear regression models postulate a linear
relationship between a scalar response and a functional covariate, and
also include a non-parametric component involving a univariate
explanatory variable. It is of practical importance to obtain
estimators for these models that are robust against high-leverage
outliers, which are generally difficult to identify and may cause
serious damage to least squares and Huber-type $M$-estimators. For
that reason,  robust estimators for semi-functional linear regression
models are constructed combining $B$-splines to approximate both the
functional regression parameter and the nonparametric component with
robust regression estimators based on a bounded loss function
and a preliminary residual scale estimator. 
Consistency and rates of convergence for the proposed estimators are derived
under mild regularity conditions. The reported numerical
experiments show the advantage of the proposed methodology
over the classical least squares and Huber-type $M$-estimators for
finite samples. The analysis of real examples  
illustrate that the
robust estimators provide better predictions for non-outlying points
than the classical ones, and that when potential outliers are removed
from the training and test sets both methods behave very
similarly.

\end{abstract}

\noi \textbf{Keywords:} $B$-splines; Functional Data Analysis; Partial Linear Models;  Robust estimation 

\noi\textbf{AMS Subject Classification:}  62G35; 62G25

\normalsize
\newpage
\section{Introduction}\label{sec:intro}

Many commonly used statistical models are either fully parametric or completely
non-parametric. On the one hand, while a reasonable parametric model results in
stable estimators and associated inferences, a misspecified one can lead to
seriously misleading and biased conclusions. On the other hand,  non-parametric
methods may avoid misspecified models, but typically result in more variable
estimators.  A particular difficulty with non-parametric models is that in many
applications they typically require multivariate smoothing which can be
seriously affected by the well-known \lq \lq curse of dimensionality\rq \rq.
This issue is even more serious when the model includes infinite-dimensional
components.

One approach to deal with this problem is to consider semi-parametric
models. 
Specifically, consider a scalar response variable $y$ and a vector of potential
covariates 
$\bw \in \real^{d}$. Partial linear regression models allow some components of
$\bw$ to enter the model in a fully non-parametric way, while the rest are
assumed to have a linear effect on $y$.  These models avoid the curse of
dimensionality problem and are easier to interpret than fully non-parametric
ones. 
An extensive review of partly linear regression models can be found in H\"ardle
\textsl{et al.}  (2000),  and H\"ardle \textsl{et al.}  (2004).   
  Here we consider the extension of these models
 to situations including  both functional and vector covariates. 
 In what follows we will use lowercase letters
to denote scalar random variables, and upper case letters for functional random elements.
   
Functional explanatory variables can be included in partial linear models
either linearly or non-parametrically.  Aneiros-P\'erez and Vieu (2006) and
Shang (2014) used a linear model for the effect of the scalar explanatory
variables and a non-parametric component for the functional covariate.  Lian
(2011) proposed a linear regression model for the infinite-dimensional
covariates $X$ and a nonparametric regression model for the other explanatory
variables via Nadaraya-Watson kernel estimators.   In this paper,
we focus on the particular case where there is one real covariate, which was 
also considered by Zhou and Chen  (2012). 
More precisely, we consider independent and identically distributed observations  
with the same distribution as the triplet $(y,X,z)$, where the response $y \in \real$  
is related linearly to the functional explanatory variable $X\in L^2(\itT)$, 
and nonparametrically to the real covariate $z$. 
In symbols: $ y =  \langle X,\beta_0 \rangle + \eta_0(z) + \sigma_0 \, \epsilon  $,
where $\langle \cdot, \cdot \rangle$ denotes the usual $L^2(\itT)$ inner product,
$\sigma_0 > 0$ is a residual scale parameter, and $\epsilon$ 
is the error term, independent of $(X,z)$. In this model the regression parameter $\beta_0$ is
assumed to be in $L^2(\itT)$, although for estimation purposes some smoothness conditions 
may be required. As usual, the unknown regression 
function $\eta_0$  is only assumed to be smooth with compact support $\itZ$. 

It is well-known that small proportions of outliers and other atypical
observations can affect seriously the estimators for these models 
although not many
robust methods exist in the literature.  
Qingguo (2015) studied $M$-estimators for 
the linear slope using a monotone score function 
and a functional principal component approximation. 
Huang \textsl{et al.} (2015) proposed  $M$-estimators 
approximating both the linear slope and the nonparametric component 
with $B-$spline bases.
These proposals have two main drawbacks: they are not scale 
equivariant since they do not consider a residual scale estimator,
and  they lack protection against high-leverage outliers. 
The lack of scale invariance may be a problem in practice since the
magnitude of the residuals that are to be considered ``large'' (outliers)
depends on their scale. This dispersion parameter 
needs to be estimated with a robust preliminary
scale estimator, as it is generally done for linear regression models (see
Maronna \textsl{et al.}, 2019).  
In finite-dimensional linear regression models, it is
well--known that using an unbounded loss function results in estimators
that cannot protect against high-leverage outliers. We note that this type of atypical
observations may also affect the estimation procedure described by Qingguo
(2015), since functional principal components are also highly sensitive to
small proportions of outliers.   
 
Our proposal overcomes these problems by adapting best practices for robust
multiple linear regression estimators to these partial linear models with functional 
covariates.  More specifically, we use $B$-splines to approximate both 
the functional regression
parameter and the nonparametric component, and apply $MM$-regression estimators
(Yohai, 1987) that are based on a bounded loss function and a preliminary
residual scale estimator. These estimators are scale equivariant, robust
against high-leverage outliers, and strongly consistent under standard
regularity conditions. Furthermore,  we derive convergence rates 
with respect to the mean squared prediction differences obtained with the 
true and estimated parameters.

We illustrate our approach with two real examples. 
We first consider hourly electricity prices in Germany 
between 1 January 2006 and 30 September 2008. The data
consist of German power prices traded at the Leipzig European Energy Exchange,
electricity demand, and eolic energy in the system. Our interest is in studying the relationship between 
hourly prices and the overall load of the system, while taking into account
the proportion of the demand that can be satisfied from wind generators, which 
follows a different price regime. 
These data were also used in Liebl (2013) in the context of electricity
price forecasting, and are available among the supplementary materials of that paper. 
Our second example is the well known Tecator data set
(see Ferraty and Vieu, 2006). This
food quality-control data was obtained from 215 samples of  finely chopped
meat with different percentages of fat, protein and moisture content.  For each
sample, a spectrometric curve of absorbances was measured using a Tecator
Infratec Food and Feed Analyzer.  
Since obtaining a spectrometric curve is faster and less
costly than the analytical procedure used to determine fat content, the
interest is in building a model to predict the fat content of a meat
sample using its protein and moisture contents  as well as its
absorbance spectrum. Boente and Vahnovan (2017) used the functional
boxplot of Sun and Genton (2011) to show the presence of atypical curves among
the second derivatives of the spectrometric curves in the Tecator data.  Thus,
reliable analyses of this data set require methods that protect against
potential outliers in the functional explanatory variables.

The rest of the paper is organized as follows. The model and our proposed
estimators are described in Section \ref{sec:estimadores}. Theoretical
assurances regarding the consistency and convergence rates of our proposal are
provided in Section \ref{sec:consis}, while in  Section \ref{sec:simu} we
report the results of a simulation study to explore their finite-sample
properties. Section \ref{sec:ejemplos} contains two real-data analyses, while 
final comments are given in Section \ref{sec:concl}. All
proofs are relegated to the Appendix.

\section{Model and estimators} \label{sec:estimadores}

The semi-functional linear regression model (see, for example, Zhou and Chen, 2012) 
assumes that the observations 
$(y_i, X_i, z_i)$, $1\le i\le n$, are independent and identically distributed
realizations of the random element $(y,X,z)$, where $y \in \real$ is the response
variable, $X$ is a stochastic process on $L^2(\itT)$, the space of
square integrable functions on the interval $\itT$, and $z \in \real$. 
The relationship between the response and the explanatory variables is given by: 
\begin{equation}
  \label{eq:plm}
  y =  \langle X,\beta_0 \rangle + \eta_0(z) + \sigma_0\, \epsilon  \, , 
\end{equation}
where $\langle \cdot, \cdot \rangle$ denotes the usual $L^2(\itT)$ inner product, 
$\beta_0 \in L^2(\itT)$ is the regression coefficient,  
$\eta_0: \itZ\to \real$  is an unknown smooth function, 
$\sigma_0 > 0$ is the unknown error scale parameter, and 
$\epsilon$ is independent of $(X,z)$. We assume that 
$\itT$ and $\itZ$ are compact intervals, and to simplify the 
notation, and without loss of generality, we will assume that $\itT=\itZ=[0,1]$. 
We allow the error distribution to have heavy tails by only requiring that 
$\epsilon$ have a symmetric distribution $G(\cdot)$ with scale parameter 
$1$.
Note that in order for $\eta_0$ to be identifiable we do
no include an intercept term in \eqref{eq:plm}.  
Just as in the finite-dimensional linear regression case, to obtain consistent robust estimators we 
need that  $\esp \|X\|^2<\infty$, where $\|X\|^2=\langle X, X\rangle$.  

To ensure that the regression coefficient $\beta_0$ in \eqref{eq:plm} is identifiable 
we will assume that the covariance operator $\Gamma$ of the stochastic process 
$X$ has infinite rank, i.e., that all its eigenvalues  $\lambda_1 \geq \lambda_2 \geq \dots$ 
are positive (Cardot \textsl{et al.}, 2003).  To see  that this condition is needed, let 
$\{\phi_k\}_{k\ge 1} \subset L^2(\itT)$ be the eigenfunctions of $\Gamma$, 
with corresponding eigenvalues $\{\lambda_k\}_{k\ge 1}$.  
The Karhunen-Lo\`{e}ve representation of $X$ is
$X=\mu + \sum_{k \ge 1} \xi_{k} \, \phi_k$, where $\mu=\esp (X)$ and the scores
$\xi_{k} \, = \, \langle X - \mu \, , \phi_k \, \rangle$ are uncorrelated
random variables with mean zero and variance $\lambda_k$. 
If $\Gamma$ has a null eigenvalue, then its kernel
 $\itN(\Gamma) \ne \{0\}$ (here $\itN(\Upsilon)$ 
denotes the kernel of the self-adjoint operator 
 $\Upsilon$, i.e., $\itN(\Upsilon)=\{x \in L^2(\itT), \Upsilon x = 0\}$). 
 Hence, for any 
$\alpha_0 \in \itN(\Gamma)$ we have $\var\left(\langle X - \mu \, , \alpha_0\, \rangle\right) =
\langle \Gamma \alpha_0, \alpha_0\rangle =0$. 
Thus, for all $\alpha_0 \in \itN(\Gamma)$, 
with probability one, 
$\langle X-\mu, \alpha_0 \rangle = 0$, so that
$\langle X-\mu, \beta_0 \rangle = \langle X-\mu,\beta_0 +\alpha_0\rangle$. 
This shows that when $\itN(\Gamma) \ne \{0\}$
the regression parameter in \eqref{eq:plm} is not identifiable. 

 To define the $B$-splines estimators, fix a desired
spline order $\ell$ and let 
$m_n^{(1)}$ and $m_n^{(2)}$ be the number of knots 
to be used to approximate 
$\beta_0$ and $\eta_0$, respectively. Recall that a spline of 
order $\ell$   is a polynomial of degree $\ell - 1$ within each  
subinterval. Then, the corresponding (normalized)
$B$-splines bases have dimensions 
$k_{n,\beta}=m_n^{(1)}+\ell$ and $k_{n,\eta}=m_n^{(2)}+\ell$,
respectively (see Corollary 4.10 of Schumaker, 1981). 
 Denote these bases by 
 $\{B_j^{(1)}: 1\leq j\leq k_{n,\beta}\}$ and $\{B_j^{(2)}: 1\leq j\leq k_{n,\eta}\}$, 
and to simplify the notation, denote their sizes with 
$p_1 = k_{n,\beta}$ and $p_2 = k_{n,\eta}$, respectively.
As usual when considering 
$B$-spline approximations, consistency results will 
be valid when $\eta_0, \beta_0  \in \itC^r([0,1])$, i.e. 
both functions $\eta_0 $ and $\beta_0$ are $r$-times continuously differentiable,
where $r\le \ell -2$ and $\ell$ is the spline order.
In particular, when cubic splines are considered, the results in 
Section  \ref{sec:consis} hold for twice continuously differentiable regression functions.

\subsection{$MM$-estimators with $B$-splines}{\label{sec:estimators}}

Robust $MM$-estimators (Yohai, 1987) are defined using two steps: first an initial robust (but possibly 
inefficient) regression estimator is used to compute a residual scale estimator, and then 
a regression $M$-estimator is calculated using 
a bounded loss function and standardized residuals. 
In what follows the loss function $\rho : \real \to \real_+$ 
will be assumed to satisfy the following
property: 
\begin{enumerate}[label=\textbf{R\arabic*}]
\item\label{ass:rho}: The function $\rho : \real \to [0, \infty)$ is
continuous, even, 
non-decreasing on $[0,+\infty)$, and such that $\rho(0)=0$. Moreover, $\lim_{u\to \infty} \rho(u)\ne 0$ and if $0 \leq u< v$ with $\rho(v) < \sup_u \rho(u)$ then $\rho(u)<\rho(v)$. When $\rho$ is bounded, we assume that $ \sup_u \rho(u)=1$.
Functions satisfying these conditions will be called $\rho$-functions (Maronna \textsl{et al.}, 2019). 
\end{enumerate}
A widely used family of $\rho$-functions is given by Tukey's bi-square:
$\rho_{\,\tuk,\,c}(t) =\min\left(1 - (1-(t/c)^2)^3, 1\right)$, where $c > 0$ is a tuning parameter 
that determines the robustness and efficiency 
properties of the associated estimators.

To define our estimators, for any vectors 
$\bb\in \real^{p_1}$ and $\ba\in \real^{p_2}$
let $r_i(\beta_{\bb}, \eta_{\ba})$, $1 \le i \le n$, 
 be the residuals with respect to the 
 corresponding spline approximations 
 $\beta_{\bb}(t)= \sum_{j=1}^{p_1} b_j \,   B_j^{(1)}(t)$ and
 $\eta_{\ba} (z)=\sum_{j=1}^{p_2} a_j \,   B_j^{(2)}(z)$:
\begin{equation} \label{eq:residuals}
r_i(\beta_{\bb}, \eta_{\ba}) \, =  \, y_i - \sum_{j=1}^{p_1} b_j \, x_{ij} - 
\sum_{j=1}^{p_2} a_j\, B_j^{(2)}(z_i) \, = \, y_i - \bb\trasp \bx_i - \ba\trasp \bB_i \, ,
\end{equation}
where $x_{ij}=\langle X_i,  B_j^{(1)}\rangle$, $\bx_i=(x_{i1}, \dots, x_{ip_1})\trasp$, 
and $\bB_i=( B_1^{(2)}(z_i), \dots,  B_{p_2}^{(2)}(z_i))\trasp$.

First, we compute an $S$-estimator of regression and its associated residual scale. 
Let $\rho_0$ be a bounded $\rho-$function and 
$s_n(\beta_{\bb}, \eta_{\ba} )$ be the $M$-scale estimator of the residuals given
as the solution to the following equation:
\begin{equation} \label{eq:s-est}
\frac{1}{n-(p_1+p_2)}\sum_{i=1}^n \rho_{0}\left(\frac{r_i(\beta_{\bb}, \eta_{\ba})}{s_n(\beta_{\bb}, \eta_{\ba})}\right) \, = \, b\, ,
\end{equation}
where $b = E( \rho_{0} (\epsilon) )$ (this is needed 
for the estimators to be consistent). 
Note that we use $1 / (n-(p_1+p_2))$ instead of $1/n$ in \eqref{eq:s-est} above 
to control the effect of a possibly large number of parameters 
($p_1 + p_2$) relative to the sample size  (see 
Maronna \textsl{et al.}, 2019). 
When $\rho_{0}$ is a Tukey's bisquare function, $\rho_{\,\tuk,\,c_0}$,
the choices $c_0 = 1.54764$ and $b = 1/2$ above yield a scale estimator  
that is Fisher-consistent when the errors have a normal distribution, 
and with a 50\% breakdown point in finite-dimensional regression models.

$S$-regression estimators are defined as the minimizers of 
the $M$-scale above:
\begin{equation} \label{eq:m-scale}
(\wbb_{\ini}, \wba_{\ini}) \ = \ \argmin_{\bb, \ba}  \, s_n(\beta_{\bb}, \eta_{\ba}) \, .
\end{equation}
The associated residual scale estimator is 
\begin{equation} \label{eq:scale_est}
\wsigma  \, =  
s_n(\beta_{\wbb_{\ini}}, \eta_{\wba_{\ini}}) \, ,
\end{equation}
 and the 
corresponding splines estimators are $
\wbeta_{\ini}(t) \,  = \, \beta_{\wbb_{\ini}}(t) \, = \, \sum_{j=1}^{p_1} \wb_{j,\ini} \,   B_j^{(1)}(t)$, and  $ \weta_{\ini}(z) \, = \, \eta_{\wba_{\ini}} (z)=\sum_{j=1}^{p_2} \wa _{j,\ini} \,   B_j^{(2)}(z) $.

Let $\rho_{1}$ be a $\rho-$function such that  $\rho_{1} \le \rho_{0}$ 
and $\sup_t\rho_1(t)=\sup_t\rho_0(t)$.
As it is well known, if $\rho_{0}=\rho_{\,\tuk,\,c_0}$ and $\rho_{1}=\rho_{\,\tuk,\,c_1}$,  
then $\rho_0\le \rho_1$ when $c_1>c_0$. We now compute an $M$-estimator using 
the residual scale estimator $\wsigma$ and 
the loss function  $\rho_{1}$:
\begin{equation}
(\wbb, \wba)  \ = \  \argmin_{\bb, \ba} \sum_{i=1}^n \rho_{1} \left ( \frac{r_i(\beta_{\bb}, \eta_{\ba}) }{\wsigma} \right )\,.
\label{eq:estfinitos}
\end{equation}
The resulting estimators of the regression function $\beta_0$ and the nonparametric component $\eta_0$ 
are given by
\begin{eqnarray}
\wbeta(t)= \sum_{j=1}^{p_1} \wb_j B_j^{(1)}(t) \, , \quad   \mbox{and }  \quad \weta(z)= \sum_{j=1}^{p_2} \wa_j B_j^{(2)}(z)  \, ,
\label{eq:estimadoresbeta}
\end{eqnarray}
where $\wbb = (\wb_1, \ldots, \wb_{p_1})\trasp$ and 
$\wba = (\wa_1, \ldots, \wa_{p_2})\trasp$.

 \subsection{Selecting the size of the $B$-spline bases}{\label{sec:BIC}}
 
The number of elements of the $B$-spline bases 
play the role of   regularization parameters in our estimation procedure, and it
 is useful to have a criterion to select them. Since standard 
 model selection methods can be highly affected by a small proportion of
 outliers, some robust alternatives have been proposed in the literature. 
 For linear regression models, see, for example, Ronchetti (1985) and
Tharmaratnam and Claeskens (2013). 

 Qingguo (2015) proposes using a criterion analogous to the 
 Schwarz information criterion (1978) (see also He \textsl{et al.}, 2002). However,
 the estimators of Qingguo (2015) do not take into account the residuals scale
 which is needed to determine which points are outliers according to the size of their residuals. 
Instead, we propose the following robust $BIC$-type criterion 
   \begin{eqnarray}
  \label{eq:bic1}
  RBIC(p_1, p_2) \ = \ \log \left (\wsigma^2 \sum_{i=1}^n \rho_1 \left(
      \frac{r_{i,p_1,p_2} }{\wsigma}
    \right) \right) +   \frac{\log n}{ n} \, (p_1+p_2) \,,
\end{eqnarray}
where $r_{i,p_1,p_2}=y_i-\langle X_i, \wbeta\rangle -\weta(z_i)$, $1 \le i \le n$, are the residuals obtained 
using bases of dimension $p_1$ and $p_2$ when computing $\wbeta$ and $\weta$, respectively, and 
$\wsigma$ is the corresponding $S-$scale. 
Note that when $\rho(x)=x^2$ the expression above reduces to the usual $BIC$ criterion. 

As is usual in spline-based procedures, in order to obtain an 
optimal rate of convergence, we let 
the number of knots increase slowly with the sample size. 
Theorem \ref{sec:consis}.2 below  shows  
that when $\beta_0$ and $\eta_0$ are twice continuously differentiable and
approximated with 
cubic splines ($\ell=4$), the rate for the 
size of the bases is almost $n^{1/5}$ (see also 
assumption \ref{ass:spacing}). Hence, 
a possible way to select $(p_1,p_2)$ is 
to search for the first local
minimum of $RBIC(p_1, p_2)$ in the range $\max(n^{1/5}/2, 4) \le p_j\le 8 + 2\, n^{1/5}$, $j=1,2$.
Note that for cubic splines the smallest possible number of knots is 4.

\subsection{Other functional regression models}{\label{sec:extensiones}}

\paragraph{Semi-functional models with varying coefficients} 
The estimators defined above can easily be extended to 
other semi-functional linear models, such as those involving varying coefficients. This 
extension will be relevant for our analysis of the Tecator data in
Section \ref{sec:ejemplo_tecator}. More specifically, consider the model 
$$
y_i \, = \, \gamma_0 + \langle X_i, \beta_0 \rangle + v_i \, \eta_0(z_i) + \sigma_0 \, \epsilon_i \, ,
$$
where $\gamma_0 \in \real$ is the intercept  
and $v_i \in \real$, $1\le i \le n$, is another explanatory variable. 
To define $MM$-estimators in this setting,  
for given bases dimensions $p_1$ and $p_2$,  define  the residuals  
as $r_i(\gamma,\beta_\bb, \eta_\ba)= y_i -\gamma- \sum_{j=1}^{p_1} b_j \, x_{ij} -
\sum_{j=1}^{p_2} a_j\, v_i B_j^{(2)}(z_i)= y_i-\gamma-\bb\trasp \bx_i -\ba \trasp (v_i\bB_i)$, where, 
 as before, $x_{ij}=\langle X_i,  B_j^{(1)}\rangle$. 
 The estimators are now defined as before, but now minimizing 
 over $(\gamma,\bb, \ba)\in \real^{1 + p_1+p_2}$ in 
 \eqref{eq:m-scale} and \eqref{eq:estfinitos}. 

\paragraph{Functional linear models}
Our proposal is also immediately applicable to functional linear models
such as $y_i=\langle X_i, \beta_0 \rangle + \bgama_0\trasp \bz_i +\sigma_0 \epsilon $, 
where $\bz_i$ are real-valued vectors of covariates. 
In this case we only need to set 
$\bx_i= (\langle X_i, B_1^{(1)}\rangle,\dots, \langle X_i, B_{p_1}^{(1)}\rangle, \bz_i)\trasp$ 
in the definition of the residuals $r_i( \beta_\bb, \eta_\ba)$ in equation \eqref{eq:residuals}. 
In the particular case that 
$\bgama_0=\bcero$, a robust estimator using the principal components basis
was given in Kalogridis and Van Aelst   (2019), while spline-based robust estimators  were 
studied in Maronna and Yohai (2013).

\paragraph{Monotone  components} 
 In  some applications, the non-parametric function $\eta_0$ or the regression function $\beta_0$ in the model \eqref{eq:plm} 
may be known to be monotone. In those cases it is preferable to take this information into account 
when computing the corresponding estimator. Neumeyer (2007) proposed the following 
method to construct such monotone 
estimators, which can be easily applied to our
$MM$-estimators based on splines. 
For any Lebesgue-measurable function $f : [a, b]\to \real$,   define the function $ \Upsilon(f) : \real \to \real$ as 
 $ \Upsilon(f) (u) =\int_a^b \indica_{\{f(z)\le u\}} dz+ a $, for any $u\in \real$, 
 where $\indica_{\cal A}$ denotes the indicator function of the set ${\cal A}$. 
Note that $\Upsilon(f)\indica_{[f(a),f (b)]}$ is the inverse of $f$ when $f$ is strictly increasing, 
and its generalized inverse $f^{-1}(u) = \inf\{z: 
f(z) > u\}$ when $f$ is non-decreasing. Furthermore, 
$\Upsilon(f)$ is always increasing and  Lebesgue-measurable. Given any function  $\eta : [0, 1] \to \real$, Neumeyer (2007) considered the   increasing modification $\eta_{\monmod} : [0, 1]\to \real$ as 
$$\eta_{\monmod}=\Upsilon\left( \Upsilon(\eta)\indica_{[\eta(0),\eta(1)]}\right) \indica_{[0,1]}\,,$$
which satisfies $\eta_{\monmod}=\eta$ for any non-decreasing function $\eta$. Hence, based on the estimators of 
$\eta_0:[0,1]\to \real$ defined in \eqref{eq:estfinitos} and \eqref{eq:estimadoresbeta}, 
a monotone estimator may be constructed as
\begin{equation}
\label{eq:etamono}
\weta_{\monmod}=\Upsilon\left( \Upsilon(\weta)\indica_{[\weta(0),\weta(1)]}\right) \indica_{[0,1]}\,.
\end{equation} 
These modified robust estimators are also strongly consistent (see Corollary \ref{sec:consis}.1) 
and have very good finite-sample properties (see Section  \ref{sec:simu}).

 \section{Consistency results}{\label{sec:consis}}

 In this section we prove that if conditions 
\ref{ass:psi} - \ref{ass:probaX*} below hold, then 
the estimators in \eqref{eq:estfinitos} are strongly consistent.  
We will assume that: 
 
\begin{enumerate}[label=\textbf{R\arabic*}]
\setcounter{enumi}{1}
 \item\label{ass:psi}: The function $\rho$ is differentiable with bounded derivative $\psi$, such that $\zeta(u)=u\psi(u)$ is bounded.
\end{enumerate}
\vspace*{-.3in}
\begin{enumerate}[label=\textbf{C\arabic*}]
 
\item\label{ass:densidad}: The random variable $\epsilon$ has a density function $g_0(t)$ that is even, non-increasing in $|t|$, and strictly decreasing for $|t|$ in a neighbourhood of $0$.  

\item \label{ass:probaX}: For almost any $z_0$,  $\prob(\langle X,\beta\rangle=a|z=z_0)<1$, for any $\beta\in L^2(0,1)$, and $a\in \real$,  $(\beta,a)\ne 0$.

\item\label{ass:beta0-eta0}:  The  true functions $\beta_0$ and $\eta_0$  are such that $\beta_0\in  \itC^r([0,1])$ and $\eta_0\in  \itC^r([0,1])$. Furthermore, their $r-$th derivative satisfies a Lipschitz condition on $[0, 1]$, with  $r\geq 1$, that is, $\eta_0, \beta_0 \in \itL_r([0,1])$ where
\begin{multline*}
\itL_r ([0,1]) \, = \, \biggl\{ g\in C^r \left([0, 1] \right):
  \big \|g^{(j)} \big \|_\infty<\infty, \;0\leq j\leq r,  \mbox{ and } 
  \sup_{z_1\ne z_2} \frac{\big |g^{(r)}(z_1) - g^{(r)}(z_2) \big|}{|z_1- z_2|}
  <\infty \biggr\} \, .
  \end{multline*}
Recall that for any continuous function $v:\real\to \real$, $\|v\|_{\infty}= \sup_{t}|v(t)|$.

\item\label{ass:spacing}: The smoothing parameters $p_1=k_{n,\beta}$ and $p_2=k_{n,\eta}$ are assumed to be of order $O(n^{\nu})$,   $ 0<\nu <1/(2r)$. 
Moreover, the   ratio of maximum and minimum spacings of knots is uniformly bounded.

\item \label{ass:probaX*}: There exists $0<c<1$ such that  $\prob(\langle X,\beta\rangle+ \eta(z)=0)< c$, for any $\beta\in   \itL_1([0,1])$, $\eta \in \itL_1 ([0,1])$,  $(\beta,\eta)\ne 0$.
\end{enumerate}
These conditions are discussed in more detail in Section \ref{sec:comentarios} below. 
 
Our first result shows the strong consistency of the scale estimators
$\wsigma=s_n(\wbeta_{\ini}, \weta_{\ini})=s_n(\beta_{\wbb_{\ini}}, \eta_{\wba_{\ini}})$
 defined in \eqref{eq:scale_est}. 
Let $S(\beta, \eta)$ be the $M$-scale functional related to the residuals $r (\beta,\eta)=y-\langle X, \beta\rangle -\eta(z)$ that is, $S(\beta, \eta)$ satisfies
$$\esp \rho_{0}\left(\frac{r(\beta,\eta)}{S(\beta, \eta)}\right)=b\,.$$
For simplicity, we will assume that $b$ has been chosen so that 
$\esp \rho_{0}(\epsilon)=b$. In this case we have that
$\sigma_0=S(\beta_0, \eta_0)=\argmin S(\beta, \eta)$, and
the scale estimators are strongly consistent.

\noi \textbf{Proposition \ref{sec:consis}.1.} \textsl{Assume that the                                                                                                                                                                                                                                                                                                                                                                                                                                                                                                                                                  function $\rho_{0}$ is bounded such that $\|\rho_0\|_{\infty}=1$ and satisfies \ref{ass:rho} and \ref{ass:psi}. Then, if $\esp\left(\|X\| \right)<\infty$ and    \ref{ass:densidad},   \ref{ass:beta0-eta0} and \ref{ass:spacing} hold,  we have that $\wsigma \convpp  \sigma_0=S(\beta_0, \eta_0)$. }

 Theorem \ref{sec:consis}.1 states the main result in this section, that is, the uniform strong consistency of the proposed estimators. 
 
\noi \textbf{Theorem \ref{sec:consis}.1.} \textsl{Let $\rho_1$ be a bounded  function with $\|\rho_1\|_{\infty}=1$ satisfying   \ref{ass:rho} and \ref{ass:psi}. Furthermore, let
$$M(\beta,\eta, \sigma)=\esp \,\rho_1\left(\frac{y - \langle X, \beta\rangle -\eta(z)}{\sigma}\right) \, , $$ 
where $M(\beta_0,\eta_0,\sigma_0)= b_{\rho_1}<1$. Assume that  \ref{ass:densidad}  to \ref{ass:spacing} hold, $\esp\|X\|^2<\infty$ 
and that  \ref{ass:probaX*} holds with $c<1-b_{\rho_1}$. 
If, in addition, $\wsigma \convpp  \sigma_0$, 
then $\|\wbeta -\beta_0\|_{\infty}+\|\weta -\eta_0\|_{\infty}\convpp 0$.
}

\vskip0.1in

The following corollary shows that this consistency is maintained when $\eta_0$ is monotone and the estimator
$\weta$ is modified as described in Section \ref{sec:extensiones}. This is 
a direct consequence of Theorem \ref{sec:consis}.1 above and Theorem 3.1 in  Neumeyer (2007). 
 
\noi \textbf{Corollary \ref{sec:consis}.1.} \textsl{Let $\rho_1$ be a bounded  function with $\|\rho_1\|_{\infty}=1$ satisfying   \ref{ass:rho} and \ref{ass:psi}. Assume that  \ref{ass:densidad}  to \ref{ass:spacing} hold,  $\esp\|X\|^2<\infty$    and that  \ref{ass:probaX*} holds with $c<1-b_{\rho}$ and $b_{\rho}=M(\beta_0,\eta_0,\sigma_0) <1 $. 
 Then, if $\weta_{\monmod}$ is the monotone modified estimator in \eqref{eq:etamono}
based on $\weta$ and 
$\wsigma \convpp  \sigma_0$, we have that  $ \|\weta_{\monmod} -\eta_0\|_{\infty}\convpp 0$.}

\subsection{Rates of Consistency}                
                                     
In this section we find the rate of convergence of the proposed estimators
when  measuring the distance between two pairs of functions
$\theta_1 = (\beta_1, \eta_1)$ and $\theta_2 = (\beta_2, \eta_2)$  through the mean square error of their prediction differences, that is, 
through $\pi^2( \theta_1, \theta_2) = \esp \left[\langle X, \beta_1-\beta_2\rangle +\eta_1(z)-\eta_2(z)\right]^2$.  Note that 
when \ref{ass:probaX} holds $\pi$ is indeed a distance.
 
For that purpose, we will need the following additional assumption,  where, as above, for any vectors of coefficients $\bb\in \real^{p_1}$ and $\ba\in \real^{p_2}$, 
we write 
$\beta_{\bb}(t)= \sum_{j=1}^{p_1} b_j \,   B_j^{(1)}(t)$,
 and  $\eta_{\ba} (z)=\sum_{j=1}^{p_2} a_j \,   B_j^{(2)}(z)$. Conditions ensuring that \ref{ass:cotainf} holds are given in Lemma \ref{sec:dem-teo-32}.1  in the Appendix.
\begin{enumerate}[label=\textbf{C\arabic*}] 
\setcounter{enumi}{5}
\item\label{ass:cotainf} There exists a neighbourhood $\itV$ of $\sigma_0$
with closure  $\overline{\itV}$ strictly included in $ (0,\infty)$, and 
constants $\epsilon_0>0$ and $C_0 > 0$, such that  
$M(\theta, \sigma)-M(\theta_0, \sigma)\ge C_0\,\pi^2(\theta,\theta_0)$, for any  $\theta=(\beta_{\bb}, \eta_{\ba})$ such that $\|\beta_{\bb} -\beta_0\|_{\infty}+\|\eta_{\ba} -\eta_0\|_{\infty}\le  \epsilon_0$ and any $\sigma\in \itV$.
\end{enumerate}

\noi \textbf{Theorem \ref{sec:consis}.2.} \textsl{Let $\rho_1$ be a bounded  function with $\|\rho_1\|_{\infty}=1$ satisfying   \ref{ass:rho} and \ref{ass:psi}. Assume that  \ref{ass:densidad}  to \ref{ass:spacing} hold,  
that $\esp\|X\|^2<\infty$, that  \ref{ass:probaX*} holds with $c<1-b_{\rho}$ and $b_{\rho}=M(\beta_0,\eta_0,\sigma_0) <1 $,  
 that \ref{ass:cotainf} holds, and  that $\wsigma \convpp  \sigma_0$. Let 
$\gamma_n$ be any sequence that satisfies $\gamma_n = O(n^{r \nu})$ and 
$\gamma_n \sqrt{\log(\gamma_n)} = O(n^{(1 - \nu)/2})$, then 
$\gamma_n\;\pi(\wtheta,  \theta_0)=O_\prob(1)$, where $\wtheta=(\wbeta, \weta)$. Hence, if $\nu=1/(1+2r)$ in \ref{ass:spacing}, 
one  can choose 
$\gamma_n = O(n^{r/(1+2r)}/\sqrt{\log(n)})$ or $\gamma_n = O(n^{r/(1+2r) - \delta})$, 
for $\delta > 0$ arbitrarily small, 
where the latter yields
a convergence rate, in terms of the prediction distance $\pi$, that is arbitrarily close to the optimal one.}

\noi  \textbf{Remark \ref{sec:consis}.2.} This theorem allows us to derive the 
order of convergence of $\|\weta  -\eta_0 \|_{\infty}$, when  
  $X$ and $z$ are independent and $\esp(X)=0$,  as follows. Note   that 
$$\pi^2(\theta_1,  \theta_2)= \esp \left[\langle X, \beta_1-\beta_2\rangle +\eta_1(z)-\eta_2(z)\right]^2= \esp \left[\langle X, \beta_1-\beta_2\rangle\right]^2 +\esp\left[\eta_1(z)-\eta_2(z)\right]^2\,.$$
Hence,  from Theorem \ref{sec:consis}.2, we get that $\gamma_n^2 \esp\left[\weta (z)-\eta_0(z)\right]^2=O_\prob(1)$. Furthermore, from the proof of Theorem \ref{sec:consis}.2,   there exists $ \wteta(z)=\sum_{j=1}^{p_2} \widetilde{a}_j \,   B_j^{(2)}(z)  $   such that
 $\|\wteta-\eta_0\|_{\infty}=O(n^{-r\,\nu})$ and $\gamma_n^2 \esp\left[\weta (z)-\wteta(z)\right]^2=O_\prob(1)$. Using that $\weta (z)-\wteta(z)\in \itM_{p_2}^{(2)}$ and Lemma 7 of Stone (1986), we obtain that, for some positive constant $A>0$ independent of the sample size, $\|\weta  -\wteta \|_{\infty}^2\le A p_2 \esp\left[\weta (z)-\wteta(z)\right]^2  $ which entails that
 $ p_2^{-1/2} \gamma_n \|\weta  -\wteta \|_{\infty} =O_\prob(1)$. Assume now that $\nu=1/(1+2r)$ and $r\ge 1$ and  
 $\gamma_n = O(n^{r/(1+2r) - \delta})$, for $0<\delta< (r-1/2)/(1+2r)$. Taking into account that $p_2=O(n^\nu)$, we conclude that 
 $ n^{\omega}  \|\weta  -\wteta \|_{\infty} =O_\prob(1)$, with $\omega=(r-1/2)/(1+2r)-\delta$, leading to $ n^{\omega}  \|\weta  -\eta_0 \|_{\infty} =O_\prob(1)$. When $\eta_0$ is monotone, this rate is also inherited by the 
 monotone modification $\weta_{\monmod}$.

\noi  \textbf{Remark \ref{sec:consis}.3.}  It is worth mentioning that 
consistency and rates of convergence for the $MM$-estimators defined in 
Section \ref{sec:extensiones} for the varying coefficients 
model $y_i=\gamma_0+\langle X_i, \beta_0 \rangle + v_i \eta_0(z_i)+\sigma_0 \epsilon_i$
can be derived similarly when $\esp [v^2]<\infty$.
 
\subsection{Comments on assumptions \ref{ass:psi} and \ref{ass:densidad} to \ref{ass:probaX*}}{\label{sec:comentarios}}
Assumption \ref{ass:psi} is an additional smoothness condition on the function $\rho$ which is standard 
in the robustness literature.
Assumptions \ref{ass:densidad} to \ref{ass:spacing} refer to the error distribution (to ensure Fisher-consistency), to the smoothness of the regression parameter and the nonparametric component, as well as to the order at which  the dimension of the bases increase. These  assumptions are  standard when using spline approximations. Assumption \ref{ass:probaX} guarantees that $(\beta_0,\eta_0)$ are the unique minimizers of $M(\beta, \eta,\sigma)$ (see Lemma \ref{sec:appen}.1), which is a standard condition needed to obtain consistent regression estimators. Furthermore,   \ref{ass:probaX*} is the functional version of assumption (A.3) in Yohai (1987) adapted to partial linear models.

 A sufficient condition for \ref{ass:probaX*} to hold is that $\prob(\langle X,\beta\rangle+ \eta(z)=0)=0$, for any $\beta\in  L^2(0,1)$,  $\eta \in \itL_1 ([0,1])$,  $(\beta,\eta)\ne 0$. Hence, it is necessary that the kernel of the covariance operator of $X$ be equal to $\{0\}$. Specifically, the Karhunen-Lo\`{e}ve expansion of $X$ cannot have finitely many terms. 
Note that when the covariance operator, $\Gamma$, of $X$ has finite rank $k$, then $\prob\left(\langle X-\esp(X), \phi_j\rangle =0\right)=1$, for $j>k$, where $\phi_j$, $j \ge 1$ are the eigenfunctions of $\Gamma$ associated to the $j$-th eigenvalue $\lambda_j$, with $\lambda_1\ge \lambda_2\ge \dots$,  which implies that \ref{ass:probaX*} does not hold. Furthermore,  $\beta_0$ is not identifiable since $\beta_0+\phi_j$ with $j > k$ also satisfies model \eqref{eq:plm}.
 
Denote as $\Gamma_{z_0}$ the covariance operator of $X|z=z_0$, that is, 
 $$\Gamma_{z_0}= \esp \left\{\left[X-\esp(X|z=z_0)\right]\otimes\left[X-\esp(X|z=z_0)\right]|z=z_0\right\}\,.$$
  Then,  assumptions \ref{ass:probaX}  and   \ref{ass:probaX*} hold  when, for almost all $z_0$, the kernel $ \itN(\Gamma_{z_0})$ of  $\Gamma_{z_0}$ equals $\{0\}$  which is analogous to the 
  assumptions of Huang \textsl{et al.} (2015). 
  To see this, assume that $ \itN(\Gamma_{z_0})=\{0\}$ and denote as $\mu_{z_0}=\esp(X|z=z_0)$, then $\Gamma_{z_0}= \esp \left\{\left[X-\mu_{z_0}\right]\otimes\left[X-\mu_{z_0}\right]|z=z_0\right\}$. We will show that  \ref{ass:probaX} holds. Note that $\langle \beta, \Gamma_{z_0} \beta\rangle= \esp \left[\left(\langle \beta, X-\mu_{z_0}\rangle\right)^2 \;|z=z_0\right]$ so that $\langle \beta, \Gamma_{z_0} \beta\rangle=0$ if and only if  $\prob\left(\langle \beta, X-\mu_{z_0}\rangle=0 \;|z=z_0\right)=1$. Assume that \ref{ass:probaX} does not hold, then there exists $\beta\in L^2(0,1)$  and $a\in \real$,  $(\beta,a)\ne 0$ such that $\prob(\langle X,\beta\rangle=a|z=z_0)=1$. Hence, in particular, we have that $a=\langle \mu_{z_0},\beta\rangle$, so that  $\prob(\langle X-\mu_{z_0},\beta\rangle=0|z=z_0)=1$ implying that $\langle \beta, \Gamma_{z_0} \beta\rangle=0$. Thus, using that  $\Gamma_{z_0}$ is a linear, self-adjoint and compact operator with finite trace, we obtain that $ \Gamma_{z_0}^{1/2} \beta=0$, so $\beta \in \itN(\Gamma_{z_0})$, 
  which implies that $\beta=0$ and $a=0$, and we reach a contradiction. Similar arguments 
  show that  \ref{ass:probaX*} holds. Hence, assumptions \ref{ass:probaX}  and   \ref{ass:probaX*} are weaker than requiring $ \itN(\Gamma_{z_0})=\{0\}$.  It is worth noticing that, if $\prob(\langle X,\beta\rangle=a|z=z_0)=0$, for any $\beta\in L^2(0,1)$ and $a\in \real$ with $(\beta,a)\ne 0$, then $ \itN(\Gamma_{z_0})=\{0\}$.

\section{Simulation study}{\label{sec:simu}}

\begin{figure}[tp] 
  \centering 
  \subfigure[Regression parameter $\beta_0$. ]
    {\includegraphics[width=0.4\textwidth]{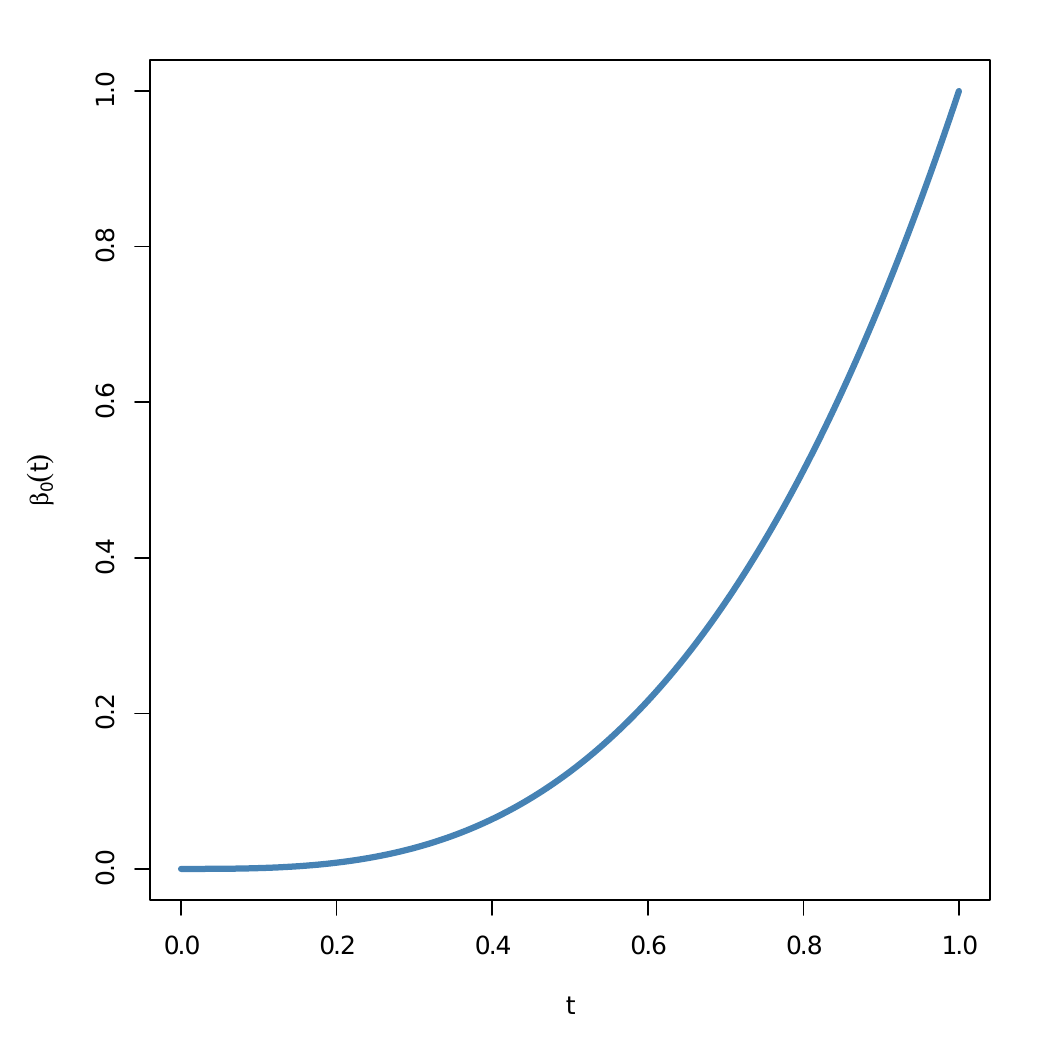}}
   \subfigure[Nonparametric component $\eta_0$.] 
  {\includegraphics[width=0.4\textwidth]{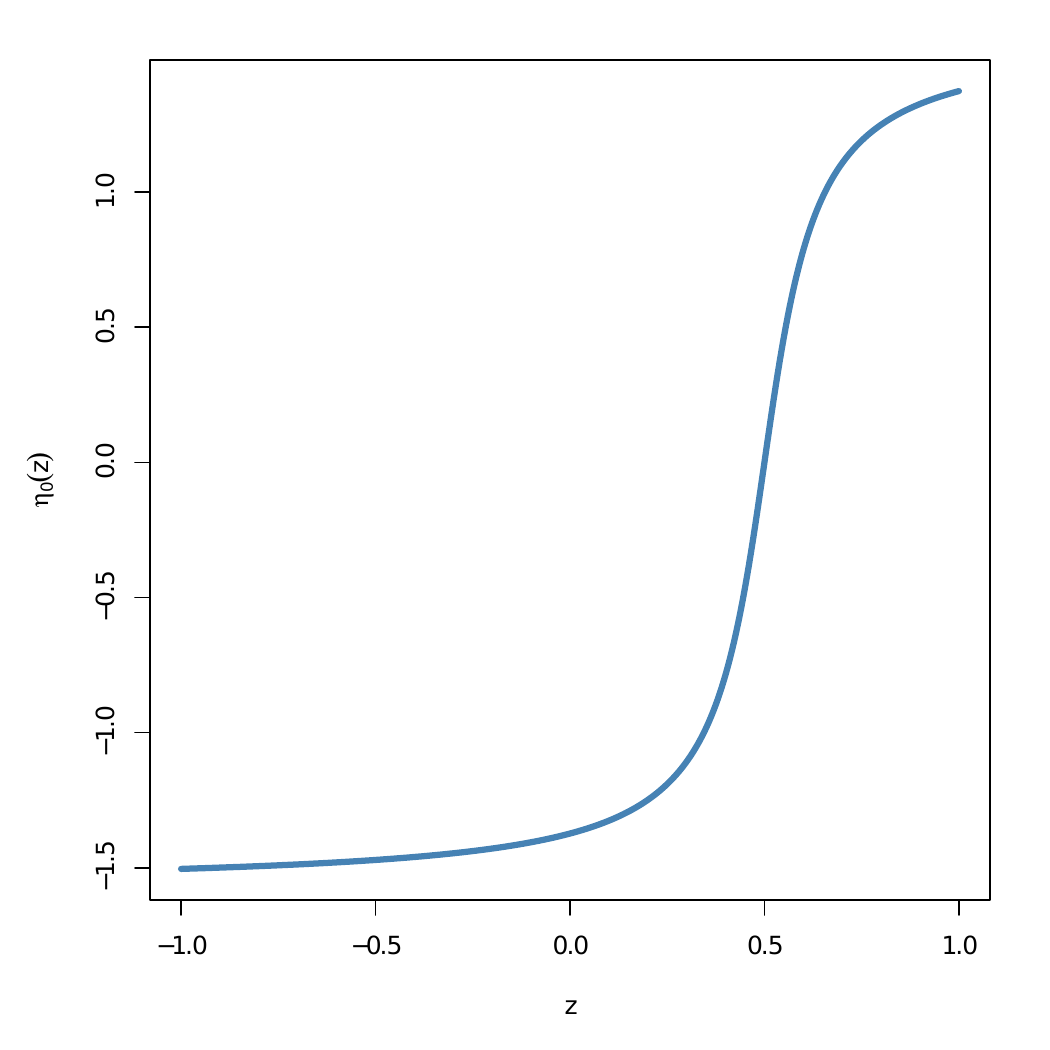}} 
   \caption{\small \label{fig:fplm-parametros-verdaderos} True parameters.}
\end{figure}

We performed a Monte Carlo study to investigate the finite-sample properties 
of our proposed estimators for the semi-functional linear regression model:
\begin{equation}
  \label{eq:monte-carlo-modelo}
  y_i = \langle \beta_0, X_i\rangle \, + \, \eta_0(z_i) \, + \, \sigma_0\,\epsilon_i \, , \quad i=1, \ldots, n \, , 
\end{equation}
with $\sigma_0=1$, $\itT=[0,1]$ and $z_i\sim \itU(-1,1)$.
The model parameters were 
$\eta_0(z) = 3 \arctan \left(10  (z -  0.5) \right)$, $\beta_0(t) = \sum_{j=1}^{50} b_{j,0} \phi_j$, 
the basis $\phi_1(t) \equiv 1$, $\phi_j(t) = \sqrt{2} \cos ((j-1)\pi t)$, $j\geq 2$, and  
the coefficients $b_{1,0} = 0.3$ and $b_{j,0} = 4(-1)^{j+1}j^{-2}$, $j \geq 2$.
The process that generates the functional covariates $X_i(t)$ was Gaussian with mean 0 and 
covariance operator 
with eigenfunctions  $\phi_j(t)$. 
For uncontaminated samples the scores  $\xi_{ij} $ were independent Gaussian 
random variables $\xi_{ij}\sim N(0,j^{-2})$, and the errors $\epsilon_i \sim N(0,1)$, 
independent from $z_i$ and $X_i$. Taking into account that $\var(\xi_{ij})\le 1/2500$ when $j > 50$, the process was approximated numerically using the first 50 terms of its
 Karhunen-Lo\`{e}ve representation. 
Figure \ref{fig:fplm-parametros-verdaderos} shows the functions  $\beta_0$ and $\eta_0$.

We compared three estimators: the classical procedure
based on least squares (\textsc{ls}), the $M$-estimators proposed by 
Huang \textsl{et al.} (2015) (\textsc{m}), and the $MM$-estimators 
(\textsc{mm}) from Section \ref{sec:estimators}. 
Since the true function 
$\eta_0$ is monotone, we also 
included the modification, $\weta_{\monmod}$, based on 
Neumeyer (2007).  
As in Huang \textsl{et al.} (2015), $M$-estimators were computed using a Huber function with tuning constant $1.345$ and no scale estimator. For the $MM$-estimators we used   a bounded $\rho-$function  $\rho_0$ to compute the initial $S$-estimators and residual  scale in \eqref{eq:m-scale}  and also a bounded $\rho_1$ for the $M$-step \eqref{eq:estfinitos}. For $j=0,1$, we choose  $\rho_j= \rho_{\,\tuk,\,c_j}$, the bisquare function, with tuning constants $c_0=1.54764$ ($b=1/2$) and  $c_1=3.444$. All calculations were performed in \verb=R=. The
 code and scripts reproducing the examples in this paper are publicly available on-line
 at \url{https://github.com/msalibian/RobustFPLM}.

For each setting 
we generated $n_R = 500$ samples of size $n = 300$ and 
used cubic splines with equally spaced knots. For the robust $MM$-estimators we  
selected the size of the spline bases ($p_1=k_{n,\beta}$ and  $p_2=k_{n,\eta}$)
by minimizing $RBIC(p_1, p_2)$ in equation \eqref{eq:bic1}
over the 2-dimensional grid $4 \leq p_1, p_2 \leq 13$.  
For the least squares estimator we used the standard BIC criterion, and
for the $M$-estimator we used the 
criterion proposed in 
Huang \textit{et al.} (2015).

To evaluate the performance of each estimator we looked at their
integrated squared bias and mean integrated squared error. These were
computed on a grid of $M = 100$ equally spaced points on $[0, 1]$ and $[-1,1]$, 
for $\wbeta$ and $\weta$, respectively. More specifically, if $\wgamma_j$
is the estimate of the function $\gamma$ obtained with the $j$-th
sample ($1 \le j \le n_R$), we compute
$$
 \mbox{Bias}^2(\wgamma)    =    \frac 1M \sum_{s=1}^M
 \left( \frac{1}{n_R} \sum_{j=1}^{n_R} \wgamma_j(t_s) - \gamma (t_s) \right )^2 \,, 
$$
and
$$
\mbox{MISE}(\wgamma)    =    \frac 1M \sum_{s=1}^M  \frac{1}{n_R} \sum_{j=1}^{n_R} \left( \wgamma_j(t_s) - \gamma (t_s) \right )^2 \,,
$$
where $t_1\le \dots \le t_M$ are equispaced points on the domain $\itI$ of $\gamma$. 
These are numerical approximations to 
$$
\int_\itI \biggl(\frac{1}{n_R} \sum_{j=1}^{n_R} \wgamma_j(t) - \gamma (t) \biggr)^2 dt \, , 
$$
and
$$
\frac{1}{n_R} \, \sum_{j=1}^{n_R} \int_{\itI} \biggl(\wgamma_j(t)-\gamma(t)\biggr)^2 \,dt \, ,
$$
respectively.
To alleviate the concern that $\mbox{Bias}^2$ and $\mbox{MISE}$ may be heavily
influenced by numerical errors at or near the boundaries of the grid, 
we follow He and Shi (1998) and also consider trimmed versions of the above computed
without the $q$ first and last points on the grid:
 \begin{eqnarray*}
 \mbox{Bias}_{\trim}^2(\wgamma)  & = &  \frac 1{M-2q} \sum_{s=q+1}^{M-q}
 \left( \frac{1}{n_R} \sum_{j=1}^{n_R} \wgamma_j(t_s) - \gamma (t_s) \right )^2 \, ,
 \\
 \mbox{MISE}_{\trim}(\wgamma)  & = &  \frac 1{M-2q} \sum_{s=q+1}^{M-q}  
 \frac{1}{n_R} \sum_{j=1}^{n_R} \left( \wgamma_j(t_s) - \gamma (t_s) \right )^2 \, ,
 \end{eqnarray*}
 We chose $q=[M\times 0.05]$ which uses the central 90\% interior points in the grid. 
 Table \ref{tab:tabla-2-neu-Bspl}  reports the  squared bias and   $\mbox{MISE}$ and 
 their trimmed counterparts for samples without outliers. 
 We note that the boundary effect is more pronounced for the 
 estimators of $\beta_0$, but it is present for $\weta$ as well.  
 Based on this observation, in what follows,  we report the trimmed measures. 
 \begin{table}[ht!]
  \centering
   \renewcommand{\arraystretch}{1.2}
\begin{tabular}{r rr   rr  rr}
\hline  
&\multicolumn{2}{c}{$\wbeta$} &\multicolumn{2}{ c}{$\weta$} &\multicolumn{2}{ c}{$\weta_{\monmod}$}\\
\hline
& Bias$^2$ & MISE & Bias$^2$ & MISE & Bias$^2$ & MISE \\
\hline
\textsc{ls} & 0.0126 & 0.1528 & 0.0197 & 0.0778 & 0.0221 & 0.0531 \\
\textsc{m} & 0.0123 & 0.1544 & 0.0248 & 0.0850 & 0.0261 & 0.0582 \\
\textsc{mm} & 0.0121 & 0.2039 & 0.0175 & 0.0871 & 0.0216 & 0.0593 \\ \hline  
& Bias$^2_{\trim}$ & MISE$_{\trim}$ & Bias$^2_{\trim}$ & MISE$_{\trim}$ & Bias$^2_{\trim}$ & MISE$_{\trim}$ \\
\hline
\textsc{ls} & 0.0018 & 0.0865 & 0.0194 & 0.0615 & 0.0185 & 0.0442 \\
\textsc{m} & 0.0018 & 0.0869 & 0.0242 & 0.0680 & 0.0226 & 0.0493 \\
  \textsc{mm} & 0.0017 & 0.1215 & 0.0173 & 0.0674 & 0.0171 & 0.0479 \\
\hline  
\end{tabular}
\caption{ \small \label{tab:tabla-2-neu-Bspl} 
    Integrated squared bias  and mean integrated squared errors over $n_R = 500$ clean samples of size $n = 300$. The top 
    3 rows report the Monte Carlo estimates of these measures using a grid of 100 equispaced points, and the bottom three
    correspond to their trimmed versions using the 90\% inner grid points. }
\end{table}

We considered two contamination scenarios.
The first one contains outliers in the response variables and is expected to affect 
mainly the  estimation of $\eta_0$. The second one
includes high-leverage  outliers in the functional explanatory variables, 
as in the Tecator example (see Section \ref{sec:ejemplo_tecator}), which 
typically affect the estimation of the 
linear regression parameter $\beta_0$. 
Specifically, we constructed our samples as follows:
\begin{itemize}
\item Scenario $C_{1,\mu}$: here only the regression errors are contaminated
in order to produce ``vertical outliers''. Their distribution $G$ is given by  
$G(u)=0.9 \, \Phi(u)+0.1\, \Phi\left((u-\mu)/0.5\right)$, with $\Phi$ the 
standard normal distribution function.

\item Scenario $C_{2,\mu}$: in these settings we introduce high-leverage outliers 
by contaminating the functional 
covariates $X_i$ and the errors simultaneously. 
Outliers in the $X_i$'s are generated 
by perturbing the distribution of the second score 
in the Karhunen-Lo\`{e}ve representation of the process. 
Specifically, 
we sample $v_i\sim Bi(1, 0.10)$ and then:
  \begin{itemize}
  \item if $v_i = 0$, let $\epsilon_i^{(c)}=\epsilon_i$ and $X_i^{(c)}=X_i$;
  \item if $v_i=1$, let $\epsilon_i^{(c)} \sim N( \mu, 0.25)$ and
    $X_i^{(c)}=\sum_{j=1}^{50} \xi_{ij}^{(c)} \phi_j(t)$, with
    $\xi_{ij}^{(c)}\sim N(0,j^{-2})$ for $j\ne 2$ and
    $\xi_{i2}^{(c)}\sim N(\mu/2, 0.25) $. 
   \end{itemize} 
 The responses are generated as
  $y_i^{(c)} = \langle \beta_0, X_i^{(c)}\rangle + \eta_0(z_i) +
  \epsilon_i^{(c)}$. 
 \end{itemize}
 
\begin{figure}[ht!]
  \begin{center} 
    \subfigure[Some trajectories $X_i(t)$ under $C_0$.]{
      \includegraphics[width=0.45\textwidth]{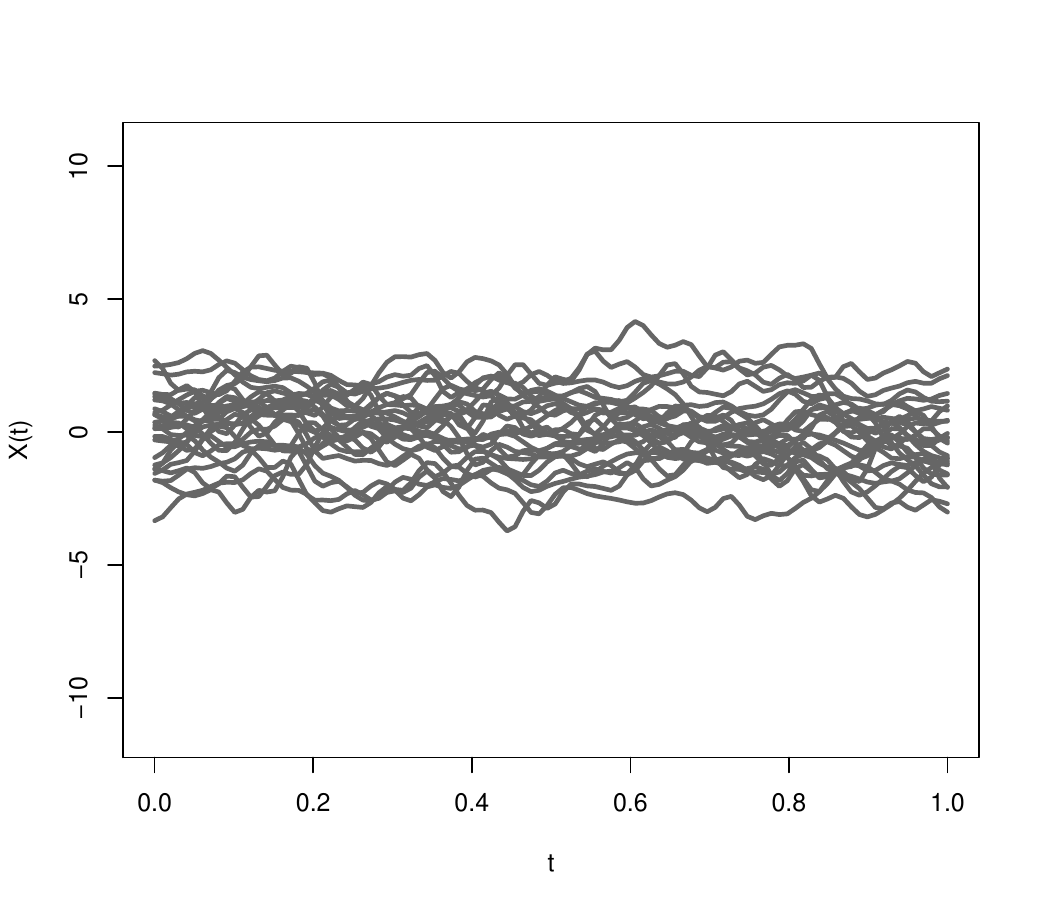}}
     \subfigure[Some trajectories $X_i(t)$ under $C_{2,12}$.]{
      \includegraphics[width=0.45\textwidth]{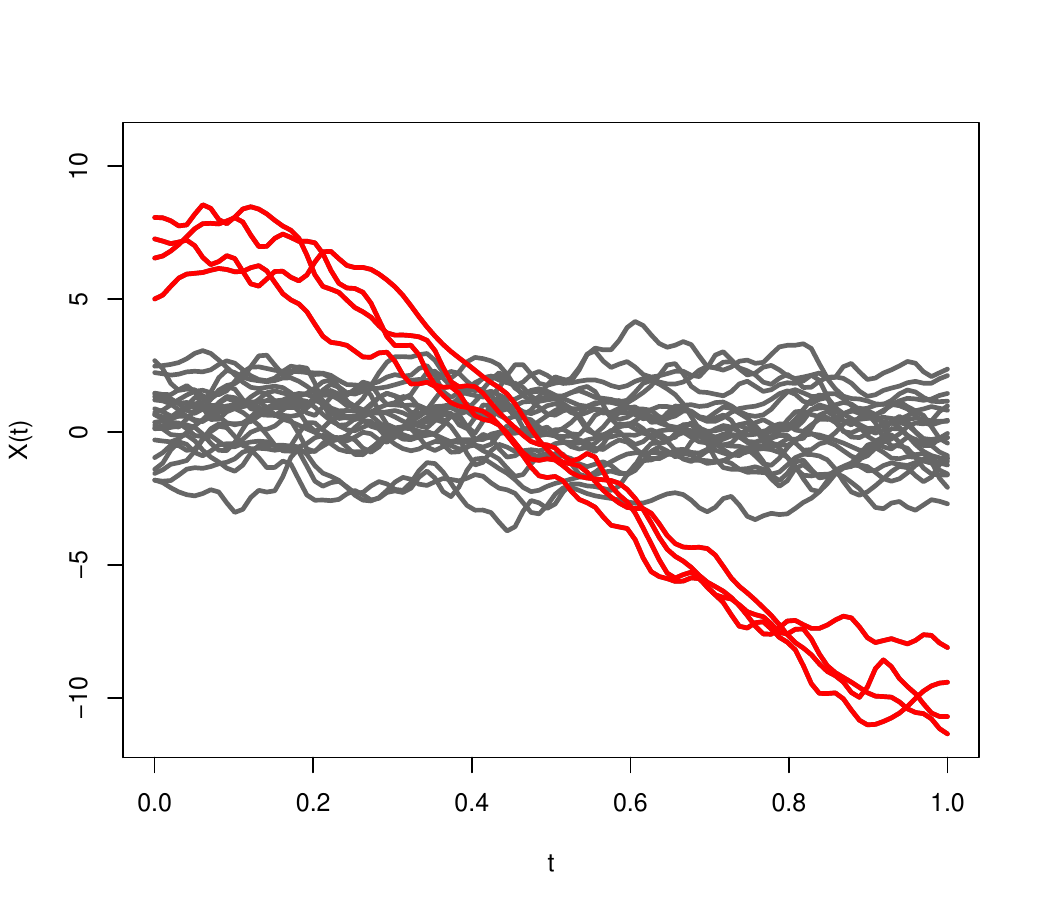}}
   \end{center}
  \caption{\small \label{fig:fplm-trayectoria} 25 trajectories $X_i(t)$ with and without contamination.}
\end{figure}

Both contamination settings above depend on the parameter $\mu \in \real$. 
In this experiment we looked at the following values of $\mu$: 
8, 10, 12, 14 and 16. They produce a range of contamination 
scenarios ranging from mild to severe. 
As an illustration of the type of outliers generated with the second setting above, 
Figure \ref{fig:fplm-trayectoria} shows 25 randomly chosen functional covariates $X_i(t)$, 
for one sample generated under $C_0$ (with no outliers) and one 
obtained under $C_{2,12}$.

\begin{figure}[tp]
 \begin{center}
 \newcolumntype{M}{>{\centering\arraybackslash}m{\dimexpr.05\linewidth-1\tabcolsep}}
   \newcolumntype{G}{>{\centering\arraybackslash}m{\dimexpr.4\linewidth-1\tabcolsep}}
\begin{tabular}{M GG}
 & {\small   $C_{1,\mu}$} & {\small   $C_{2,\mu}$}   \\[-4ex]
 $\wbeta$ &  \includegraphics[scale=0.38]{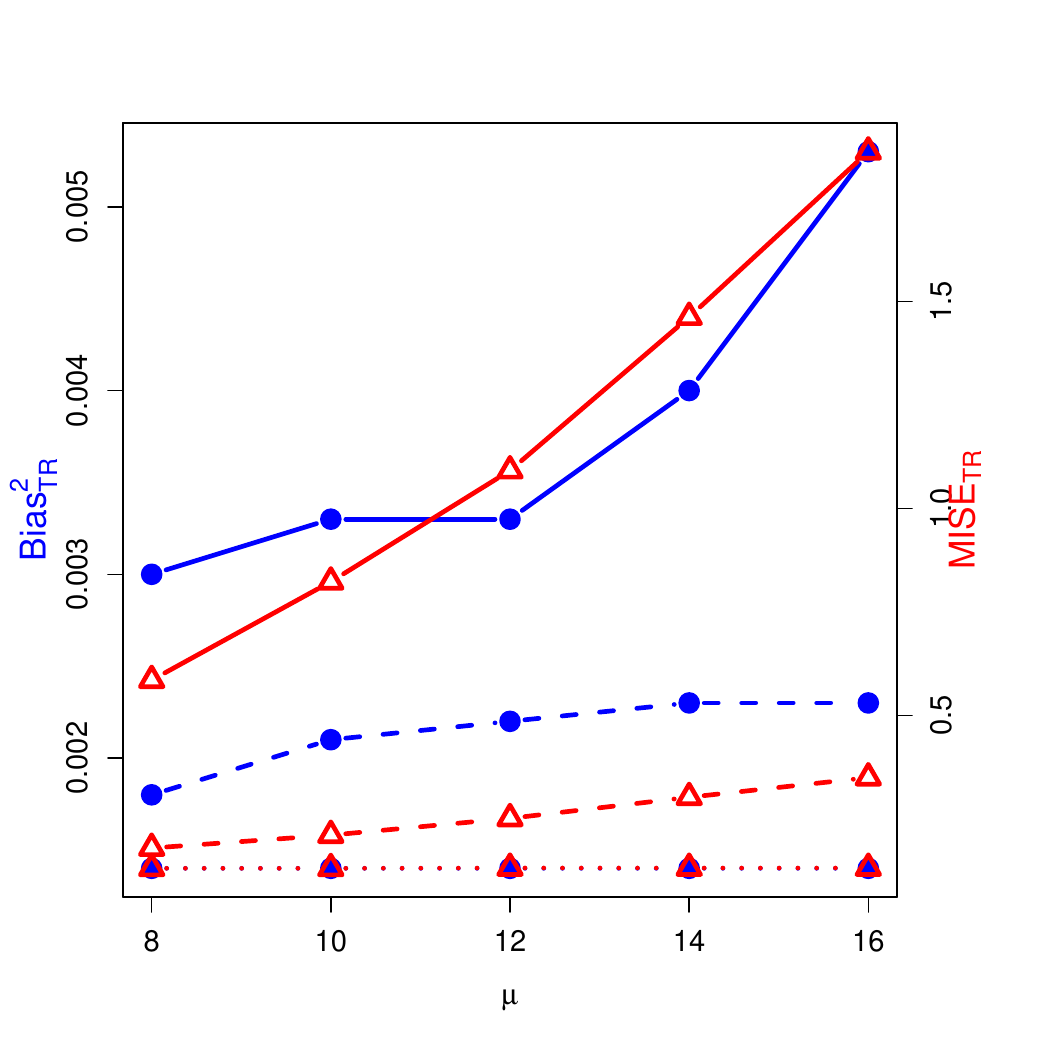} 
&  \includegraphics[scale=0.38]{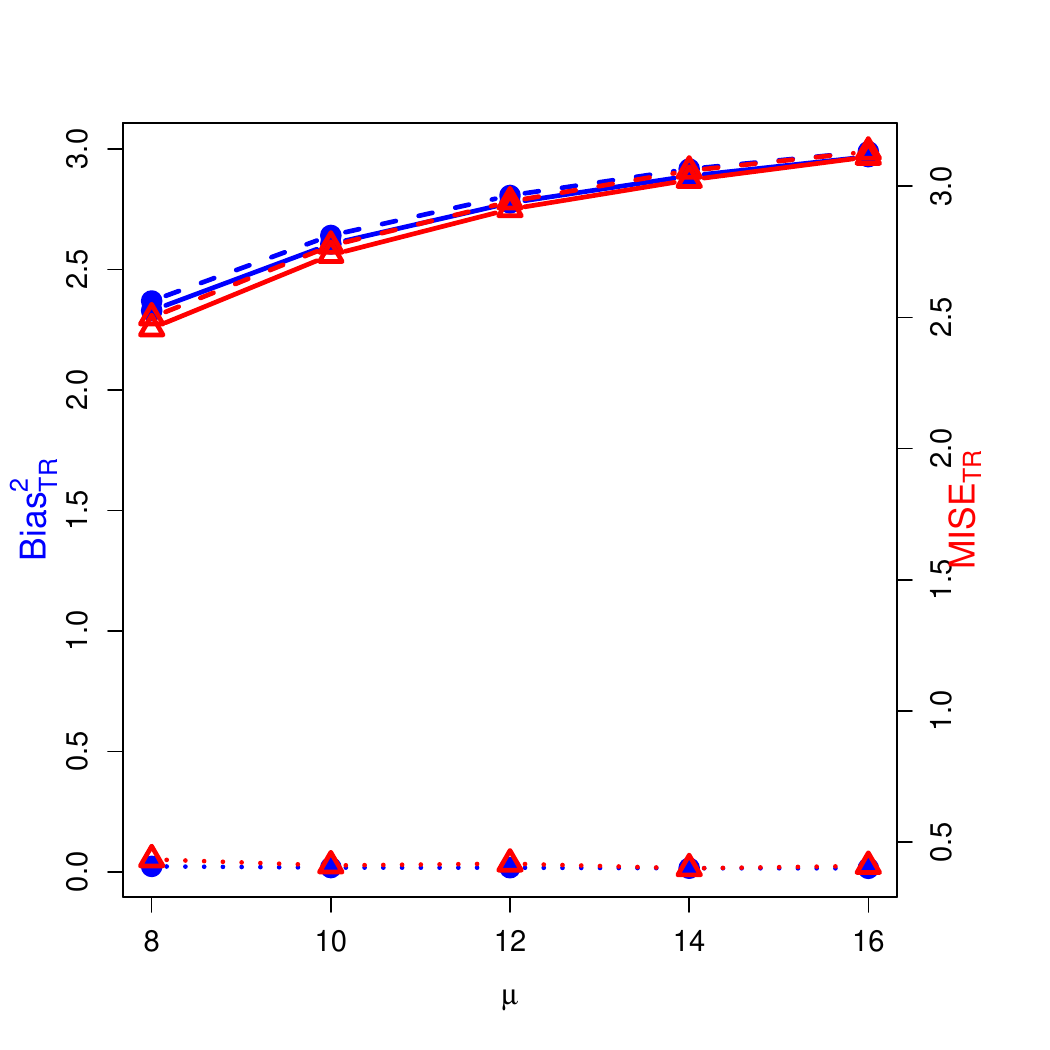}
 \\[-5ex]
$\weta$ &  \includegraphics[scale=0.38]{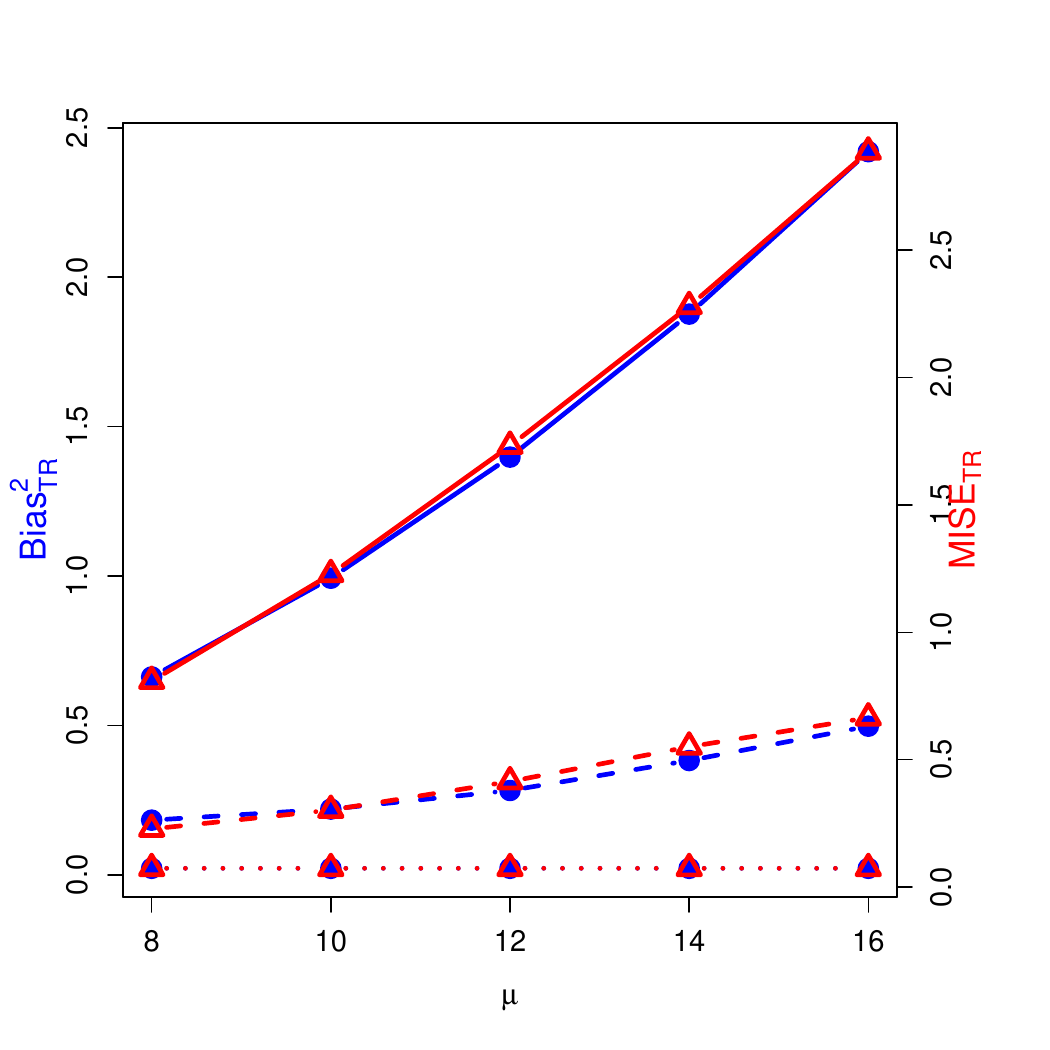} 
&  \includegraphics[scale=0.38]{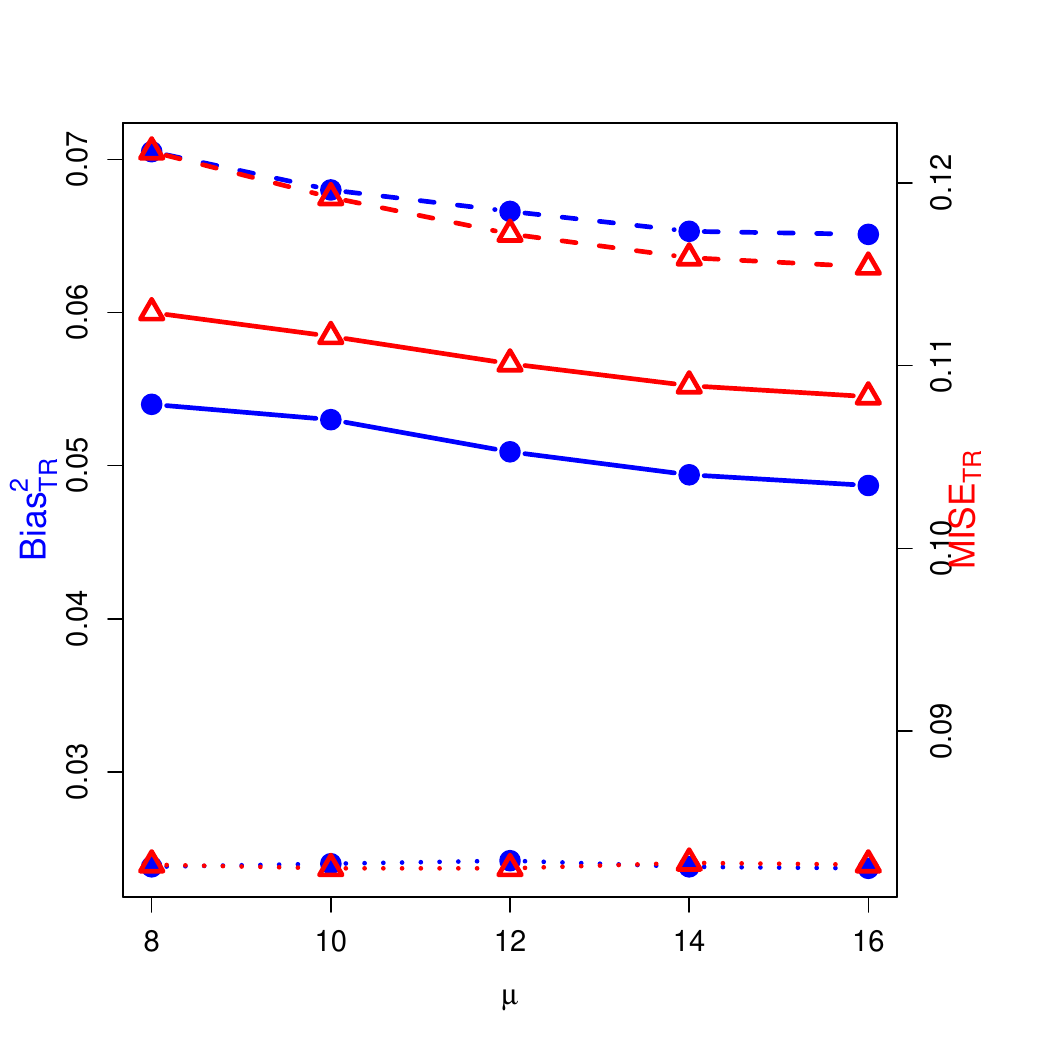}
\\ [-5ex]
 $\weta_{\monmod}$ &  \includegraphics[scale=0.38]{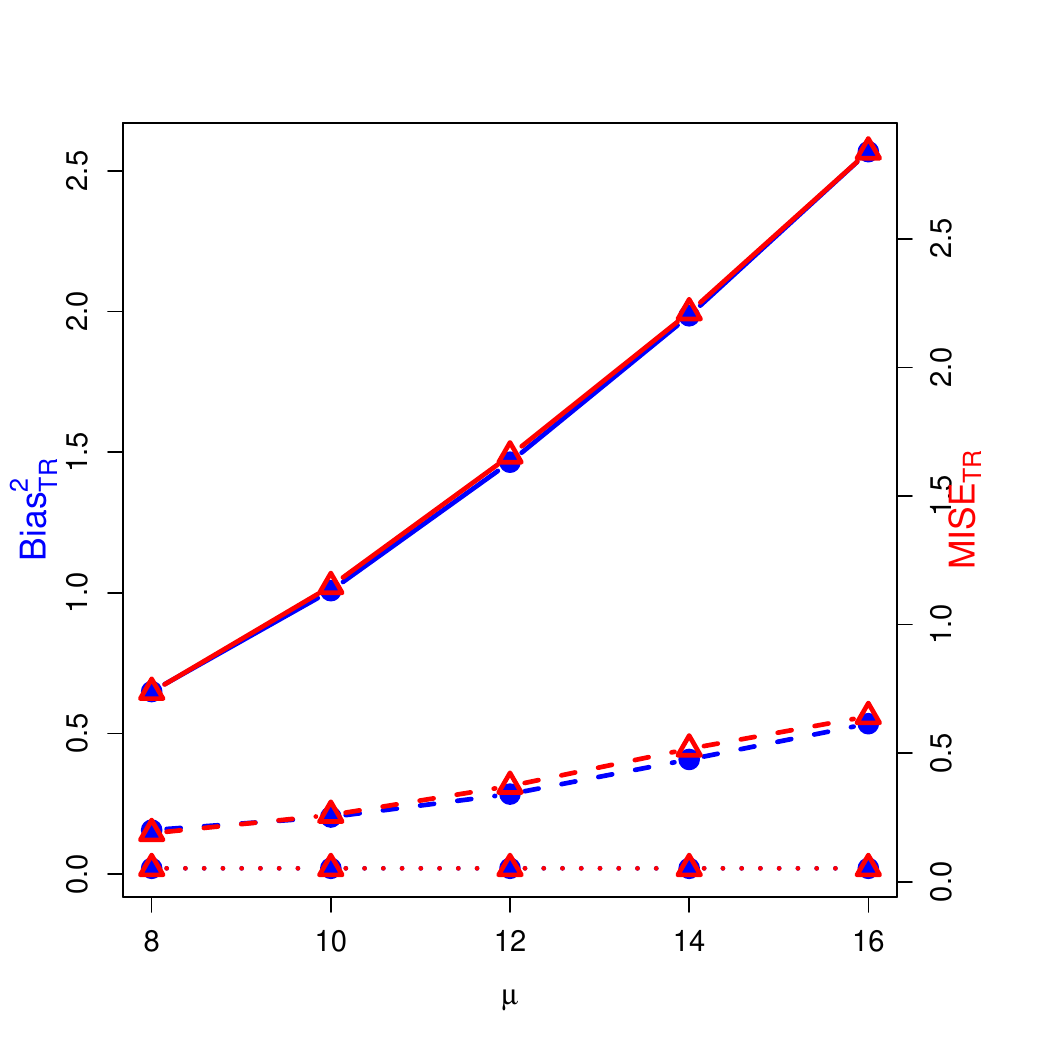} 
 &  \includegraphics[scale=0.38]{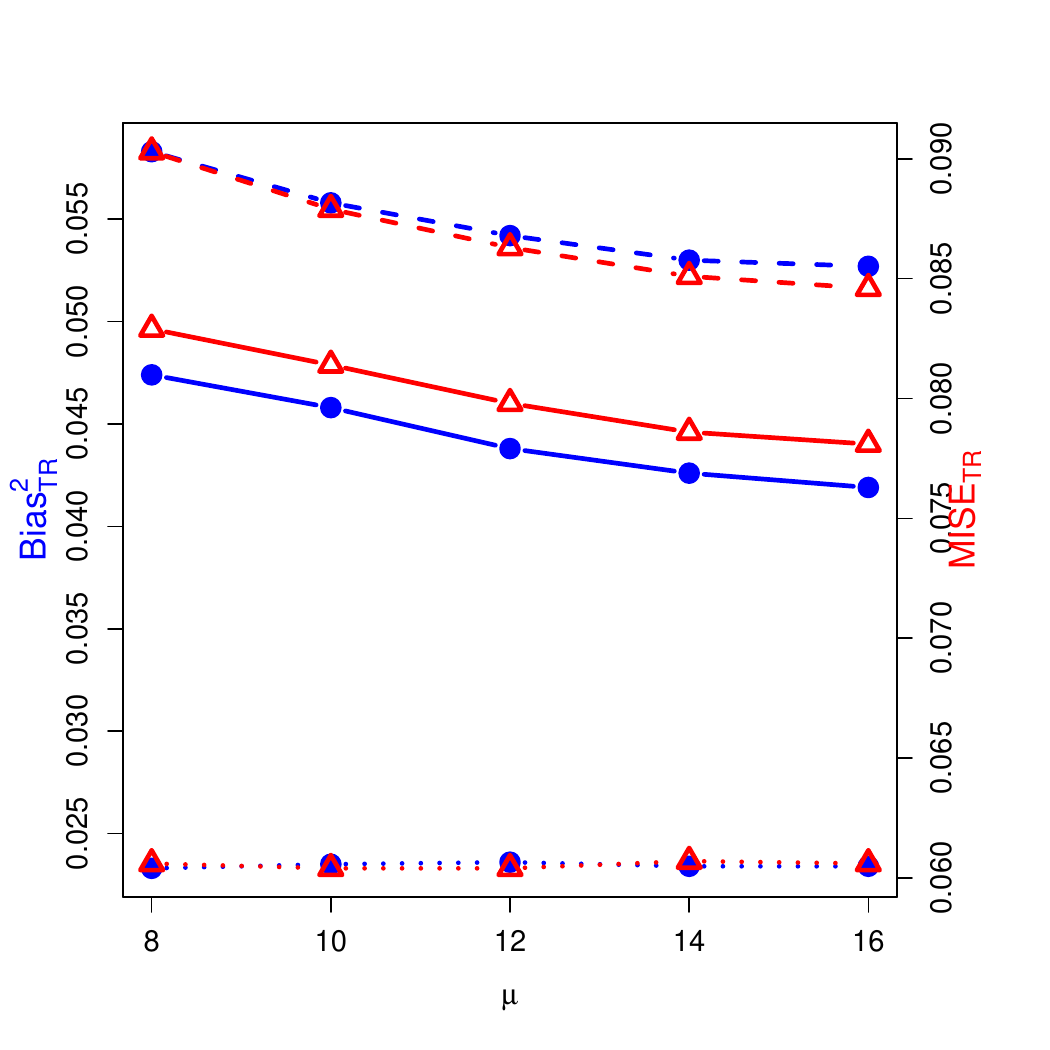}
 \end{tabular}
\caption{\small \label{fig:BIAS-MISE}  Plots of the trimmed squared bias and MISE of the estimators of $\beta_0$ and $\eta_0$ 
as a function of $\mu$ for both contamination scenarios: results for $C_{1,\mu}$ are in the first column of plots,
while the second one contains those of $C_{2,\mu}$.  The solid, dashed and dotted lines correspond to the least squares, 
the $M$- and $MM$-estimators, respectively. The squared bias is indicated with circles, and the MISE with triangles. }
\end{center} 
\end{figure}

The plots in Figure \ref{fig:BIAS-MISE} summarize the effect of the contamination scenarios
for different values of $\mu$. Each plot corresponds to one contamination scenario 
and one parameter estimator. Within each panel, the solid, dashed and dotted lines correspond to 
the measures for the least squares, the $M$- and $MM$-estimators, respectively.    
There are two lines per estimation method: the one with  triangles shows the 
trimmed MISE, and one with solid circles indicates the corresponding trimmed
bias squared. 
 
\begin{figure}[tp]
 \begin{center}
 \footnotesize
 \renewcommand{\arraystretch}{0.2}
 \newcolumntype{M}{>{\centering\arraybackslash}m{\dimexpr.01\linewidth-1\tabcolsep}}
   \newcolumntype{G}{>{\centering\arraybackslash}m{\dimexpr.35\linewidth-1\tabcolsep}}
\begin{tabular}{M GGG}
 & $\wbeta_{\ls}$ & $\wbeta_{\eme}$ & $\wbeta_{\eme\eme}$ \\
$C_{0}$ 
&  \includegraphics[scale=0.35]{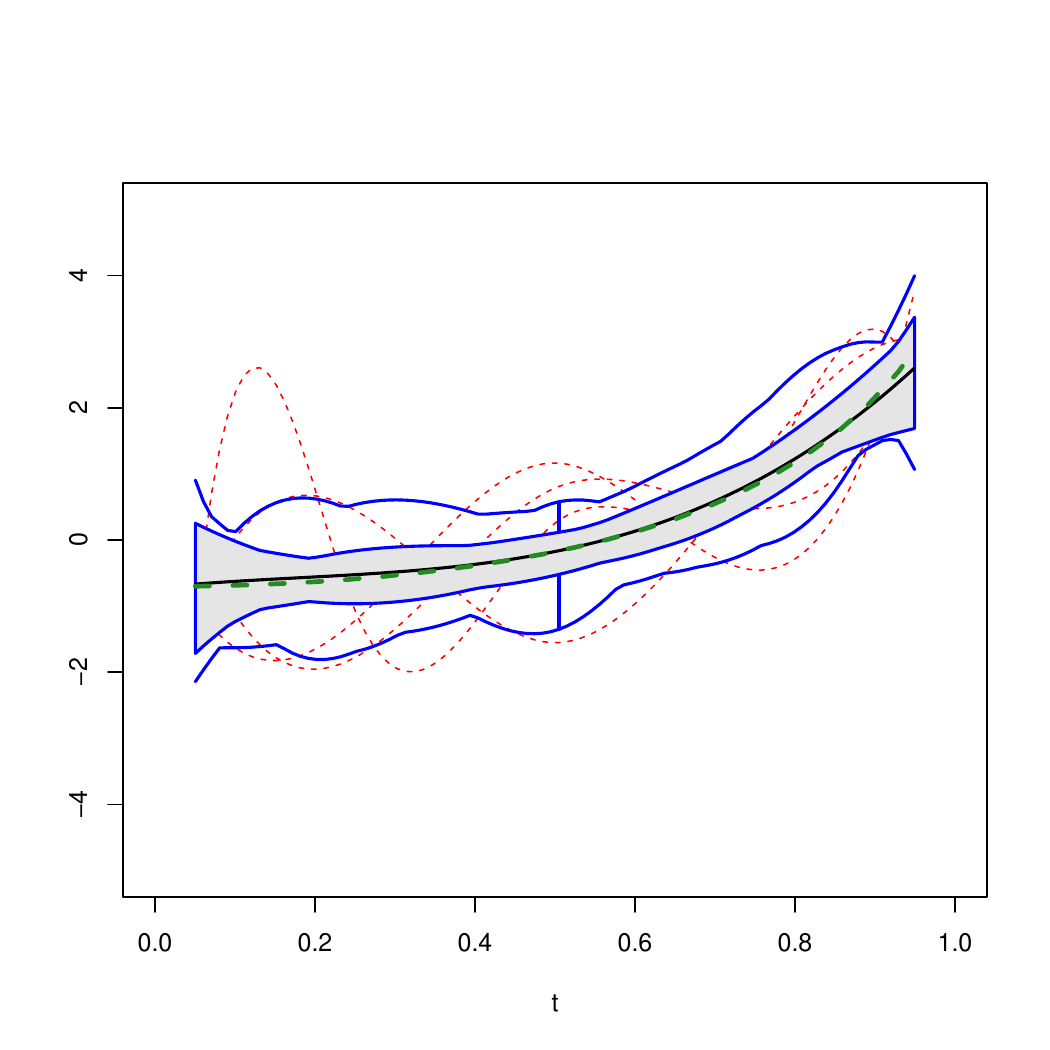} 
&  \includegraphics[scale=0.35]{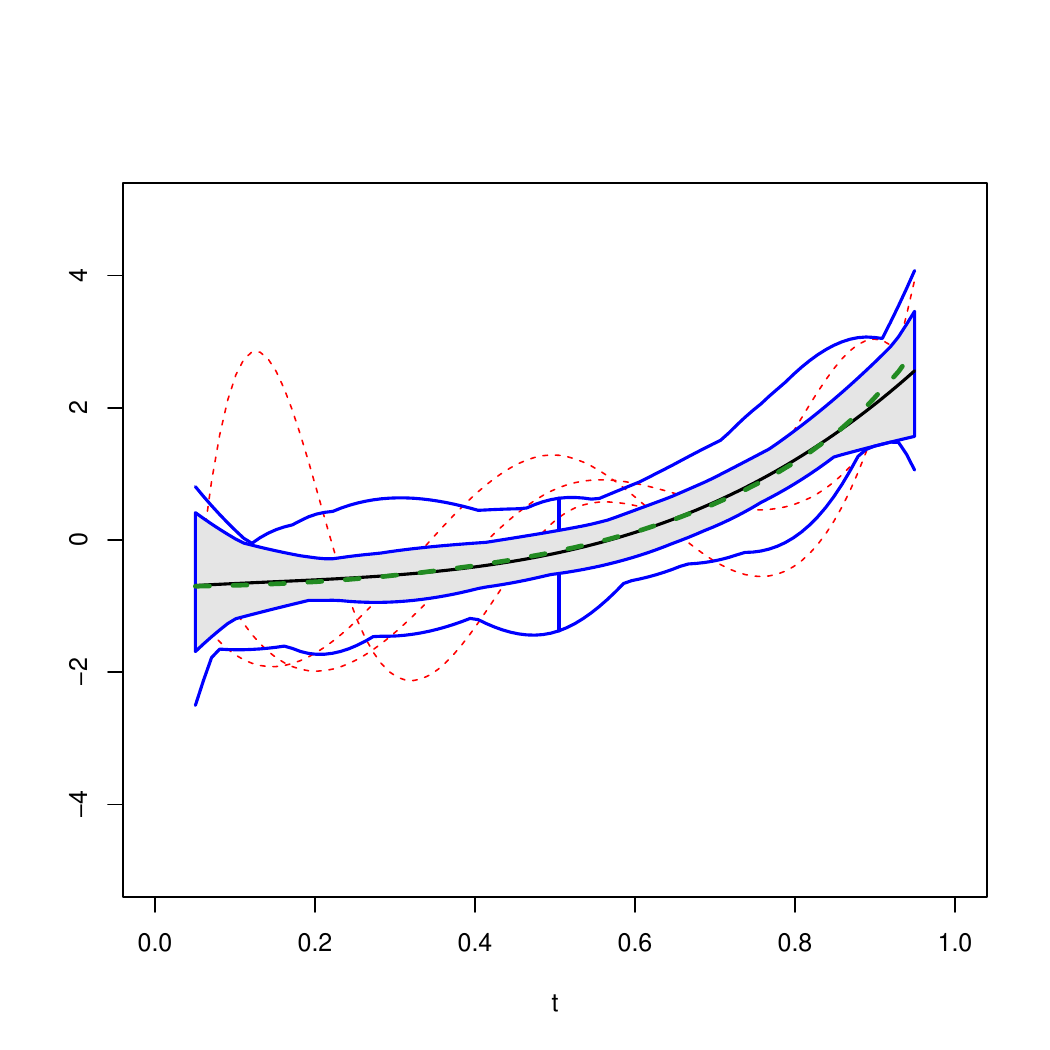} 
&  \includegraphics[scale=0.35]{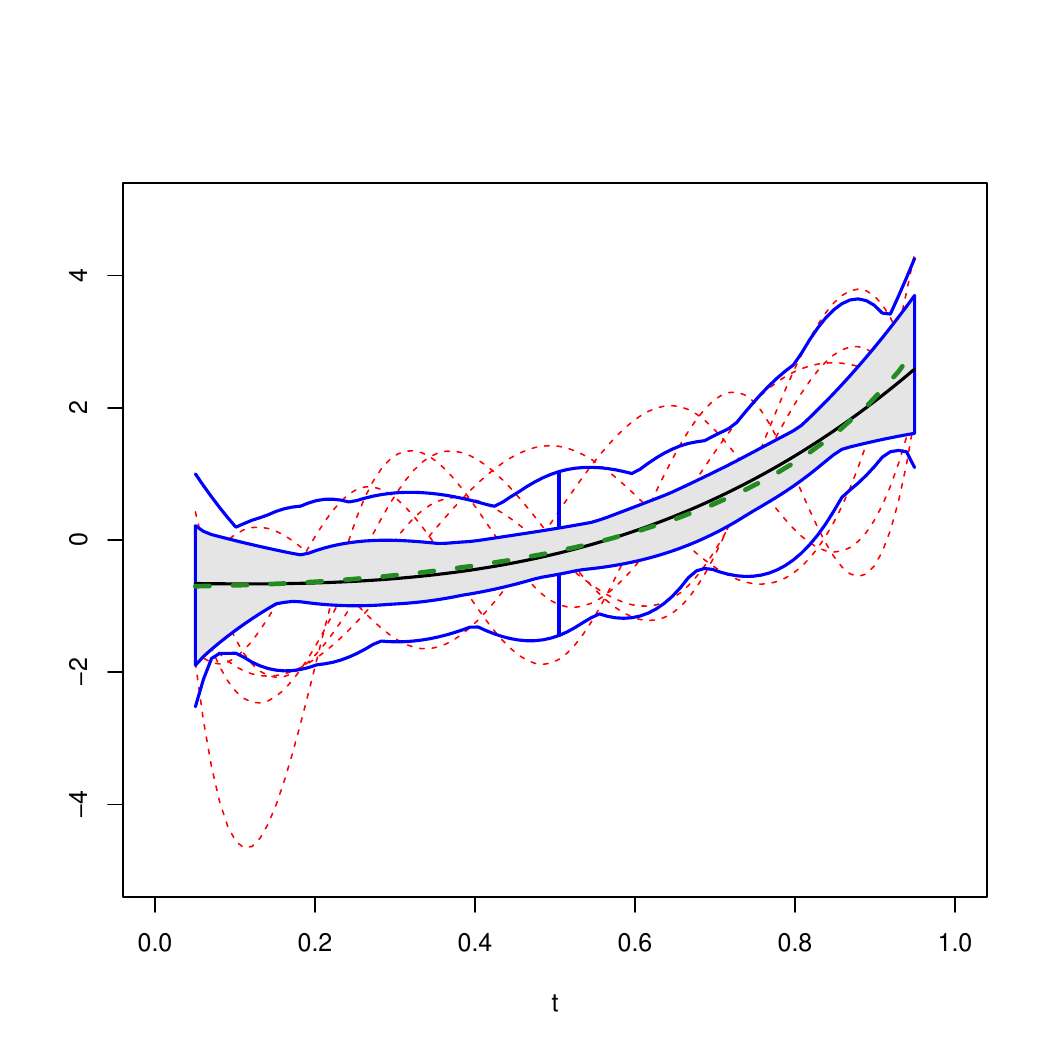} 
\\
$C_{1, 12}$ &  
\includegraphics[scale=0.35]{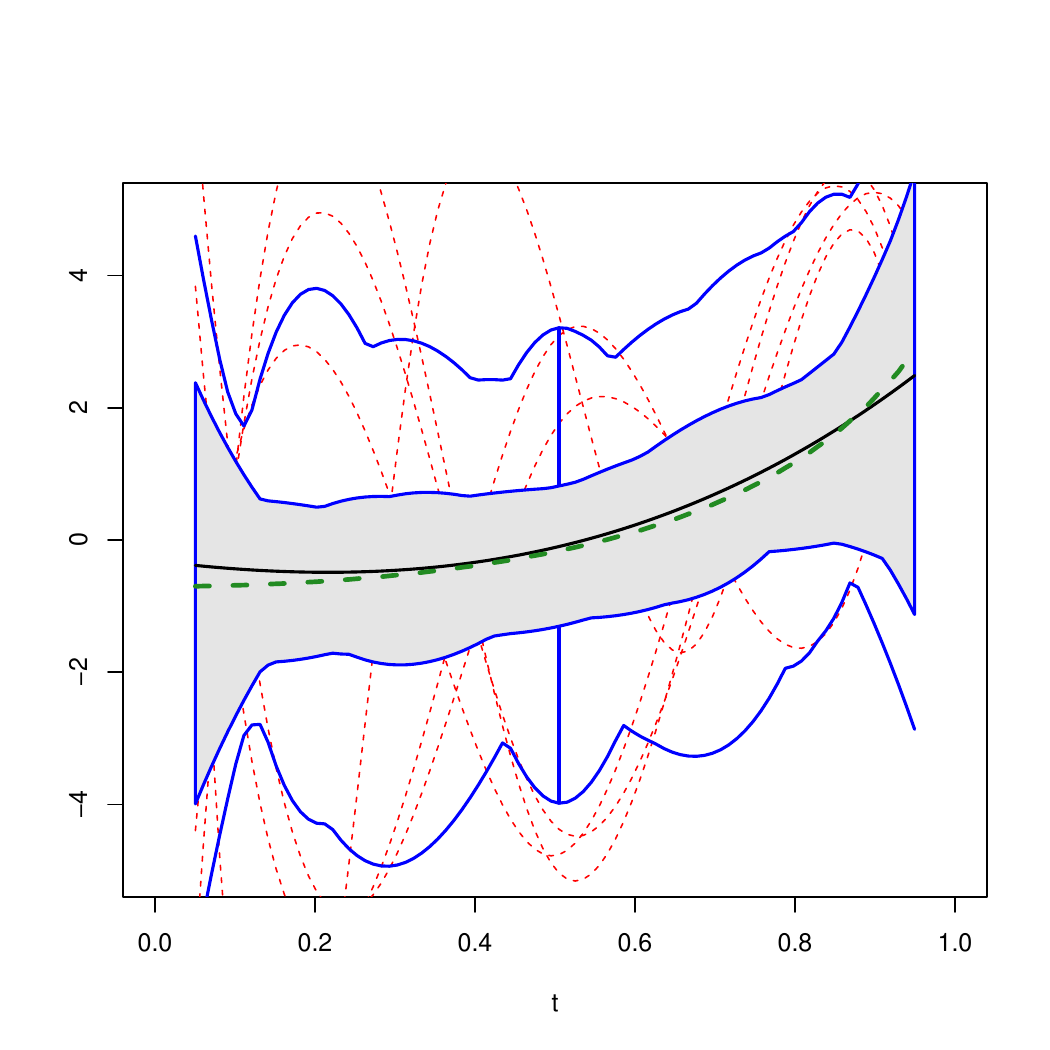}
 &  \includegraphics[scale=0.35]{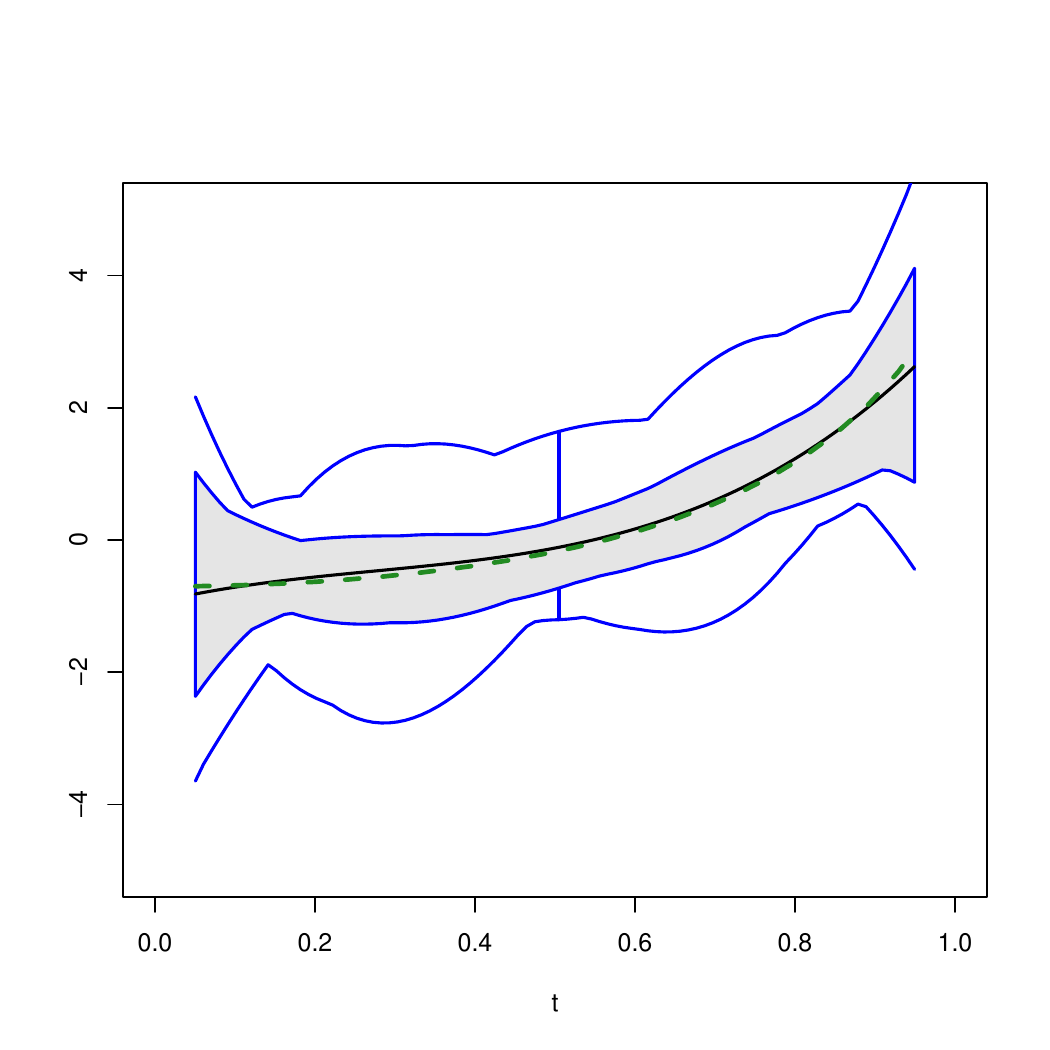}
  &  \includegraphics[scale=0.35]{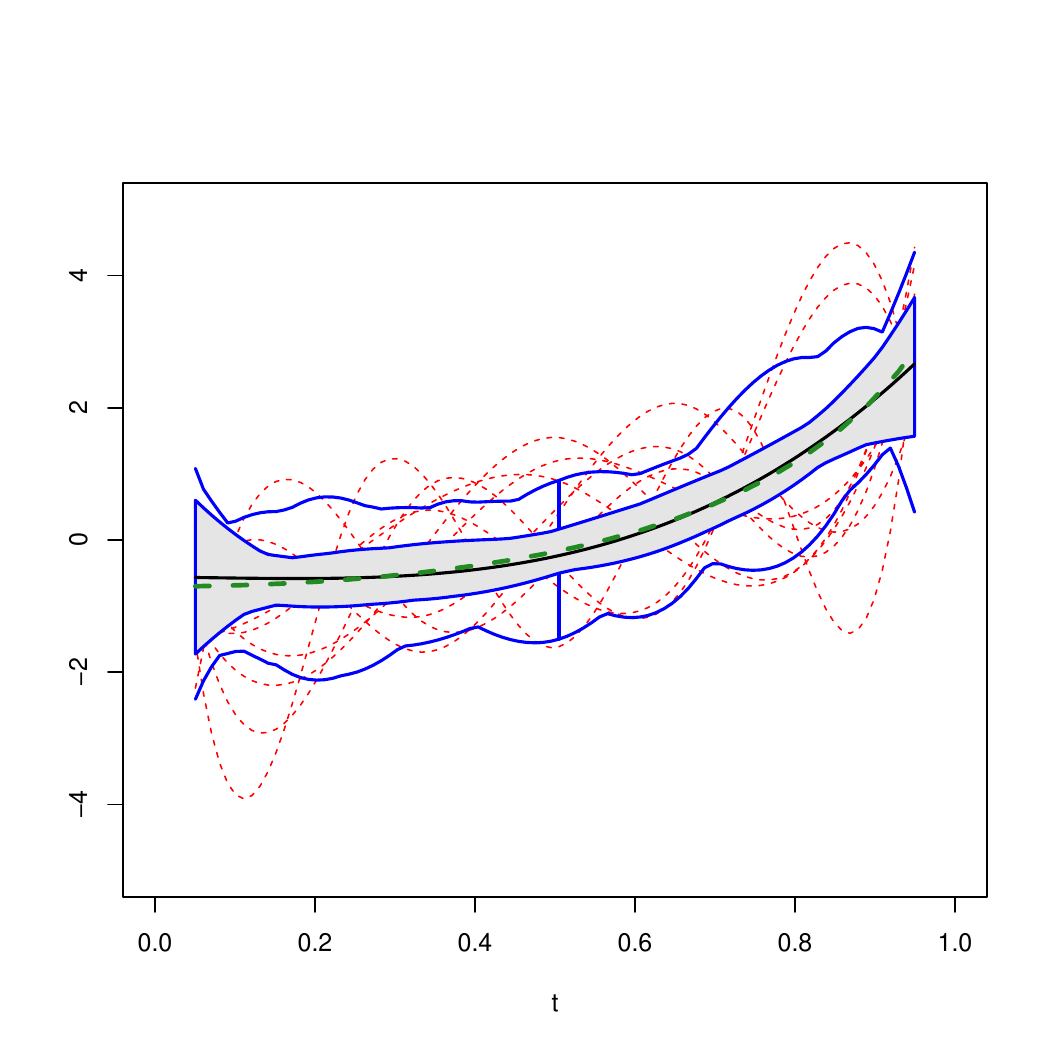}
  \\
   
$C_{2, 12}$ 
&  \includegraphics[scale=0.35]{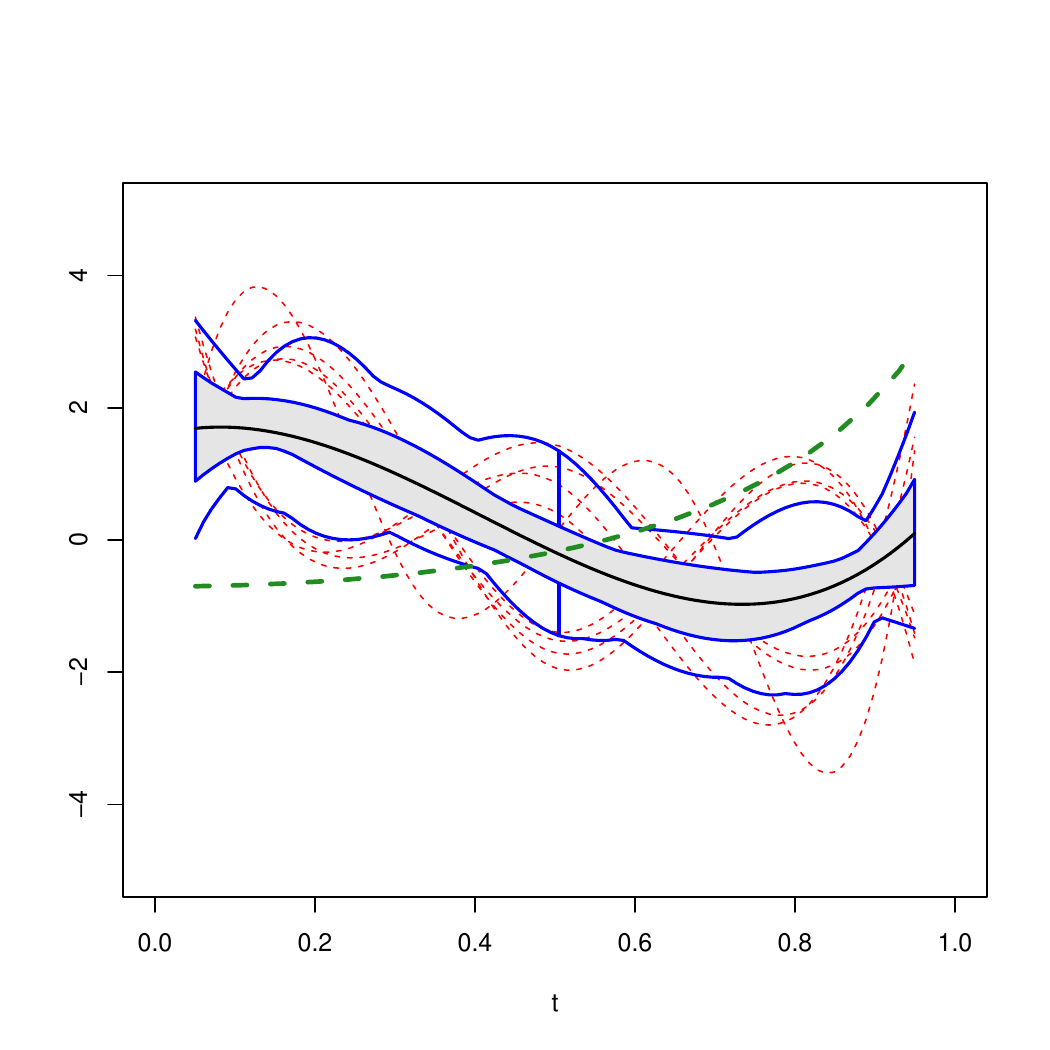}
&  \includegraphics[scale=0.35]{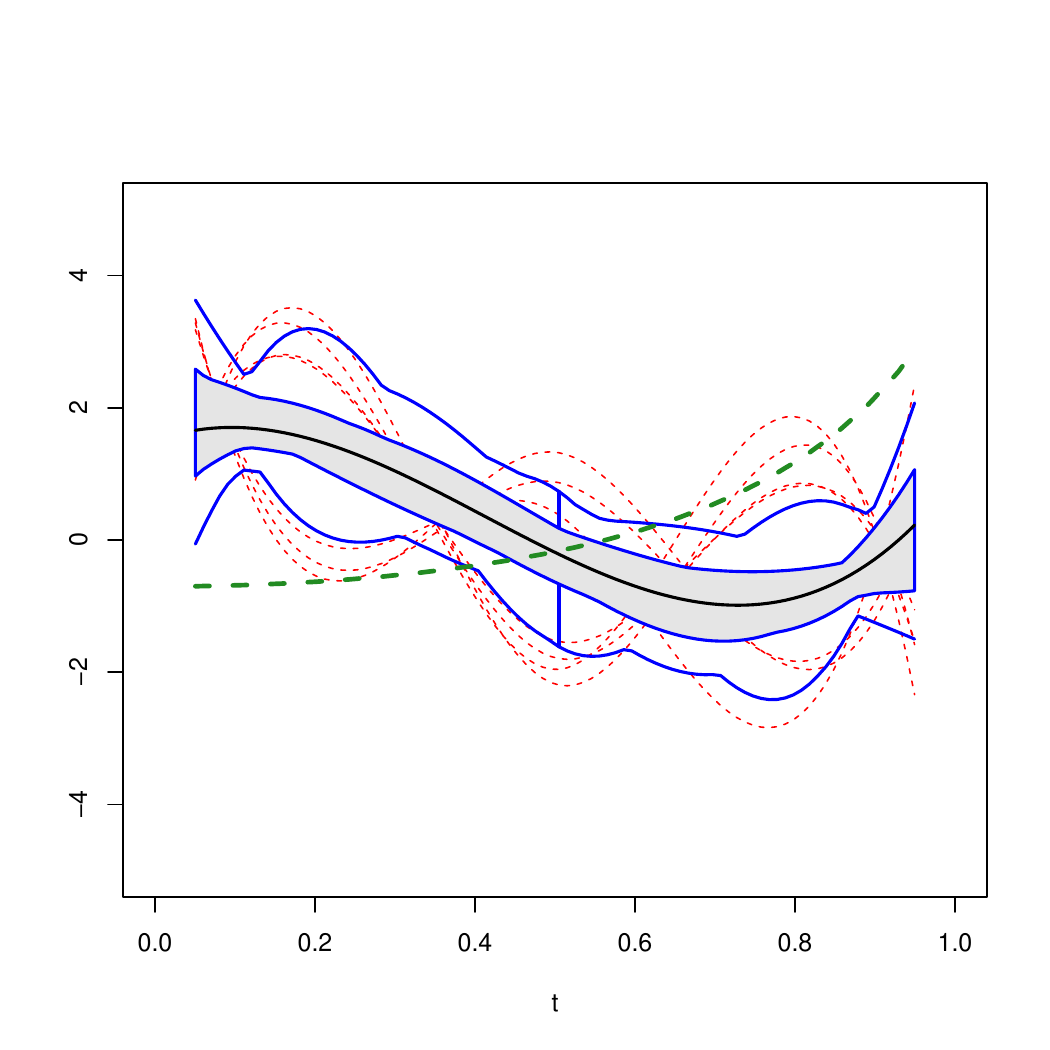}
&  \includegraphics[scale=0.35]{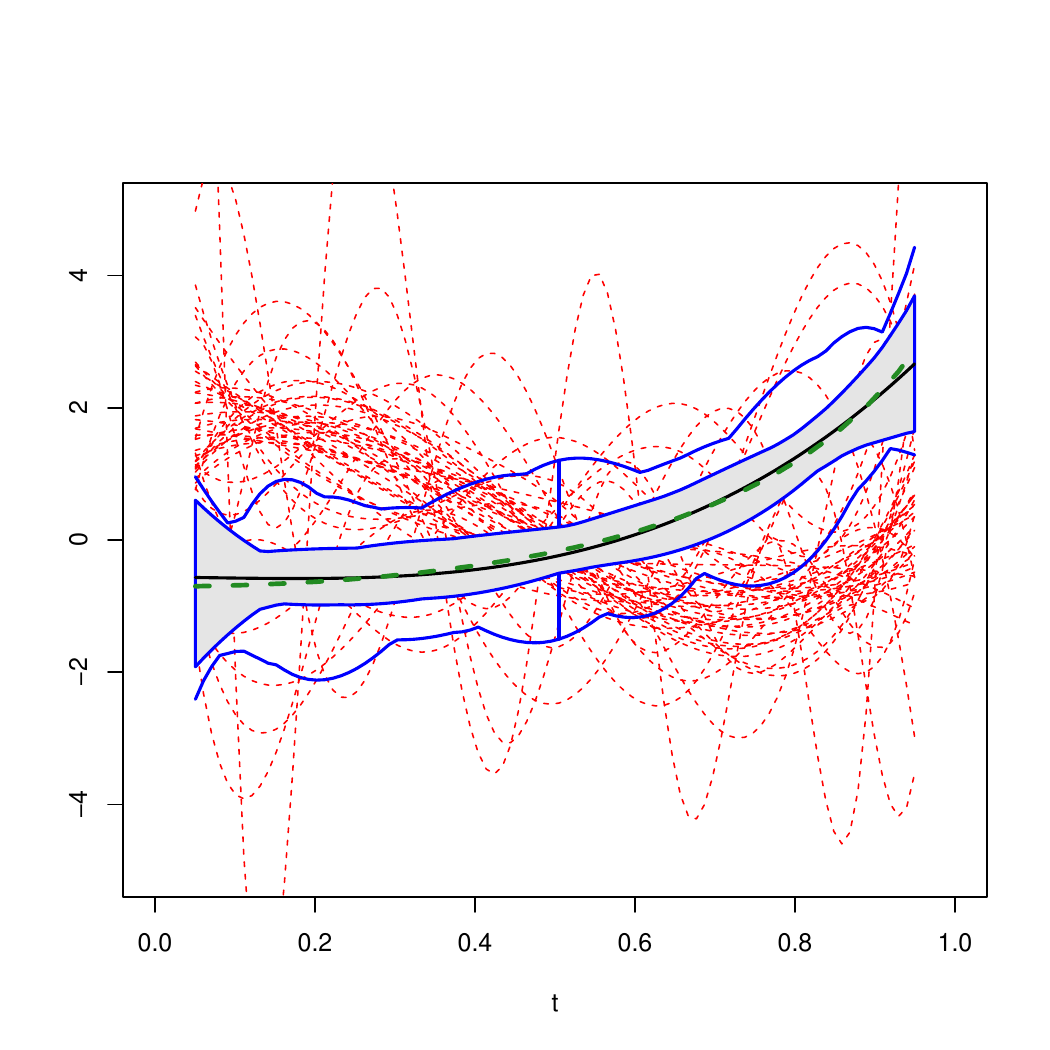}
\end{tabular}
\caption{\small \label{fig:wbeta}  Functional boxplot of the estimators for $\beta_0$. 
The true function is shown with a green dashed line, while the black solid one is the central 
curve of the $n_R = 500$ estimates $\wbeta$. Columns correspond to estimation 
methods while rows to three contaminations settings.}
\end{center} 
\end{figure}

\begin{figure}
 \begin{center}
 \footnotesize
 \renewcommand{\arraystretch}{0.2}
 \newcolumntype{M}{>{\centering\arraybackslash}m{\dimexpr.01\linewidth-1\tabcolsep}}
   \newcolumntype{G}{>{\centering\arraybackslash}m{\dimexpr.35\linewidth-1\tabcolsep}}
\begin{tabular}{M GGG}
 & $\weta_{\ls}$ & $\weta_{\eme}$ & $\weta_{\eme\eme}$ \\
$C_{0}$ 
&  \includegraphics[scale=0.35]{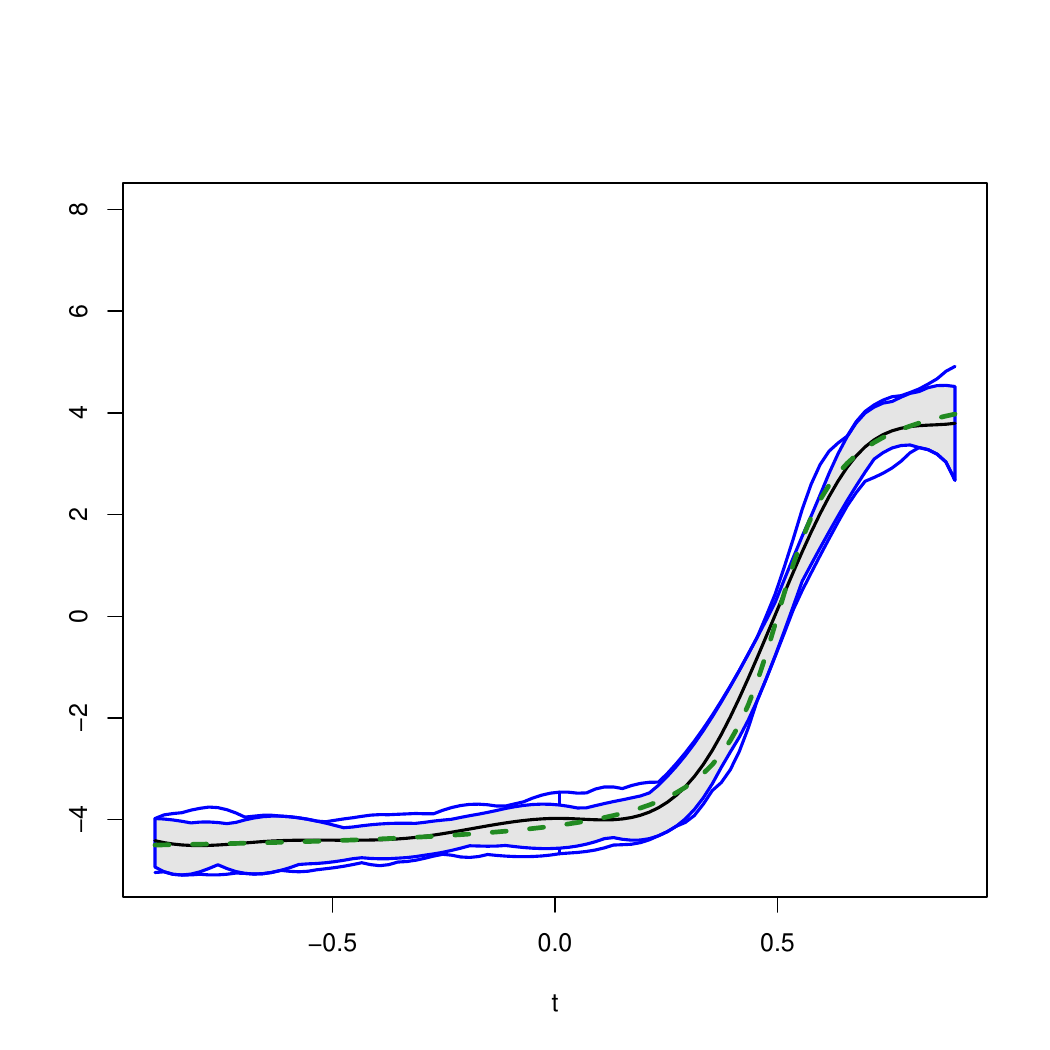} 
&  \includegraphics[scale=0.35]{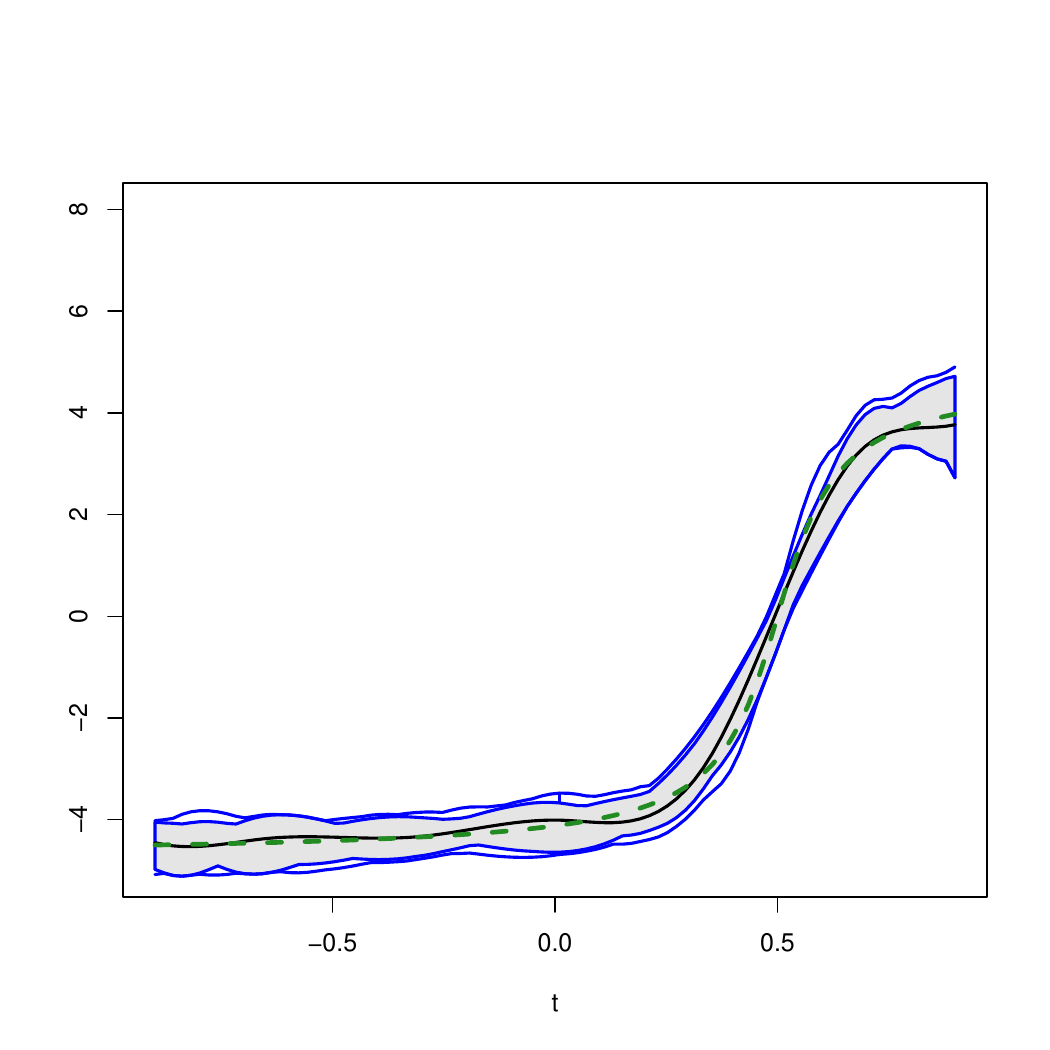} 
&  \includegraphics[scale=0.35]{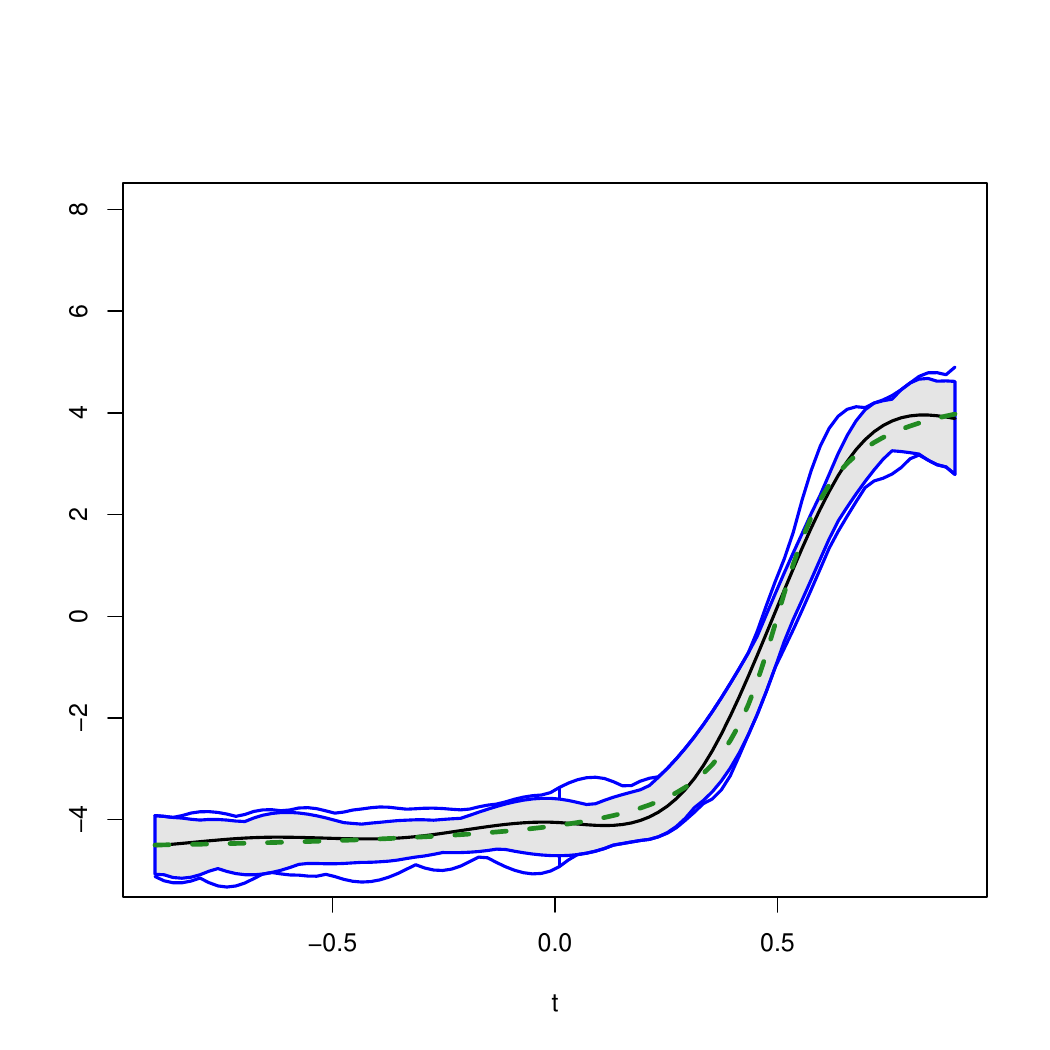} 
\\
$C_{1, 12}$ &  \includegraphics[scale=0.35]{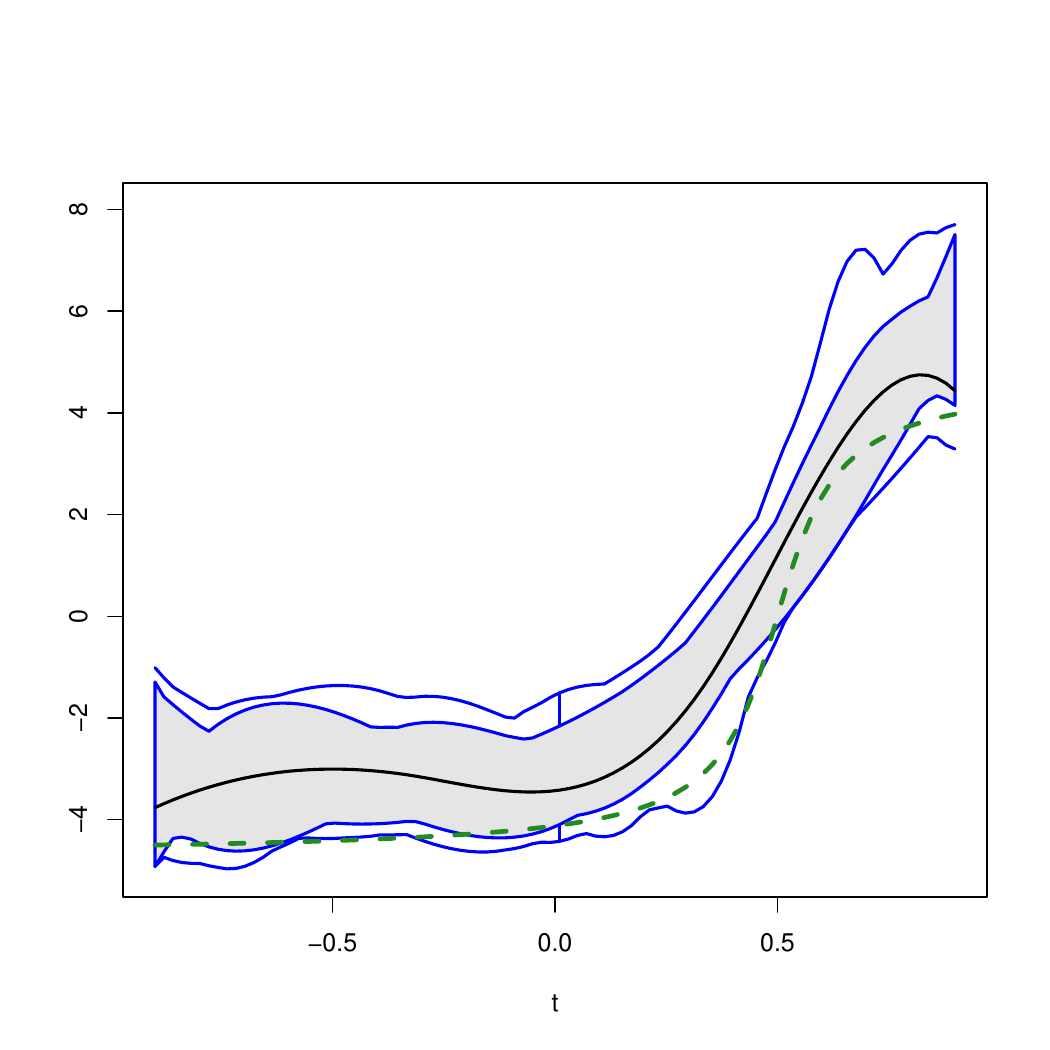}
 &  \includegraphics[scale=0.35]{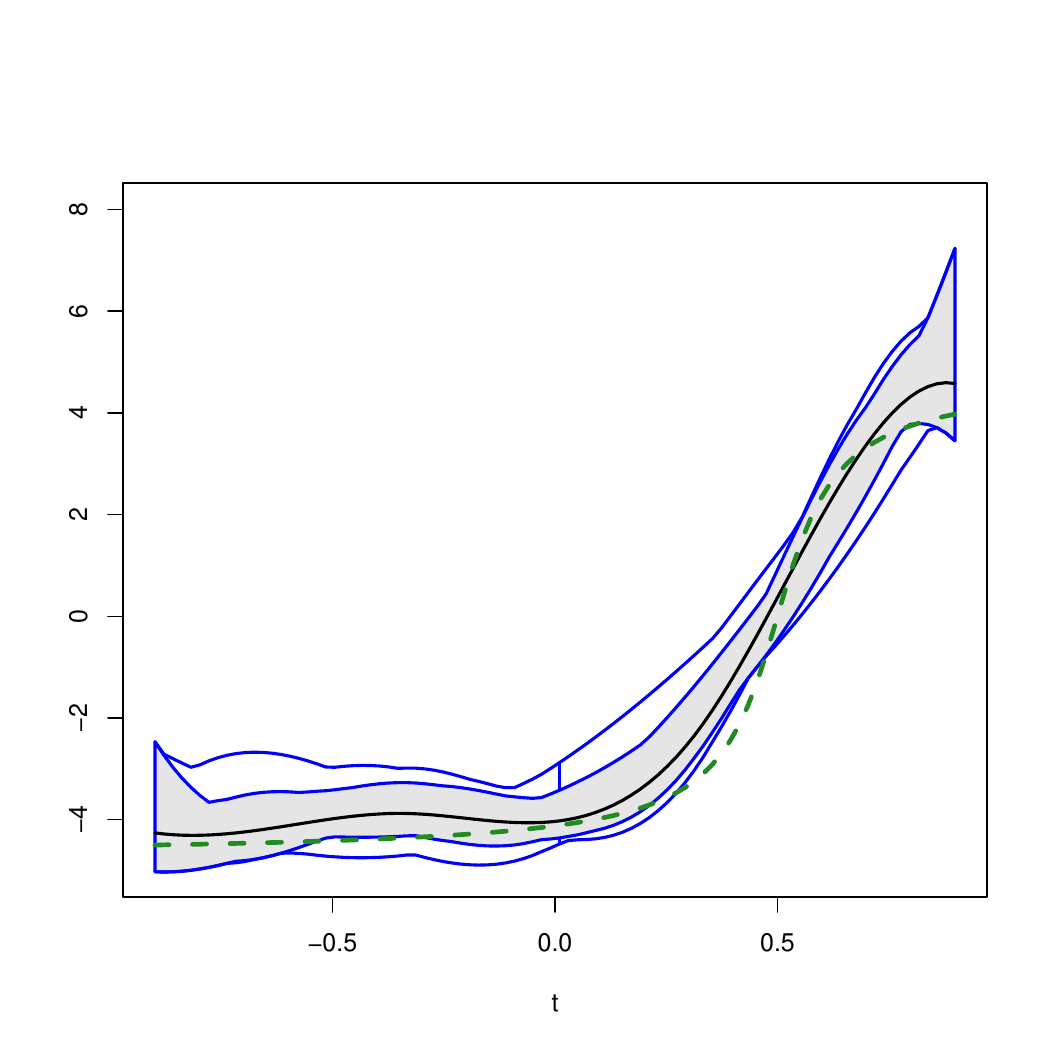}
  &  \includegraphics[scale=0.35]{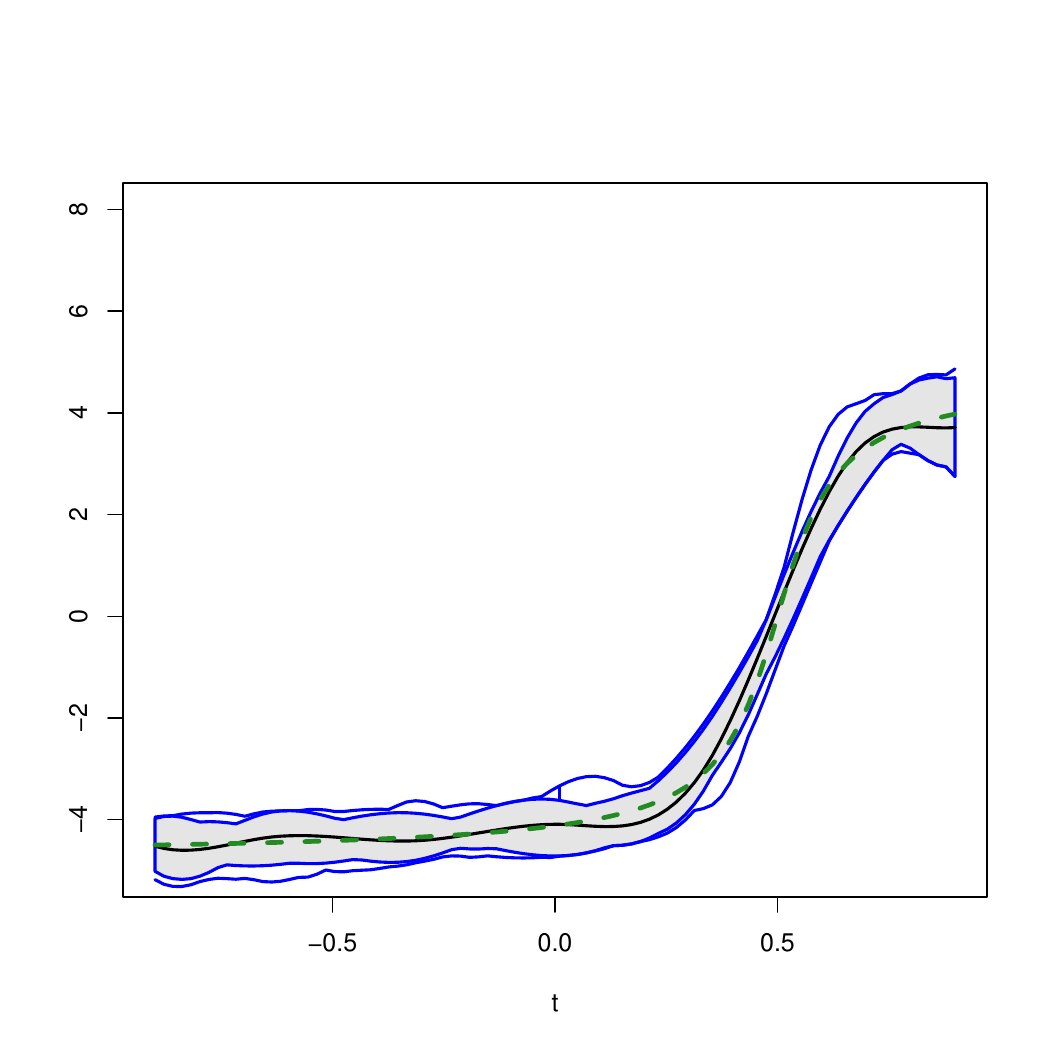}
  \\
   
$C_{2, 12}$ 
&  \includegraphics[scale=0.35]{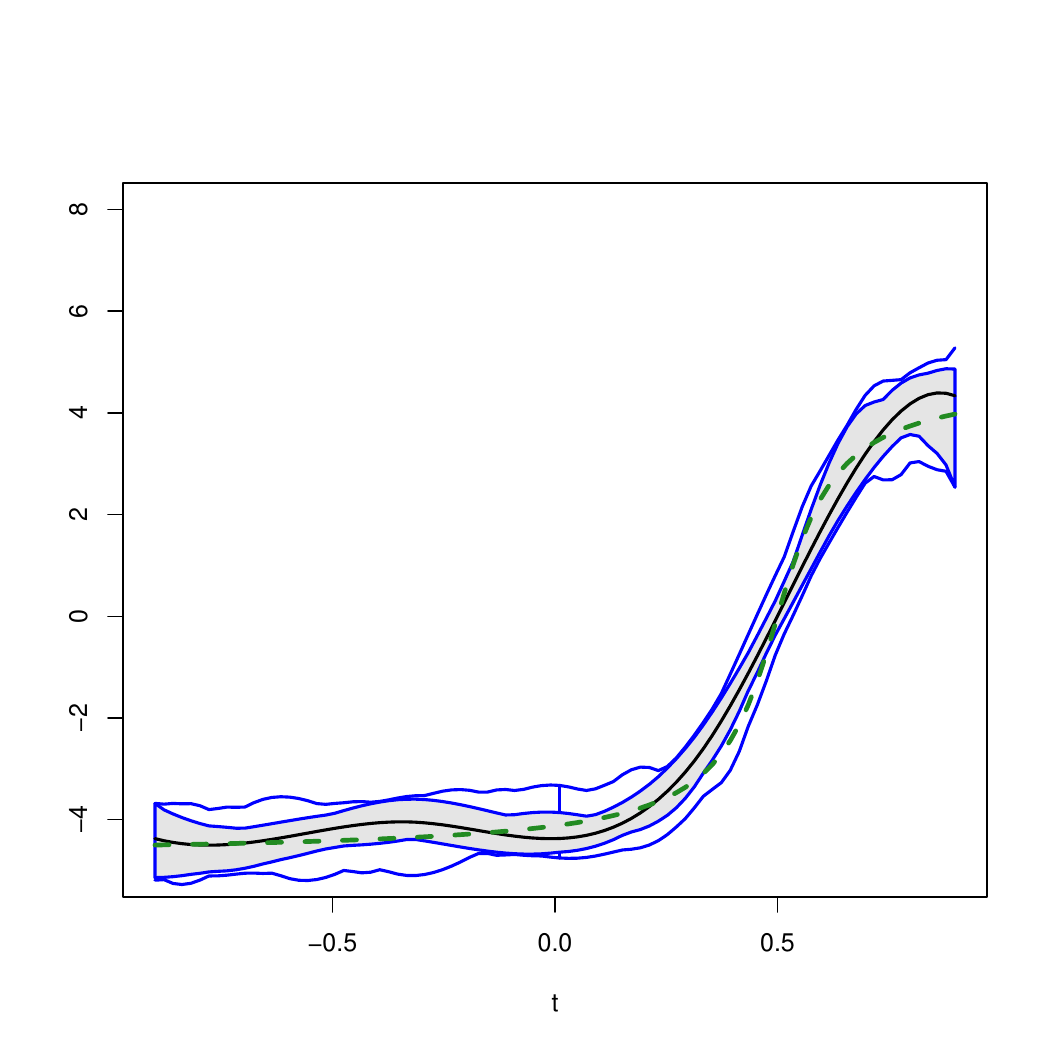}
&  \includegraphics[scale=0.35]{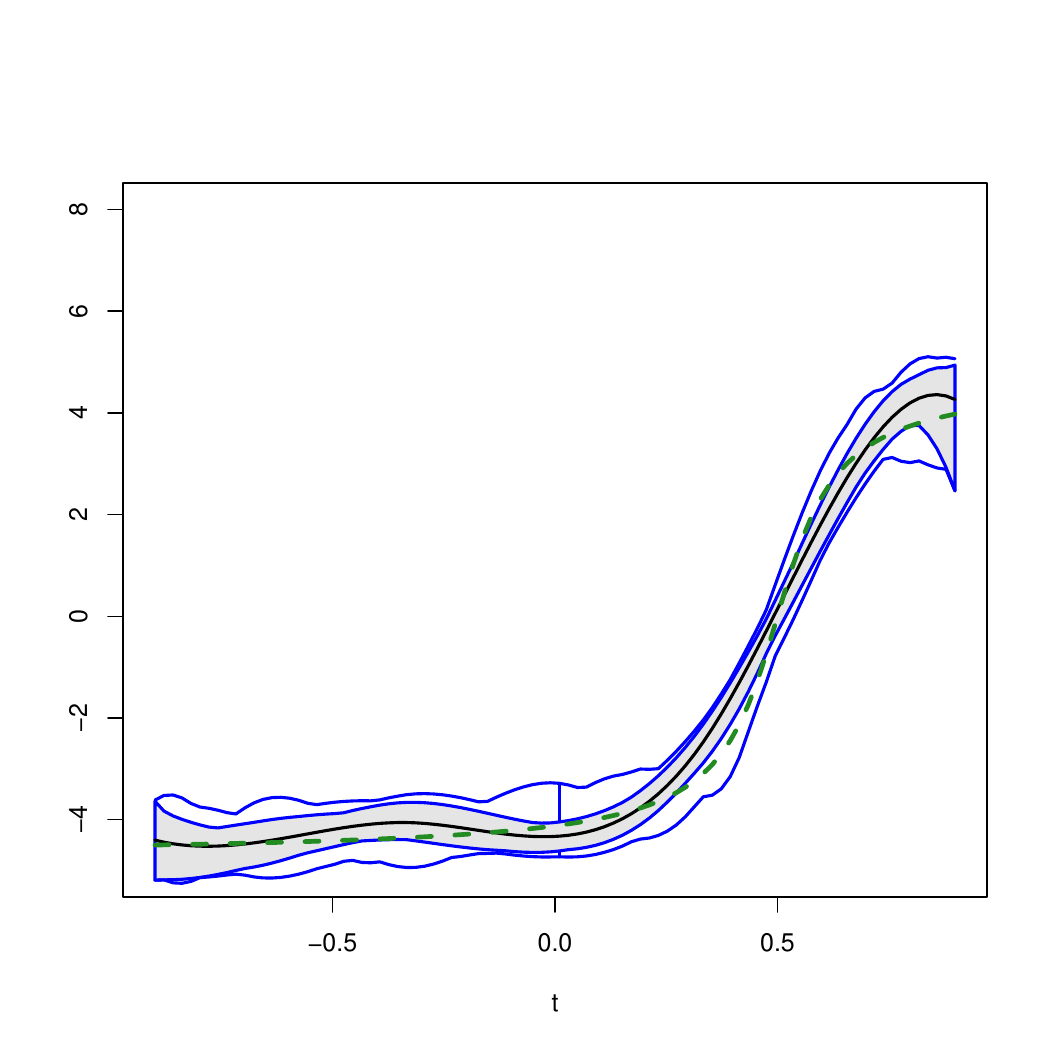}
&  \includegraphics[scale=0.35]{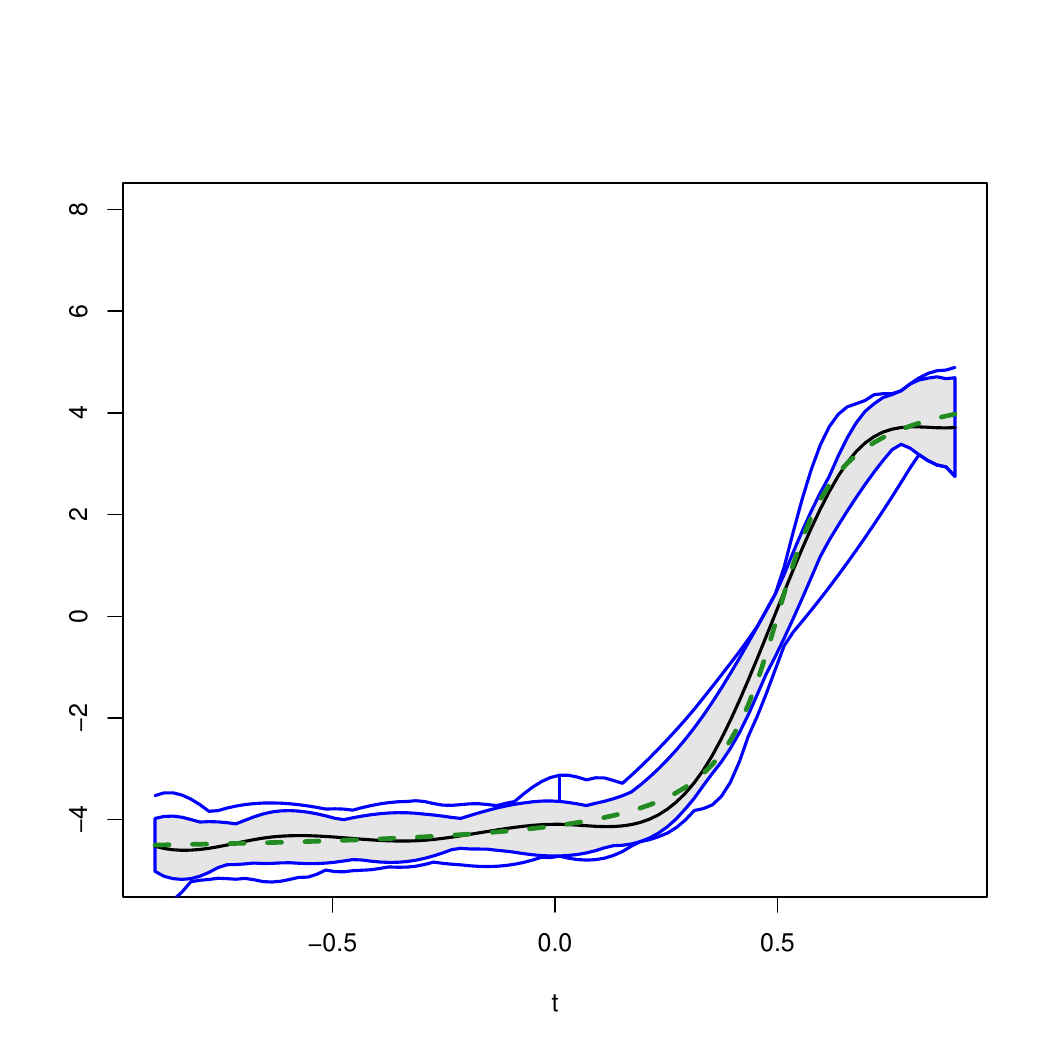}
\end{tabular}
\vskip-0.1in  \caption{\small \label{fig:weta}  Functional boxplot of the estimators for $\eta_0$. 
The true function is shown with a green dashed line, while the black solid one is the central 
curve of the $n_R = 500$ estimates $\wbeta$. Columns correspond to estimation 
methods while rows to three contaminations settings.}
\end{center} 
\end{figure}

\begin{figure}
 \begin{center}
 \footnotesize
 \renewcommand{\arraystretch}{0.2}
 \newcolumntype{M}{>{\centering\arraybackslash}m{\dimexpr.01\linewidth-1\tabcolsep}}
   \newcolumntype{G}{>{\centering\arraybackslash}m{\dimexpr.35\linewidth-1\tabcolsep}}
\begin{tabular}{M GGG}
 & $\weta_{\ls, \monmod}$ & $\weta_{\eme, \monmod}$ & $\weta_{\eme\eme, \monmod}$ \\
$C_{0}$ 
&  \includegraphics[scale=0.35]{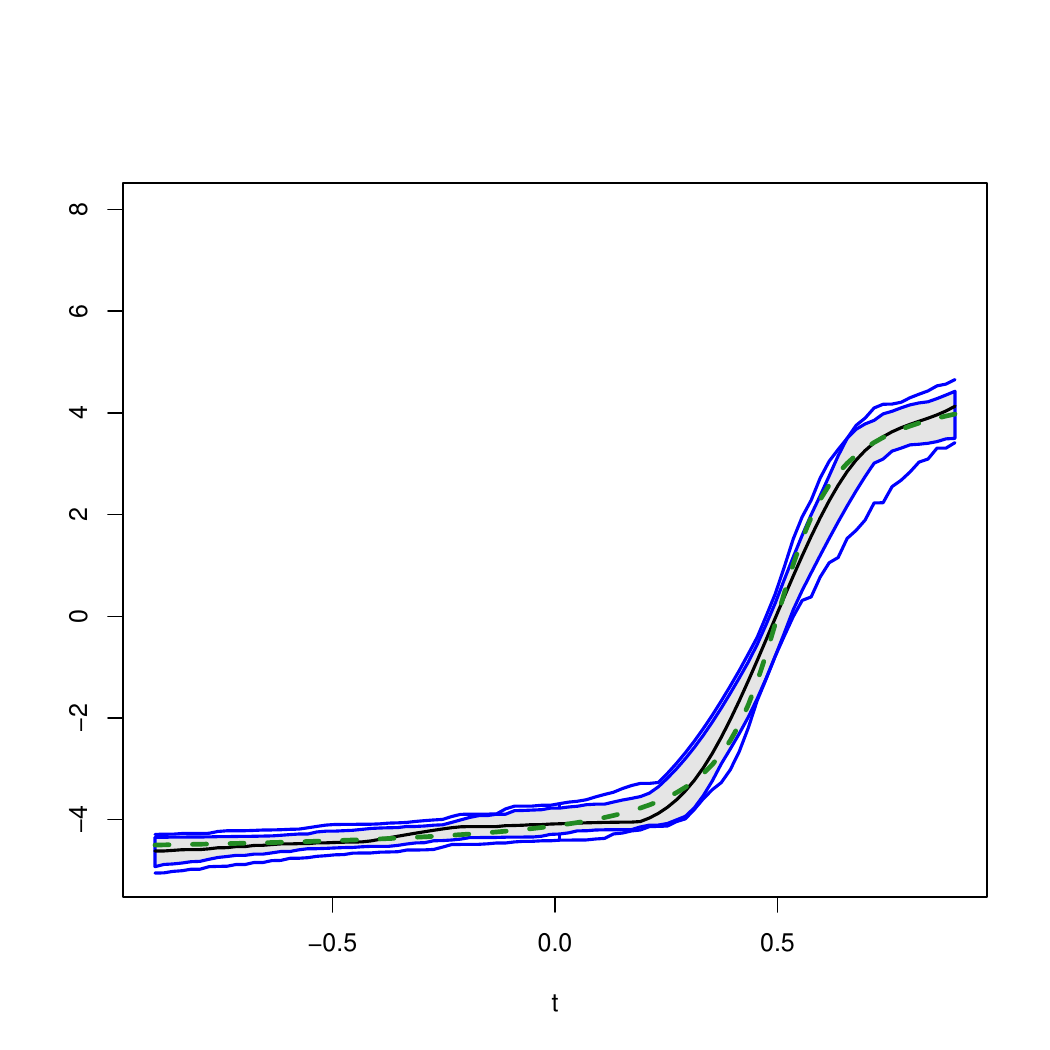} 
&  \includegraphics[scale=0.35]{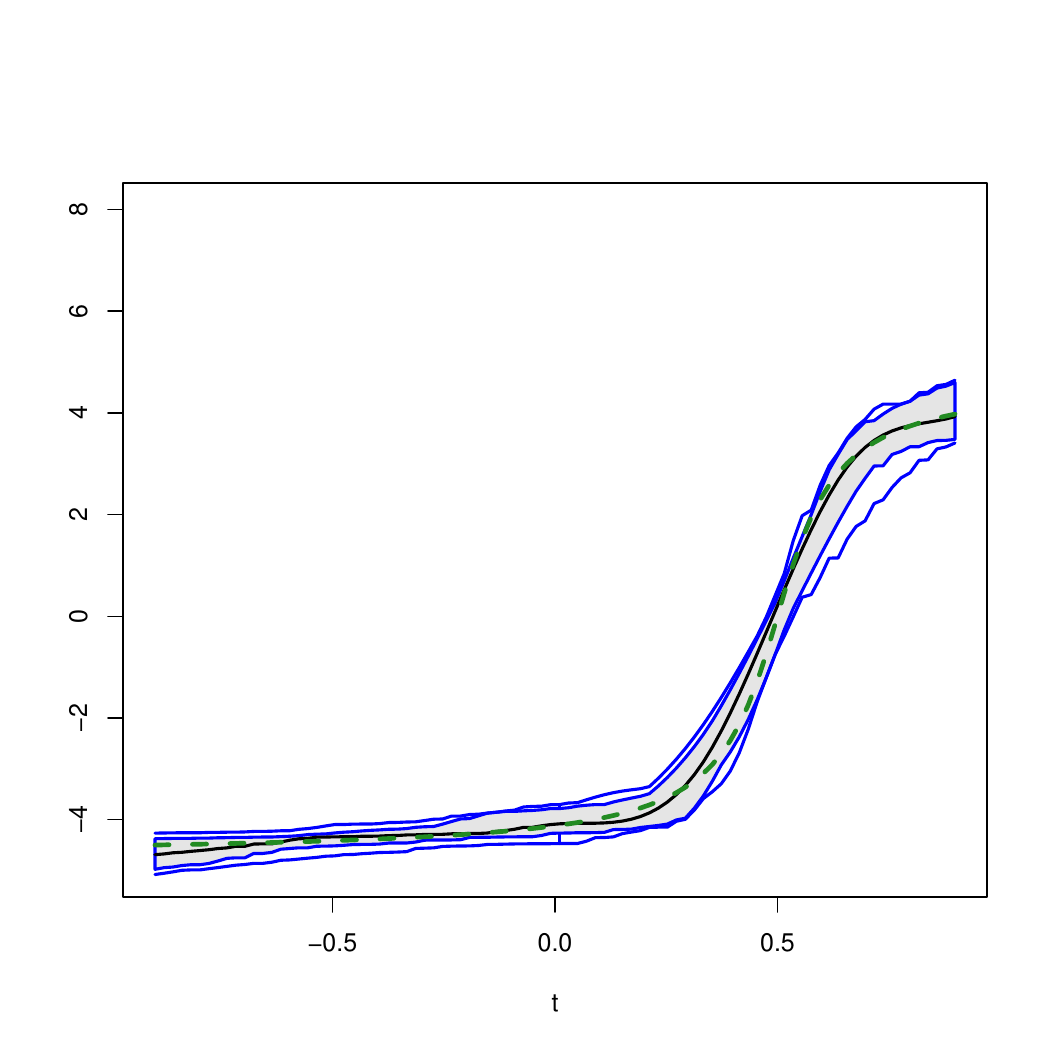} 
&  \includegraphics[scale=0.35]{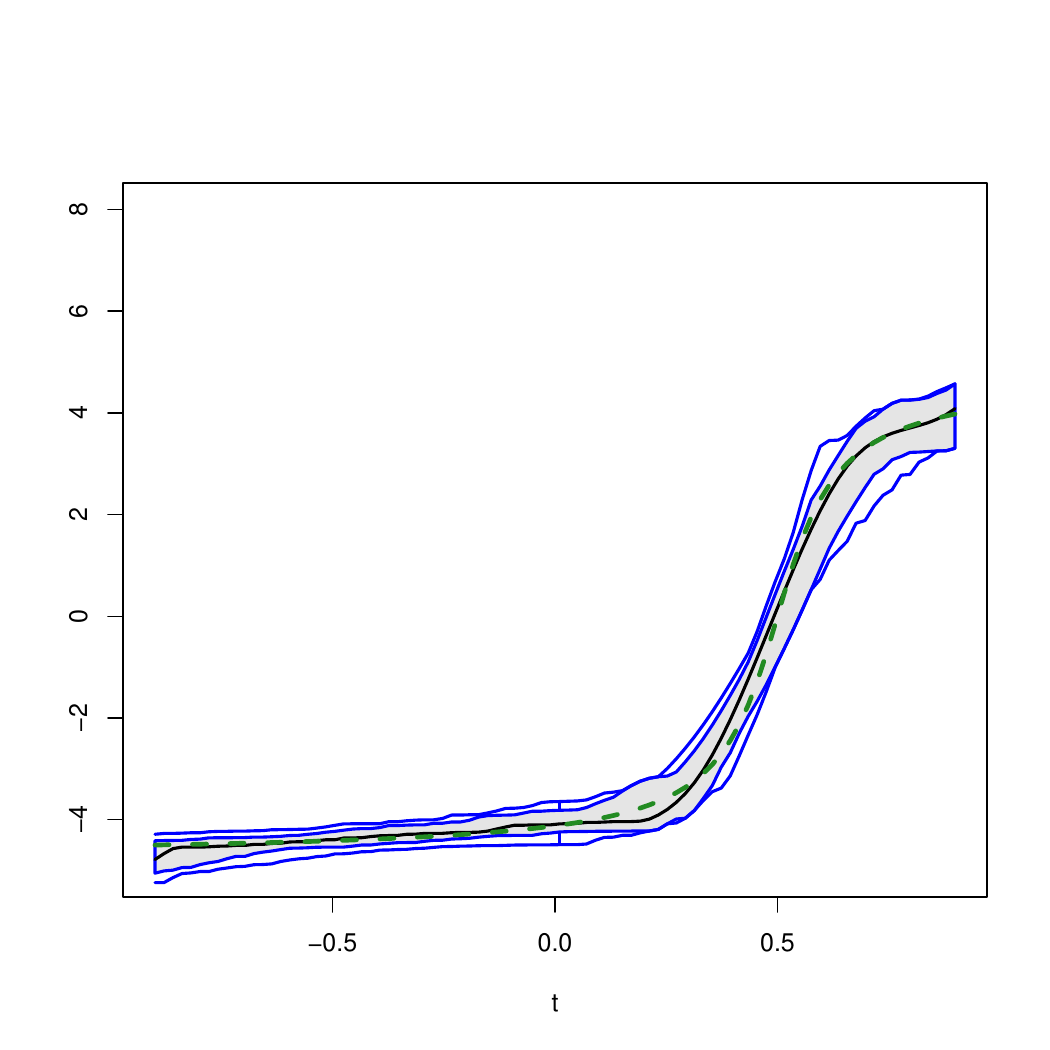} 
\\
$C_{1, 12}$ &  \includegraphics[scale=0.35]{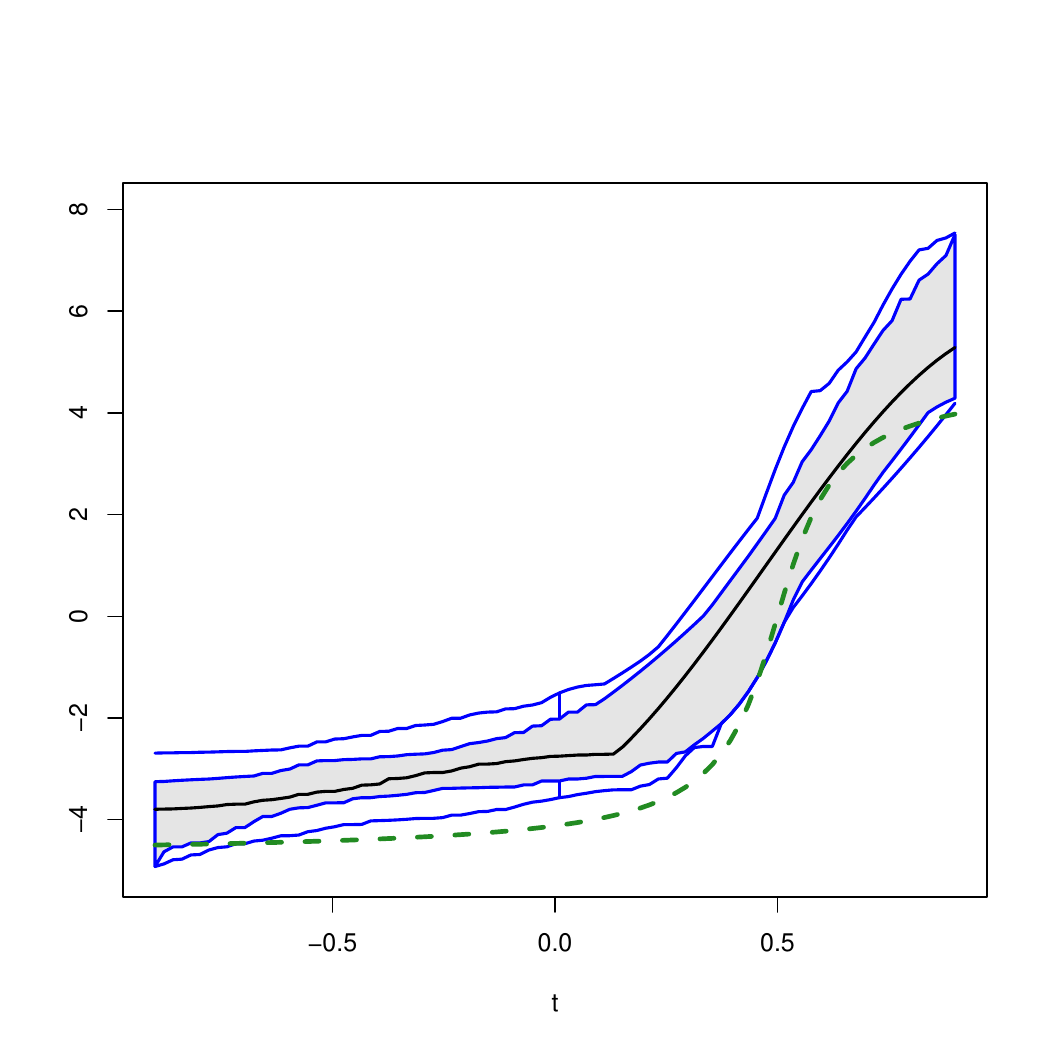}
 &  \includegraphics[scale=0.35]{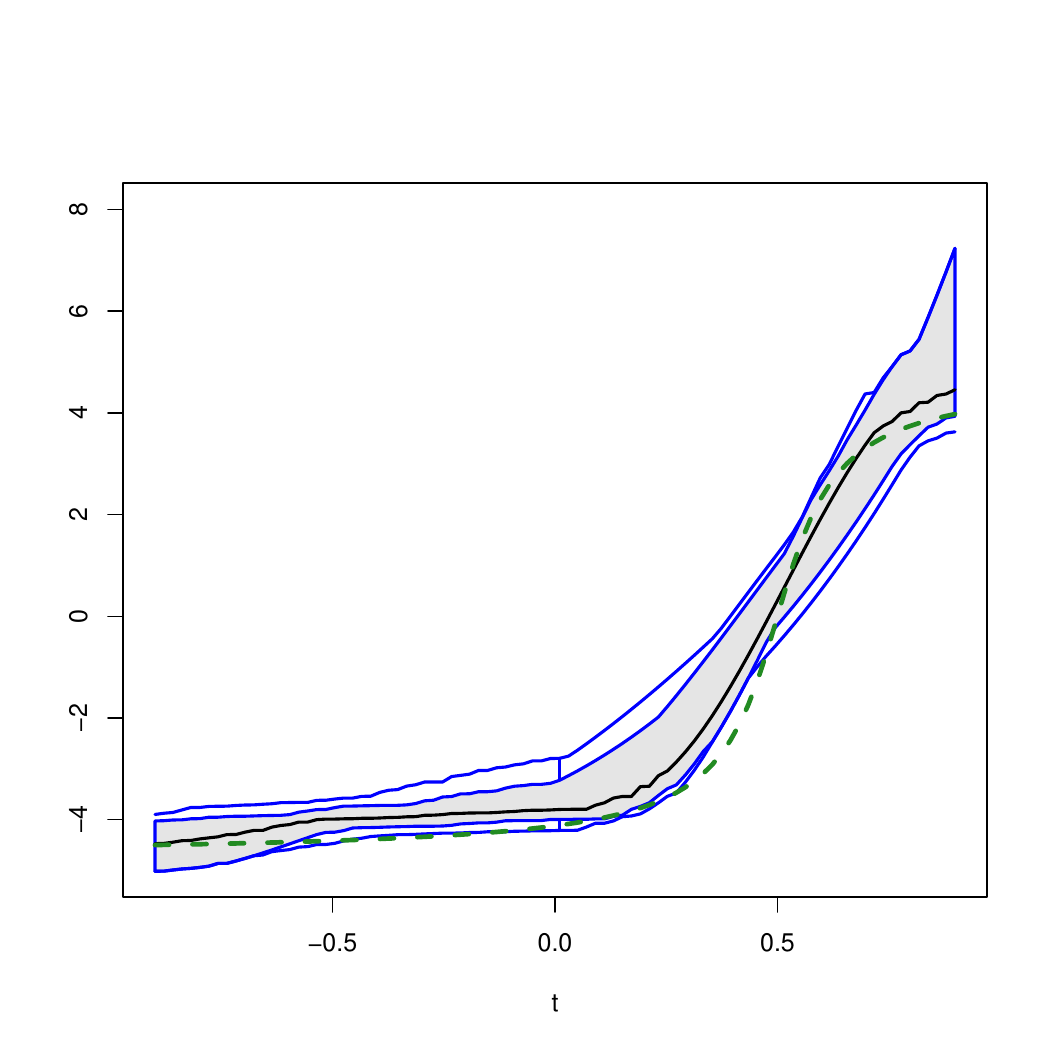}
  &  \includegraphics[scale=0.35]{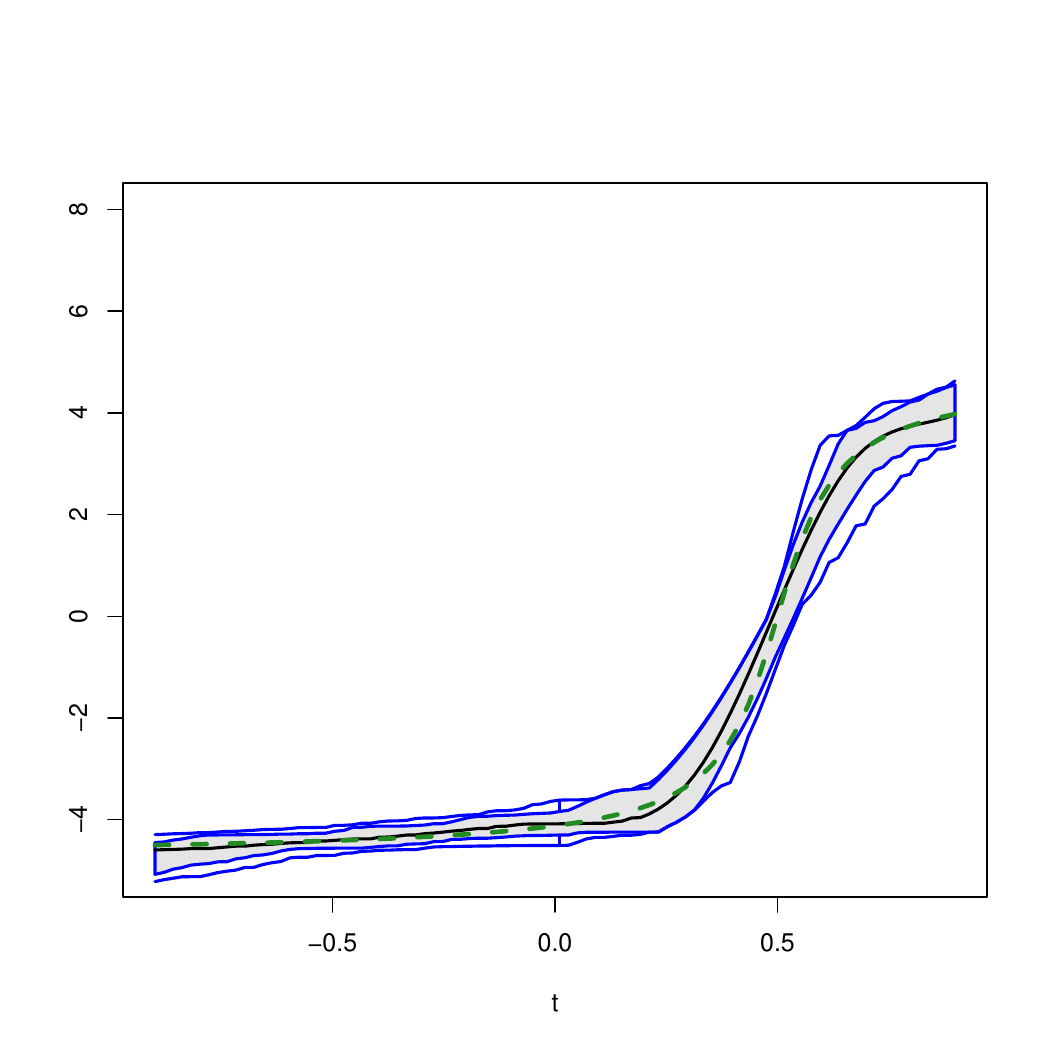}
  \\
   
$C_{2, 12}$ 
&  \includegraphics[scale=0.35]{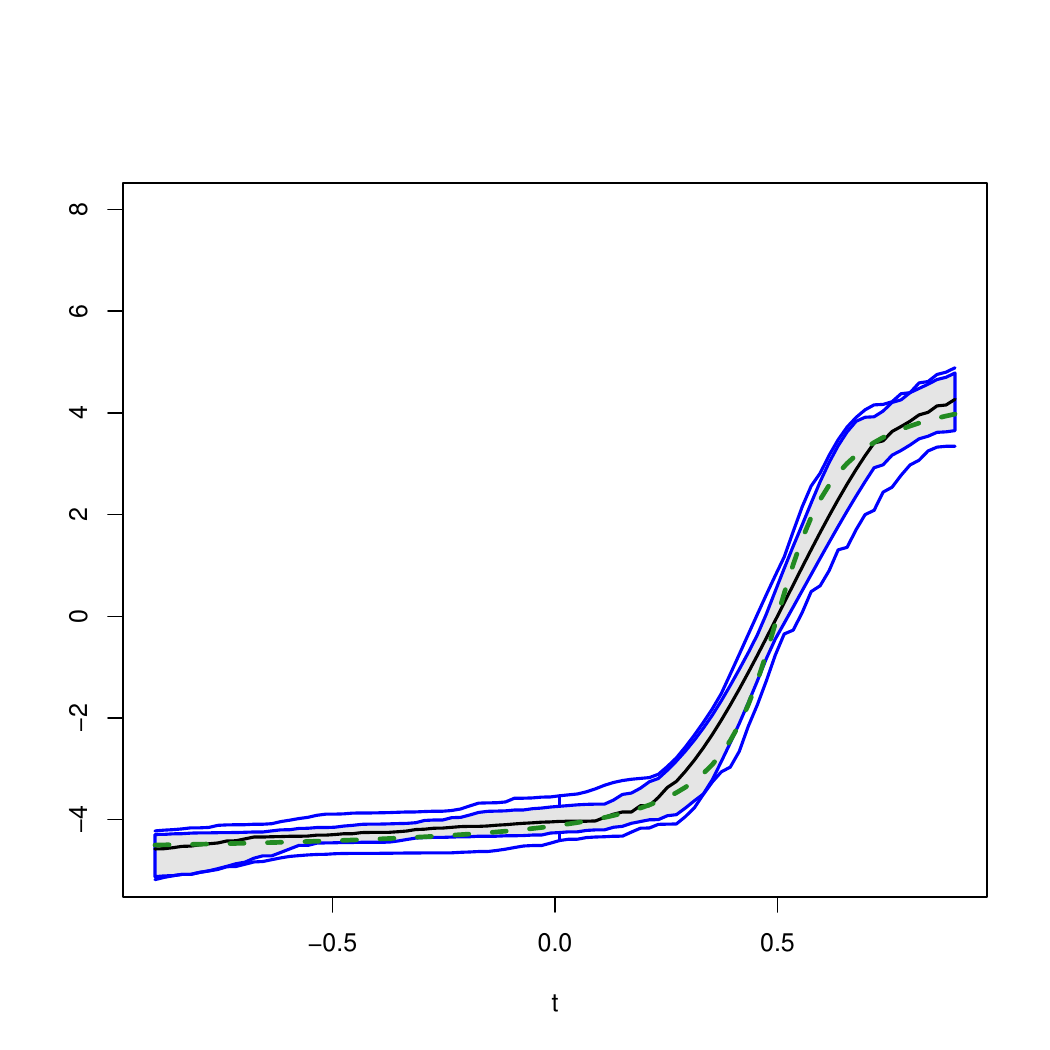}
&  \includegraphics[scale=0.35]{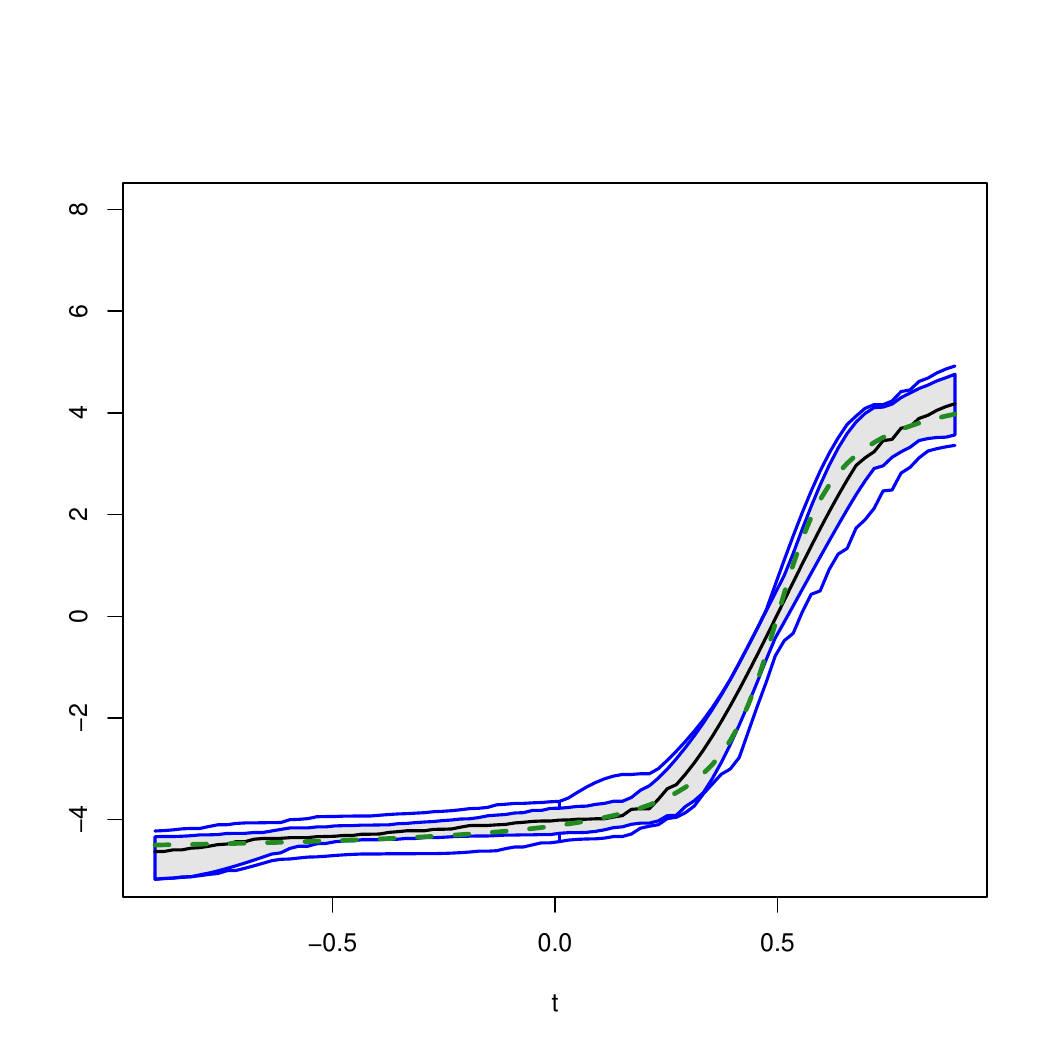}
&  \includegraphics[scale=0.35]{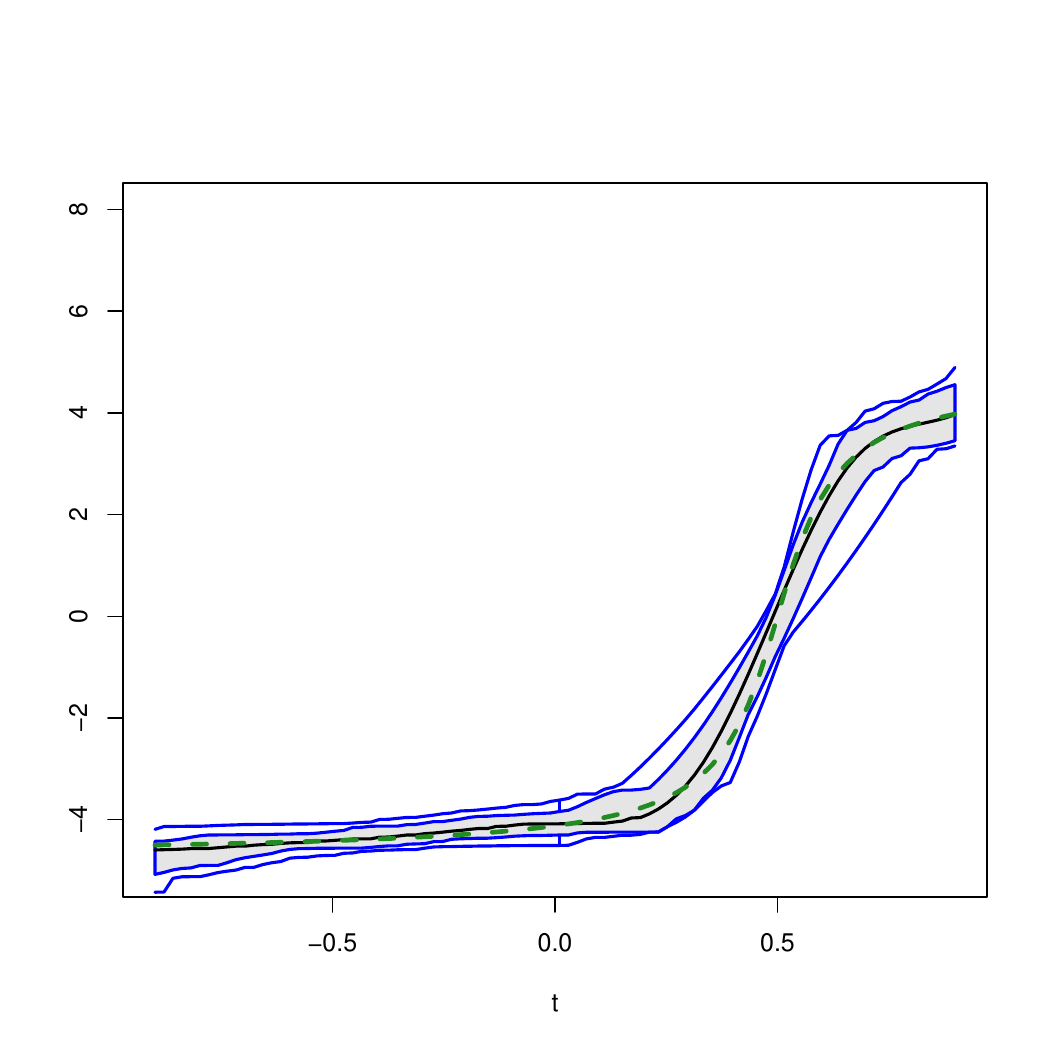}
\end{tabular}
\vskip-0.1in  \caption{\small \label{fig:weta-iso}  Functional boxplot of the monotone estimators for $\eta_0$. 
The true function is shown with a green dashed line, while the black solid one is the central 
curve of the $n_R = 500$ estimates $\wbeta$. Columns correspond to estimation 
methods while rows to three contaminations settings.}
\end{center} 
\end{figure}

In order to also explore visually the performance of these estimators, 
Figures \ref{fig:wbeta} to \ref{fig:weta-iso} contain functional 
 boxplots (Sun and Genton, 2011) for the $n_R = 500$ realizations of the different estimators
 for $\beta_0$ and $\eta_0$ under three
 contamination settings. As in standard boxplots, the central box
 of these functional boxplots represents the 50\% inner band of curves, the solid black line indicates the 
 central (deepest) function and the dotted red lines indicate outlying 
 curves (in this case: outlying estimates $\wbeta_j$ or $\weta_j$ for some $1 \le j \le n_R$). 
 We also indicate the target (true) functions $\beta_0$ and $\eta_0$ with 
 a dark green dashed line. 
To avoid boundary effects, we show here the different estimates $\wbeta_j$ or $\weta_j$
evaluated on the interior points of a grid of 100 equispaced points. 
In addition, to facilitate comparisons between contamination cases and estimation methods, 
the scales of the vertical axes are the same for all panels within 
each Figure.  
 
As expected, when the data do not contain outliers, all estimators behave similarly to each other 
(see Table \ref{tab:tabla-2-neu-Bspl}). When estimating the regression coefficient 
$\beta_0$, the
less efficient robust $MM$-estimator naturally results in higher MISE's. However, 
this efficiency loss
is much smaller for the estimators of $\eta_0$.
The serious damage caused to the least squares estimators by a small proportion of outliers (10\%) 
can be seen clearly in Figure \ref{fig:BIAS-MISE} (solid lines). The integrated squared bias and the 
MISE of the least squares estimators of $\beta_0$ and $\eta_0$  
are consistently much higher than those of the robust $MM$-estimators.
The ways in which the different oultiers affect the classical estimators for $\beta_0$ 
can be seen in Figure \ref{fig:wbeta}.
Note that under $C_{1, 12}$ the classical $\wbeta$ becomes highly variable, but mostly 
retains the same shape of the true $\beta_0$, which lies within the central box.  
However, with high-leverage outliers (as in $C_{2, 12}$) the estimator becomes completely uninformative,
and does not reflect the shape of the true regression coefficient $\beta_0$. 
The effect of outliers on the classical estimator for 
$\eta_0$ can be seen in Figure \ref{fig:weta} (and Figure \ref{fig:weta-iso} for
its monotone modification). We see that vertical outliers cause a vertical bias in the least squares 
estimator $\weta$, so that the central region of the functional boxplot fails to contain the 
true function $\eta_0$ for much of its domain.  As expected, the effect of high-leverage 
outliers on $\weta$ is less marked, but certain upwards bias is apparent. 

It is interesting to note that the $M$-estimators behave similarly to the classical ones. 
Vertical outliers result in more variable functional regression $M$-estimators 
$\wbeta$, although this increase is less pronounced than what we saw for the least squares estimators. 
High-leverage outliers are very damaging to these estimators.
In particular, note from Figure \ref{fig:BIAS-MISE} that in this case their integrated squared bias and MISE 
for $\wbeta$ are almost the same as those for the least squares estimator. 
We can also see this in Figure \ref{fig:wbeta}
where the $M$-estimators do not resemble the true function at all. 
Similar conclusions hold for the $M$-estimators for $\eta_0$. Vertical outliers produce an upward shift on the 
$\weta$'s, and a slight increase in variability, although this is less pronounced than what
happened with classical estimators. The behaviour of the $M$-estimator
 $\weta$ with high-leverage outliers is
similar to that of the least squares estimators, although notably their 
 integrated squared bias and MISE is worse than those of the least squares estimators 
 (Figure \ref{fig:BIAS-MISE}). 

In contrast, the $MM$-estimators display a remarkably stable behaviour across contamination
settings. Their bias and MISE curves in Figure \ref{fig:BIAS-MISE} show that the $MM$-estimators
for $\beta_0$ and $\eta_0$ are highly robust against both types of contamination scenarios considered here. 
If we look at the behaviour of these estimators in Figure \ref{fig:wbeta} we note that the central box and the 
``whiskers'' for the $MM$-estimators remain almost constant in all three simulation
scenarios (clean data, vertical outliers and high-leverage outliers), in sharp constrast to what happens to the other
estimators considered here. The number of affected $MM$ replicates for $\wbeta$ is higher for $C_{2, 12}$ than it is for $C_{1, 12}$, but even in the former case  
this happened for only 45 of the 500 $\wbeta$'s. 
We expect some effect on the 
estimators under this type of particularly damaging contamination, and we note that the 
robust proposal is the only one that can resist it in the vast majority of samples. 
The results in Figure \ref{fig:weta} tell the same story, but less strickingly so. The $MM$-estimators 
for $\eta_0$ are almost unaffected by the different types of outliers, and the 
functional boxplots remain very similar to each other.

\begin{table}[ht!]
\centering
   \renewcommand{\arraystretch}{1.2}
   \begin{tabular}{r rr   rr  rr}
\hline  
&\multicolumn{2}{c}{$\wbeta$} &\multicolumn{2}{ c}{$\weta$} &\multicolumn{2}{ c}{$\weta_{\monmod}$}\\
\hline
  Max over $C_{1, \mu}$ & Bias$^2_{\trim}$ & MISE$_{\trim}$ & Bias$^2_{\trim}$ & MISE$_{\trim}$ & Bias$^2_{\trim}$ & MISE$_{\trim}$ \\ \hline
  \textsc{cl} & 0.0053 & 1.8616 & 2.4201 & 2.8861 & 2.5686 & 2.8384 \\
  \textsc{m} & 0.0023 & 0.3499 & 0.4974 & 0.6651 & 0.5350 & 0.6438 \\
  \textsc{mm} & 0.0014 & 0.1317 & 0.0218 & 0.0739 & 0.0209 & 0.0529 \\  \hline
    Max over $C_{2, \mu}$ & Bias$^2_{\trim}$ & MISE$_{\trim}$ & Bias$^2_{\trim}$ & MISE$_{\trim}$ & Bias$^2_{\trim}$ & MISE$_{\trim}$ \\ \hline
  \textsc{cl} & 2.9685 & 3.1119 & 0.0540 & 0.1129 & 0.0474 & 0.0829 \\
  \textsc{m} & 2.9891 & 3.1318 & 0.0705 & 0.1217 & 0.0583 & 0.0903 \\
  \textsc{mm} & 0.0229 & 0.4341 & 0.0242 & 0.0828 & 0.0236 & 0.0607 \\
   \hline
\end{tabular}
\caption{Values of the trimmed summary measures for the worst contamination 
settings under each scenario.} \label{table:maxsummaries-trimmed}
\end{table}

Table \ref{table:maxsummaries-trimmed} 
reports the maximum values of Bias$^2_{\trim}$ and MISE$_{\trim}$ over $\mu$ for
the two contamination settings $C_{1,\mu}$ and $C_{2,\mu}$.  
Regarding the behaviour of the estimators of the functional regression parameter $\beta_0$, 
high-leverage outliers ($C_{2,\mu}$) are more damaging for the classical and $M$-estimators 
than ``vertical'' ones ($C_{1,\mu}$). The increase in square bias shows that the 
estimation is completely distorted for these estimators.  
Note that 
the trimmed squared bias  increases more than 1000 times and the MISE$_{\trim}$ more than 30 times with respect to those reported in Table \ref{tab:tabla-2-neu-Bspl}.
In contrast, the classical estimator of $\eta_0$ is only slightly affected by $C_{2,\mu}$, since both the  MISE$_{\trim}$ are increased by a factor of at most 2.5 with respect to the ones obtained for clean samples. 
Vertical outliers, however, produce increases of  more than $40$ times when $\mu=16$ where the maximum is attained (see Figure \ref{fig:BIAS-MISE}). 
Even though the $M$-estimators of Huang \textsl{et al.} (2015) are affected by both contaminations, 
they deteriorate less than the classical ones. 
In particular, when estimating $\eta_0$, the $M$-estimator at least triples the $\MISE_{\trim}$  under $C_{1,\mu}$ with respect to that obtained for clean samples (the worst effect is observed when $\mu=16$, reaching 10 times the value under $C_0$). 
Finally, the $MM$-estimators are quite stable under the considered contaminations.  In particular, under $C_{1,\mu}$ the values of Bias$^2_{\trim}$of $\weta_{\eme}$ are between 8 and 20 times larger than those of  $\weta_{\eme\eme}$, even for their monotone counterparts (see Figure \ref{fig:BIAS-MISE}), while the   increase of the  $\MISE_{\trim}$ of $\weta_{\eme}$ varies between 3 and   9. Regarding the performance of the estimators of $\beta_0$, under $C_{1,\mu}$, the differences between the $MM$-estimators and the $M$-estimators are less pronounced than those between the $MM$- and the classical one.  
Under $C_{2,\mu}$, the  worst $\MISE_{\trim}$ of $\wbeta_{\eme\eme}$ is multiplied less than 
5 times with respect to that obtained under $C_0$. However, 
it is still only a sixth of the $\MISE_{\trim}$ for the classical and $M$-estimator, 
which suffer from a huge bias. In all cases, the $\MISE_{\trim}$ of $\weta_{\monmod}$ 
is smaller than that of $\weta$.

\section{Real data examples} \label{sec:ejemplos}

\subsection{German electricity prices} \label{sec:ejemplo_germany}

In our first example we look at the relationship between 
hourly electricity prices and the overall load of the German energy 
system. As discussed in Liebl (2013), such analysis needs to consider that 
eolic energy prices in this type of markets 
follow a different price regime. 
The data consist of hourly electricity prices in Germany 
between 1 January 2006 and 30 September 2008, as
traded at the Leipzig European Energy Exchange, German 
electricity demand (as reported by the European Network of Transmission System
Operators for Electricity), and the amount of 
eolic energy in the system (taken from the EEX Transparency
Platform). The data set is available from the on-line supplementary materials of 
Liebl (2013). Weekends, holidays and other non-working days were removed
from the dataset. 
Our model is 
 \begin{equation}
  y =  \langle X, \beta_0 \rangle + \eta_0(z) + \sigma_0 \epsilon \, ,
  \label{eq:electricity-model}
\end{equation}
where $y$ is the daily average hourly energy demand, 
$X$ is the curve of energy prices (as a function of time) observed hourly, and $z$ is the
mean hourly amount of wind-generated electricity in the system for that day. 
As usual, $\epsilon$ is a random variable centered at zero, and independent from 
$X$ and $z$. The shape 
of the function $\beta_0$ can be used to identify times of the day
when hourly prices are informative regarding the overall system demand. 

In addition to our proposed robust $MM$-estimators, we also computed the classical least squares and
the $M$-estimators of Huang \textsl{et al.\!}\! (2015). 
The robust $MM$-estimators were calculated using 
 the same $\rho$-functions as in our simulation study and we
 selected the size of the splines bases with the $RBIC$ criterion \eqref{eq:bic1}.
 Following Huang \textsl{et al.\!}\! (2015), 
 the $M$-estimator was computed using a Huber function with tuning constant 
 equal to $1.345$ and no scale estimator.  
 
\begin{figure}[ht!]
\centering
\subfigure[Estimates of $\beta_0$]{\includegraphics[width = .4 \textwidth]{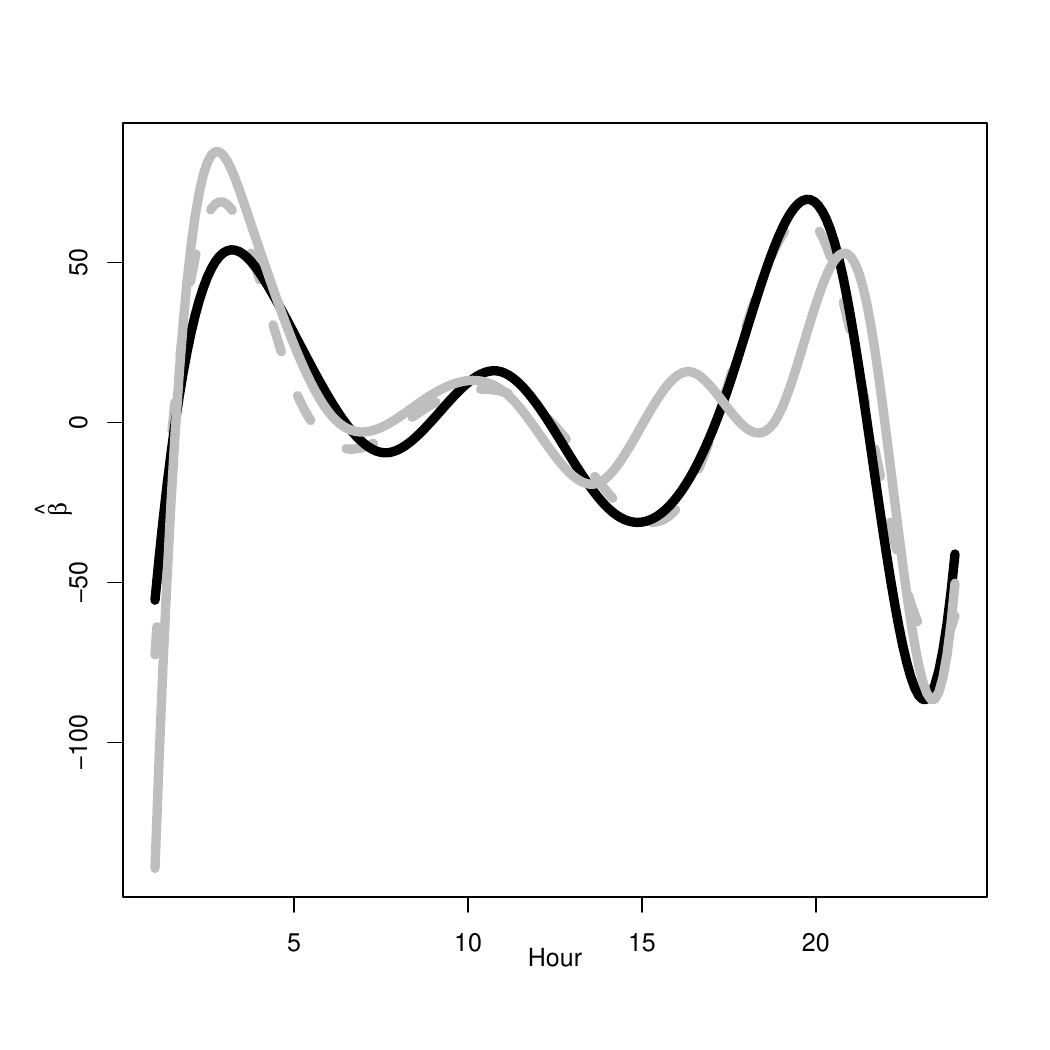} \label{fig:german-betas}}
\subfigure[Estimates of $\eta_0$]{\includegraphics[width = .4 \textwidth]{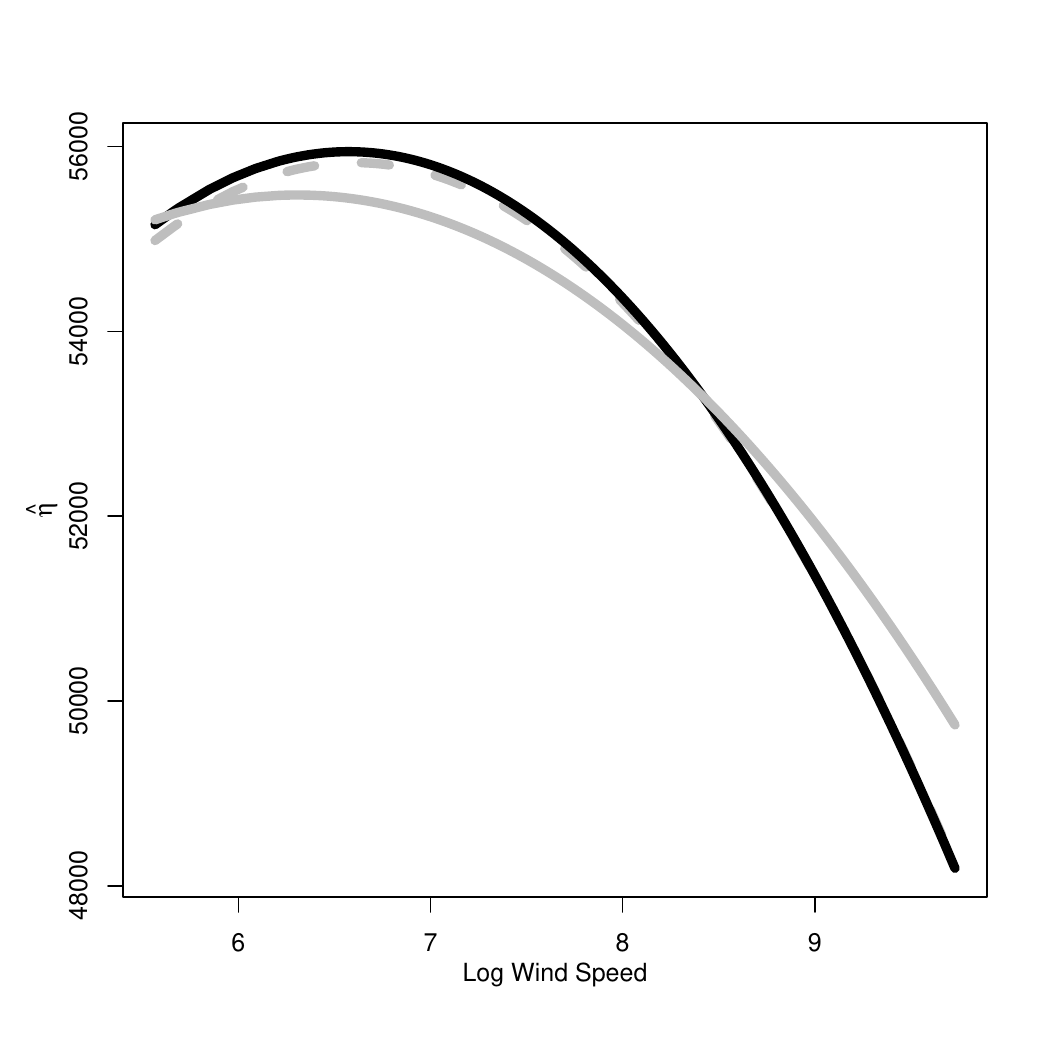} \label{fig:german-etas}}
\caption{\small  German Electricity: Estimates of $\beta_0$ and $\eta_0$. The  black  line corresponds to the $MM$-fit, while the  solid and dashed gray ones correspond to the least squares   computed with the whole   training set and  without the outliers, respectively.
The $M$-estimators were almost identical to the least squares ones, and not included in
this plot.}
\end{figure}

The estimators for $\beta_0$ and $\eta_0$ 
are shown in 
Figures \ref{fig:german-betas} and \ref{fig:german-etas},
respectively. Solid black lines are used for the $MM$-estimator,
and solid gray ones for the least squares one. 
The $M$-estimators were indistinguishable from the classical ones and 
so we did not 
include them in these plots. 
Comparing the  $MM$ and classical estimators for $\beta_0$, 
we note that the robust fit identifies two ``peak'' times 
(around 4am and 8pm) and two ``slump'' times around 3pm and 11pm, 
where prices have a larger (in magnitude) association with 
the daily average load in the system. 
However, the least squares fit appears to not include the early afternoon prices 
as important (note that the magnitude of the function $\hat{\beta}_0$ is smaller 
than that of the robust estimator between 2pm and 8pm). 
On the other hand, although the estimators for  
$\eta$ are slightly different,
their shapes are rather consistent with each other.  

We next identified potential outliers in the data by using a boxplot of the residuals from the
robust $MM$-fit. The dashed lines in Figures \ref{fig:german-betas} and \ref{fig:german-etas}
correspond to the classical fit computed without these possible atypical observations. 
We note that the classical estimators computed without these potential 
outliers are very close to the
robust ones. In other words, the robust estimator behaves similarly to 
the classical one if one were able to manually remove suspected outliers. 
 
\subsection{Tecator}{\label{sec:ejemplo_tecator}}

The Tecator data set was
analysed in Ferraty and Vieu (2006),
Aneiros-P\'erez and Vieu (2006), Shang (2014) and Huang \textsl{et al.} (2015), 
and it is available in the package \texttt{fda.usc} 
(Febrero-Bande and Oviedo de la Fuente, 2012). 
See also  \url{http://lib.stat.cmu.edu/datasets/tecator}. 
These data contain measurements taken on samples from finely chopped meat with different 
percentages of fat, protein and moisture content. 
Each observation consists of a spectrometric curve, $\itX_i$, which corresponds to 
the absorbance measured on an equally spaced grid of 100 wavelengths between 850 and 1050nm. 
The goal of the analysis is to predict the fat content ($y$) 
using the spectrometric curve ($\itX$) and the variables 
water ($v$) and protein contents ($z$).

 Huang \textsl{et al.} (2015) compared several models in terms of their 
 predictive properties. They used the second derivative $X$ of the 
 spectrometric curve as the functional covariate, which enters
 the model linearly, while the variables $(z, v)$ appear either through
 an additive non-parametric component or 
 a varying coefficient model
 \begin{equation}
  y = \gamma_0+\langle X, \beta_0 \rangle + v \, \eta_0(z) + \sigma_0 \epsilon \, ,
  \label{eq:fplm-tecator-modelo}
\end{equation}
where $\gamma_0 \in \real$ and $\eta_0$ is a smooth function of $z$. 

Following Aneiros-P\'erez and Vieu (2006), the sample was divided into  a training set 
(corresponding to the first $155$ observations)
and a testing one with the remaining $60$ data points. 
As in Section \ref{sec:simu}, the $M$-estimators of Huang \textsl{et al.} (2015) were computed using a Huber function with tuning constant $1.345$ and no scale estimator. 
The robust $MM$-estimators were calculated using 
 the same $\rho-$functions as in our simulation study and we
 selected the size of the splines bases with the $RBIC$ criterion \eqref{eq:bic1}.
To compare the predictions obtained with the
different estimators 
we  computed the mean and median  square prediction errors on the test set:
$$ 
  MSPE = \frac{1}{n_{\itJ}} \sum_{j \in \itJ} 
  \frac{(y_i-\wy_i)^2}{s^2_{\itJ}} \quad\mbox{and}\quad MedSPE 
  = \frac{\mbox{median} (y_i-\wy_i)^2}{s^2_{\itJ}} \, ,
 $$
where $\itJ$ contains the indices of the observations in the test set, 
$n_{\itJ}$ denotes its size, 
and $s_{\itJ} = \mad_{j \in \itJ}(y_j)$. 

The first three columns 
and two rows of Table \ref{tab:fplm-tecator-MSPEclean} report the mean and median square prediction errors for
the classical, $M$ and $MM$-estimators. Although the least squares and $M$ fits  have   lower 
mean squared prediction errors than that of the robust $MM$ one, their larger median suggests that 
most prediction errors may in fact be smaller for the robust estimator, but
that a few outliers may be present in the test set. 
 \begin{table}[ht!]
\centering
\begin{tabular}{c c c c c}
\hline
                            & \textsc{ls}  & \textsc{m} & \textsc{mm} & \textsc{ls}$^{-\;out}$ \\ \hline
$1000\times MSPE$           & 2.52    & 2.44  &  4.56   &  4.83 \\ \hline
$1000\times MedSPE$         & 0.95    & 0.85  &  0.65   &  0.78 \\ \hline  
$1000\times MSPE_{\clean}$  & 1.47    & 1.42  &  1.33   &  1.40 \\ 
    \hline
  \end{tabular}
  \caption{\small \label{tab:fplm-tecator-MSPEclean}   Mean and median  square prediction errors of the classical, $M$ and $MM$-estimators labelled \textsc{ls}, \textsc{m} and \textsc{mm}, respectively.}
\end{table}

To evaluate the ability of the procedure to predict 
non-outlying observations,  
we also computed the mean squared prediction errors over non-outlying points
 in the test set:
$$
MSPE_{\clean} = \frac{1}{n_{\itJ} - \sum_{i \in \itJ} 
    \gamma_i} \; \sum_{j \in \itJ} (1-\gamma_j) \,
  \frac{(y_j-\wy_j)^2} {s_{\itJ}^2} \,,$$ 
where $\gamma_i = 1$ if the $i$-th observation was flagged as atypical,
and 0 otherwise. 
To identify potential outliers in the data
we used the boxplots of the residuals 
from the fits obtained using the $MM$-procedure both for the training and testing sets.
The mean squared prediction errors of both estimators using the non-outlying 
points in the test set ($MSPE_{\clean}$) are reported in the last row of Table  \ref{tab:fplm-tecator-MSPEclean}. Note that now the  $MSPE_{\clean}$ for the $MM$-estimators  is smaller than those of the least squares and $M$-estimators. 
The fourth column  ($\textsc{ls}^{-out}$) of that table displays the results obtained with the classical 
estimator when it was computed without the 13 potential outliers in the training set. Since the $M$-estimators of Huang \textsl{et al.} (2015) 
behaved very similarly to the classical ones from now on,  we 
only comment the  results obtained when using the least squares and the $MM$-estimators. 

\begin{figure}[ht!]
  \centering \subfigure[\label{fig:fplm-tecator-betas} Estimates of
  $\beta_0$] {
  \includegraphics[width = .45\textwidth]{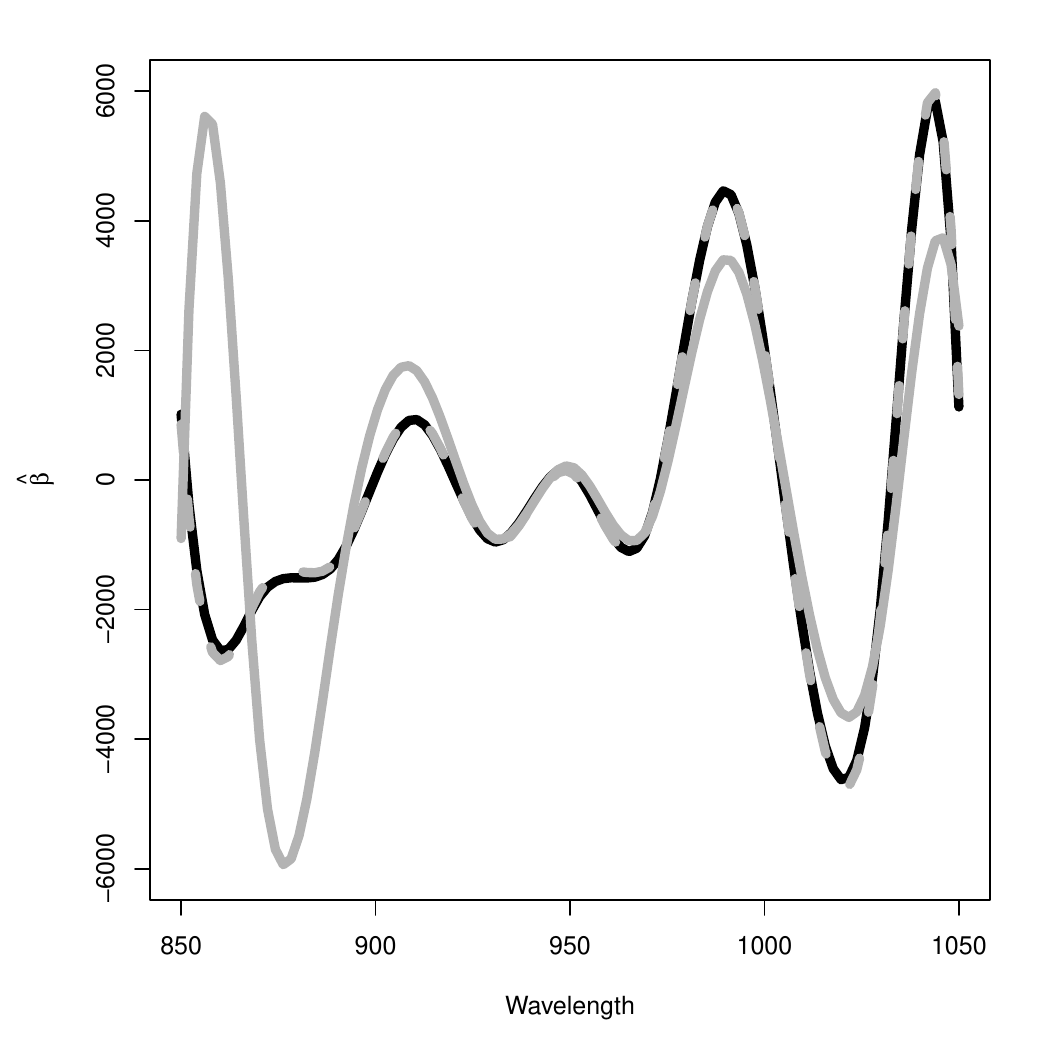}}
  \subfigure[\label{fig:fplm-tecator-etas} Estimates of $\eta_0$]
  {\includegraphics[width = .45\textwidth]{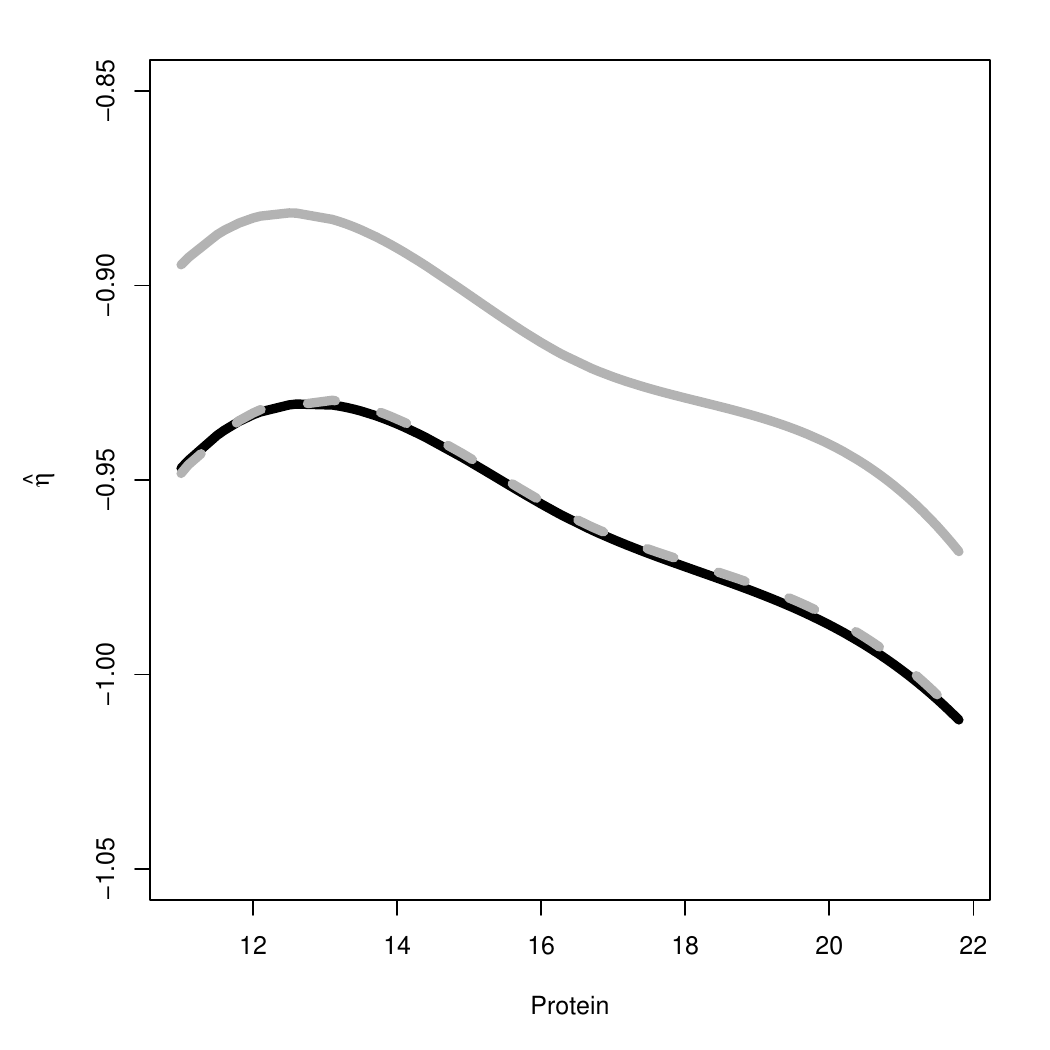}}
  \caption{\small  Tecator: Estimates of $\beta_0$ and $\eta_0$. The  black  line corresponds to the $MM$-fit, while the    solid and dashed gray ones correspond to the least squares   computed with the whole   training set and  without the outliers, respectively.}
\end{figure}

As in the German Electricity example, we 
note that the classical procedure trained after eliminating potential atypical observations
gives very similar results to those obtained with the  $MM$-estimator. 
The black and gray solid lines in 
Figures \ref{fig:fplm-tecator-betas} and \ref{fig:fplm-tecator-etas}
show the estimators $\weta$ and $\wbeta$ 
obtained using the classical and robust estimators, respectively. 
On both panels we also overlay (in dashed gray lines) the 
corresponding least squares estimates computed on the ``cleaned'' training set. 
In both cases it is clear that the classical estimators are seriously affected by the atypical 
training points, while the robust estimator provides estimates similar to those that are obtained 
with the classical methods after removing possible outliers. 
 
\section{Conclusion}{\label{sec:concl}}
In this paper we propose robust estimators based on $B$-splines for semi-functional linear regression models. 
Our estimators are robust against outliers in the response variable, and also in the functional 
explanatory variables. 
Furthermore, we propose a robust $BIC$-type criterion to select the optimal dimension
of the splines bases that works very well in practice. 
We prove that the estimators are strongly consistent under standard regularity
conditions, and show how they can be extended rather straightforwardly to other 
semiparametric models with functional covariates that enter the model linearly. 
A simulation study shows that our proposed estimators have good robustness
and finite-sample statistical properties. 

We apply our method to two real data sets and confirm that the robust 
$MM$-estimators remain reliable even when the training set contains
atypical observations in the functional explanatory variables. Moreover, the
residuals obtained from the robust fit provide a natural way to identify potential
atypical observations.


\section{Acknowledgements.}
The authors would like to thank Professor Ioannis Kalogridis for his careful reading of earlier versions of this manuscript and for pointing out a mistake in the proof of Theorem 3.2 which has been corrected in the current version.
This research was partially supported by    Universidad de Buenos Aires [Grant  20020170100022\textsc{ba}]  and  \textsc{anpcyt} [Grant   \textsc{pict} 2018-00740] at  Argentina (Graciela Boente and Pablo Vena), the   Ministry of Economy, Industry
and Competitiveness, Spain (MINECO/AEI/FEDER, UE) [Grant  {MTM2016-76969P}] (Graciela Boente) and by the Natural Sciences and Engineering Research Council of Canada [Discovery Grant RGPIN-2016-04288] (M. Salibi\'an Barrera).

 \setcounter{section}{0} 
\renewcommand{\thesection}{\Alph{section}}

\setcounter{equation}{0}
\renewcommand{\theequation}{A.\arabic{equation}}

\section{Appendix}
\subsection{Proofs of  Proposition \ref{sec:consis}.1 and Theorem \ref{sec:consis}.1}{\label{sec:appen}}
In what follows, $\itV$ stands for a neighbourhood of  $\sigma_0$ with closure  $\overline{\itV}$ strictly included in $ (0,\infty)$.
Furthermore, recall that  we have denoted $\itL_r = \itL_r ([0,1])$, $r \geq 1$, 
the space of  functions whose $r-$th derivative satisfies a 
Lipschitz condition in $[0,1]$:
\begin{multline*}
\itL_r ([0,1])= \Bigl\{ g\in C^r \left([0, 1] \right):
  \big \|g^{(j)} \big \|_\infty<\infty, \;0\leq j\leq r \, ,    \mbox{ and } \,
  \sup_{z_1\ne z_2} \frac{\big |g^{(r)}(z_1) - g^{(r)}(z_2) \big|}{|z_1- z_2|}
  <\infty \Bigr\} \, .
  \end{multline*}
  We use the following norm
$$\|f\|_{\itL_r}=\max_{1\le j\le r} \big \|f^{(j)} \big \|_{\infty}+ \sup_{x \neq y,
  x,y \in (0,1)} \frac{\big |f^{(r)}(x)-f^{(r)}(y) \big|}{|x-y|}\,,$$
where $f^{(j)}$ stands for the $j$-th derivative of $f$. The unit ball in $\itL_r$ will be denoted as 
$\itV_1^{(r)}=\{f\in \itL^r([0,1]): \|f\|_{\itL_r} \le 1 \}$. 

We begin by stating some Lemmas that will be used in the proofs of Proposition \ref{sec:consis}.1 and Theorem \ref{sec:consis}.1.
Lemma \ref{sec:appen}.1 entails that the functional related to the considered estimators are indeed Fisher--consistent, which is a condition needed to ensure that we are estimating the target quantities. 

\vskip0.2in
\noi \textbf{Lemma \ref{sec:appen}.1.} \textsl{Assume that \ref{ass:densidad}  holds  and let $\rho$ be a  function satisfying \ref{ass:rho}. Then, we have that, for any $\sigma>0$,
\begin{itemize}
\item[a)] $M(\beta,\eta, \sigma)\ge M(\beta_0,\eta_0, \sigma)$, where
$$M(\beta,\eta, \sigma)=\esp \,\rho\left(\frac{y - \langle X, \beta\rangle -\eta(z)}{\sigma}\right)\,.$$ 
\item[b)] If in addition   \ref{ass:probaX} holds, $(\beta_0,\eta_0)$ is the unique minimizer of $M(\beta,\eta, \sigma)$. 
\end{itemize}}

\vskip0.2in
\noi \textsc{Proof.} Lemma 3.1 of Yohai (1987) together with \ref{ass:rho}  and the fact that $\widetilde{\epsilon}=\epsilon \sigma_0/\sigma$ satisfy assumption \ref{ass:densidad}  
imply that for all $a \ne 0$, 
\begin{equation}
\esp  \left[ \rho \left( \epsilon \frac{\sigma_0}{\sigma} -a \right) \right] > 
\esp  \left[ \rho \left( \epsilon \frac{\sigma_0}{\sigma}  \right) \right]\,,
\label{eq:cotaesp}
\end{equation} 
and a) follows immediately.

To derive b), denote as $\itA_0 \, = \, \left\{(X,z) : \Phi(X,z)=\langle X,\beta-\beta_0 \rangle+ \eta (z)-\eta_0(z) = 0 \right\}$ and $a(X,z)= \Phi(X,z)/\sigma$. Then, we have  that
\begin{eqnarray*}
M(\beta,\eta, \sigma)&=&\esp \rho\left(\frac{\epsilon \sigma_0 - \Phi(X,z)}{\sigma}\right)=\esp \rho\left(\epsilon  \frac{\sigma_0}{\sigma} -\frac{ \Phi(X,z)}{\sigma}\right)\\
&=& \esp\left\{  \rho\left(\epsilon \frac{\sigma_0}{\sigma}\right) \indica_{\itA_0}(X,z)\right\} \, + \,
\esp\left\{\esp\left[ \rho\left(\epsilon \frac{\sigma_0}{\sigma} -a(X,z)\right) |(X,z)\right]\indica_{\itA_0^{c}}(X,z)\right\}\\
&=& \esp\left( \rho\left(\epsilon \frac{\sigma_0}{\sigma}\right) \right)\esp\left\{\indica_{\itA_0}(X,z)\right\} \, + \,
\esp\left\{\esp\left[ \rho\left(\epsilon \frac{\sigma_0}{\sigma} -a(X,z)\right) |(X,z)\right]\indica_{\itA_0^{c}}(X,z)\right\} \,.
\end{eqnarray*}
Note that (\ref{eq:cotaesp}) entails that, for any $(X,z)\notin \itA_0$,  
\begin{eqnarray*}
\esp\left[ \rho\left(\epsilon \frac{\sigma_0}{\sigma} -a(X,z)\right) |(X,z)=(X_0,z_0)\right]&=& \esp\left[ \rho\left(\epsilon \frac{\sigma_0}{\sigma} -a(X_0,z_0)\right) |(X,z)=(X_0,z_0)\right]\\
&=& \esp\left[ \rho\left(\epsilon \frac{\sigma_0}{\sigma} -a(X_0,z_0)\right)\right]>\esp  \left[ \rho \left( \epsilon \frac{\sigma_0}{\sigma}  \right) \right]\,,
\end{eqnarray*}
where the last equality follows from the fact that the errors are independent of the covariates.   Thus, using that $\prob(\itA_0^{c})>0$ from  \ref{ass:probaX}, we get  
\begin{align*}
M(\beta,\eta, \sigma)&=  \esp \rho\left(\epsilon \frac{\sigma_0}{\sigma}\right)  \,  \prob\left(   {\itA_0} \right)  \, + \,
\esp\left\{\esp\left[ \rho\left(\epsilon \frac{\sigma_0}{\sigma} -a(X,z)\right) |(X,z)\right]\indica_{\itA_0^{c}}(X,z)\right\} 
\\
&>  \esp \rho\left(\epsilon \frac{\sigma_0}{\sigma}\right)  \,  \prob\left(   {\itA_0} \right) \, + \,
\esp\left\{\esp\left[ \rho\left(\epsilon \frac{\sigma_0}{\sigma} \right) \right]\indica_{\itA_0^{c}}(X,z)\right\} \\
& >  \esp \rho\left(\epsilon \frac{\sigma_0}{\sigma}\right)  \,  \prob\left(   {\itA_0} \right)  +  \esp \rho\left(\epsilon \frac{\sigma_0}{\sigma}\right)  \,  \prob\left(   {\itA_0^{c}} \right)\\
&>  \esp\left( \rho\left(\epsilon \frac{\sigma_0}{\sigma}\right) \right)=M(\beta_0,\eta_0, \sigma)\,.\; \qed
\end{align*}

Let $\itM_{p_s}^{(s)}$, $s=1, 2$, denote the linear spaces
spanned by the $B$-splines bases of size $p_1$ and $p_2$:
$$
\itM_{p_s}^{(s)}=\left\{  \sum_{j=1}^{p_s} b_j \,   B_j^{(s)}(t) \, ,\, \bb \in \real^{p_s}\right\}\quad s=1,2 \, .
$$
Recall that $p_1$ and $p_2$ increase with the sample size $n$. 
The  Lemmas below will be useful to derive consistency of the proposed estimators. 
In particular, the next lemma shows that 
$$
M_n(\beta,\eta, \sigma) =  \frac 1n \sum_{i=1}^n \
\rho\left(\frac{y_i - \langle X_i, \beta\rangle -\eta(z_i)}{\sigma}\right) \, ,
$$
converges to $M(\beta,\eta, \sigma)$ with probability one, uniformly over $\sigma > 0$ and 
$\itM_{p_1}^{(1)} \times \itM_{p_2}^{(2)}$.

\vskip0.2in
\noi \textbf{Lemma \ref{sec:appen}.2.} \textsl{Let $\rho$ be a bounded function satisfying \ref{ass:rho} and \ref{ass:psi} and assume that \ref{ass:spacing} holds. Let $p_1=k_{n,\beta}$ and $p_2=k_{n,\eta}$. Then, we have that, 
\begin{itemize}
\item[a)] $\sup_{\sigma >0, \beta \in \itM_{p_1}^{(1)}, \eta \in \itM_{p_2}^{(2)}} \left|M_n(\beta, \eta, \sigma)- M(\beta, \eta, \sigma)\right| \convpp 0$.
\item[b)] Furthermore, 
$$\sup_{\sigma >0, \beta \in \itM_{p_1}^{(1)}, \eta \in \itM_{p_2}^{(2)}} \left| \frac 1{n-p_1-p_2} \sum_{i=1}^n \left[\rho\left(\frac{y_i - \langle X_i, \beta\rangle -\eta(z_i)}{\sigma}\right)-M(\beta, \eta, \sigma)\right]\right| \convpp 0\,.$$
\end{itemize}}

\vskip0.2in
\noi \textsc{Proof.} b) follows immediately from a) noting that $n/(n-p_1-p_2)\to 1$. 

To prove (a) we need to introduce some notation. 
For any measure $\qu$, $N(\epsilon, \itF_n, L_s(\qu))$ and $N_{[\;]}(\epsilon, \itF_n, L_s(\qu))$ stand  for the covering  and bracketing  numbers  of the class $\itF_n$ with respect to the distance in $ L_s(\qu)$, as defined, for instance, in van der Vaart and Wellner (1996).

Recall that we have denoted as   $\bB(z)= ( B_1^{(2)}(z ), \dots,  B_{p_2}^{(2)}(z ))\trasp$  and $\bx=(\langle X,B_1^{(1)}\rangle, \dots,$ $ \langle X,B_{p_1}^{(1)}\rangle )\trasp$ and define the class of functions
$$\itF_n=\{f(y,\bx,z)=\rho\left(\frac{y - \bb\trasp \bx - \ba\trasp \bB(z)}{\sigma}\right) , \bb \in \real^{p_1}, \ba\in \real^{p_2}, \sigma>0  \}\,,$$
To obtain a), first note that the boundedness of $\rho$ and \ref{ass:rho}, entails that the class  $\itF_n$ 
has envelope 1. Lemma S.2.1 in Boente \textsl{et al.} (2020) allows to bound, for any probability measure $\qu$, the covering number $N(2\epsilon;\itF_n ;L_1(\qu))$ as
\begin{equation}
N(2\epsilon;\itF_n ;L_1(\qu))\le  \left[K q_n\; (16e)^{q_n}\left(\frac 1\epsilon\right)^{q_n-1}\right]^2\,,
\label{eq:cotaNL1}
\end{equation}
where $q_n=2(p_1+p_2+3)-1$. 
Hence,  using that   $\log(q_n)/({p_1+p_2 + 3})<1$ and assuming without loss of generality that $K>1$, we get from \eqref{eq:cotaNL1} that
\begin{eqnarray*}
\log\left(N(2\,\epsilon, \itF_n, L_1(\qu)) \right) &\leq & \log\left[K q_n\; (16e)^{q_n}\left(\frac 1\epsilon\right)^{q_n-1}\right]^2\\
&\le & 2 \left\{\log(K)+\log q_n \; + q_n \log(16 e)+ (q_n-1)\log\left(\frac 1\epsilon\right)\right\} \\
&\le & 2 \left\{ q_n \left[\log(K)+1+  \log(16 e)+ \log\left(\frac 1\epsilon\right)\right]\right\} \\
& \le & C (p_1+p_2) \log\left(\frac 1\epsilon\right)\,,
\end{eqnarray*}
for $\epsilon<  \min((16e)^{-1}, e^{-K})$ and some constant $C$. Thus, using that from \ref{ass:spacing} $p_j=O(n^{\nu})$ with $\nu<1$, we conclude that
\begin{eqnarray*}
\frac 1n \log N(2\epsilon,\itF_n,L_1(P_n)) & \le &  C\, \frac{p_1+p_2}n  \,  \log\left(\frac 1\epsilon\right)\to 0\,,
\end{eqnarray*}
which entails,  (see, for instance, exercise 3.6 in van  der Geer, 2000 with $b_n=2$),   that 
$$\sup_{\sigma>0, \beta \in \itM_{p_1}^{(1)}, \eta \in \itM_{p_2}^{(2)}} \left|M_n(\beta, \eta, \sigma)- M(\beta, \eta, \sigma)\right| \convpp 0\,,$$
concluding the proof. \qed

\vskip0.2in
\noi \textbf{Lemma \ref{sec:appen}.3.} \textsl{Let $\rho$ be a bounded  function   satisfying   \ref{ass:rho} and \ref{ass:psi}.  Assume that  \ref{ass:densidad}  to \ref{ass:spacing} hold. Then, if in addition $\wsigma \convpp  \sigma_0$,  we have that,  $ M(\wbeta,\weta, \sigma_0)\convpp M(\beta_0,\eta_0, \sigma_0)$.  }
  
\vskip0.2in
\noi \textsc{Proof.} Recall that $p_1$ and $p_2$ stand for  $p_1=k_{n,\beta}$ and $p_2=k_{n,\eta}$ which are assumed to be of order $O(n^{\nu})$. Furthermore, we have denoted as $r_i(\beta_{\bb}, \eta_{\ba})= y_i - \bb\trasp \bx_i - \ba\trasp \bB_i$, where $\bx_i=(x_{i1}, \dots, x_{ip_1})\trasp$, $x_{ij}=\langle X_i,  B_j^{(1)}\rangle$ and $\bB_i=( B_1^{(2)}(z_i), \dots,  B_{p_2}^{(2)}(z_i))\trasp$, while
\begin{eqnarray*}
M_n(\beta,\eta, \sigma)&=& \frac 1n \sum_{i=1}^n \rho\left(\frac{y_i - \langle X_i, \beta\rangle -\eta(z_i)}{\sigma}\right)\,.
\end{eqnarray*}
Lemma \ref{sec:appen}.2 implies that
\begin{equation}
A_n=\sup_{\sigma>0, \beta \in \itM_{p_1}^{(1)}, \eta \in \itM_{p_2}^{(2)}} \left|M_n(\beta, \eta, \sigma)- M(\beta, \eta, \sigma)\right| \convpp 0\,.
\label{eq:An}
\end{equation}
On the other hand,  Lemma \ref{sec:appen}.1 entails that $M(\beta_0,\eta_0, \sigma)=\inf_{\beta,\eta} M(\beta,\eta, \sigma)$, for any $\sigma>0$, so
  $$0\le M(\wbeta,\weta, \sigma_0)-M(\beta_0,\eta_0, \sigma_0)=\sum_{i=1}^3 A_{n,i}\,,$$
  with
  \begin{eqnarray*}
  A_{n,1}&=& M(\wbeta,\weta, \wsigma)- M_n(\wbeta,\weta, \wsigma)\,,\\
  A_{n,2} &=&  M_n(\wbeta,\weta, \wsigma)- M(\beta_0,\eta_0, \sigma_0)\,,
  \\
  A_{n,3} &=& M(\wbeta,\weta, \sigma_0)- M(\wbeta,\weta, \wsigma)\,.
  \end{eqnarray*}
Note that   $|A_{n,1}|\le A_n$, which together with (\ref{eq:An}) leads to $A_{n,1}\convpp 0$. Using a Taylor's expansion of order one and assumption \ref{ass:psi}, we get that
$$|A_{n,3}|\le \|\zeta\|_{\infty} \frac{|\sigma_0-\wsigma|}{\wxi}\, ,$$
where $\wxi=\theta \sigma_0+(1-\theta)\wsigma$ is an intermediate point, which together with the fact that $\wsigma\convpp\sigma_0$ lead  to $A_{n,3}\convpp 0$.

 We will now bound $A_{n,2}$. Using \ref{ass:beta0-eta0} and \ref{ass:spacing} 
 and that $\ell \ge r+2$, we get from Corollary 6.21 in Schumaker (1981) that there exist $\wtbeta \in \itM_{p_1}^{(1)}$ and $ \wteta \in \itM_{p_2}^{(2)}$ such that
 $$\|\wtbeta-\beta_0\|_{\infty}=O(n^{-r\,\nu})\qquad \qquad \|\wteta-\eta_0\|_{\infty}=O(n^{-r\,\nu})\,.$$
Hence, using that $(\wbeta,\weta)$ minimize $ M_n(\cdot,\cdot, \wsigma)$, we obtain that
 $A_{n,2}\le M_n(\wtbeta,\wteta,\wsigma)- M(\beta_0,\eta_0, \sigma_0)=  \sum_{j=1}^3 C_{n,j} $,
where $C_{n,1}= M_n(\wtbeta,\wteta,\wsigma)-M(\wtbeta,\wteta,\wsigma)$, $  C_{n,2} = M(\wtbeta,\wteta, \sigma_0)- M(\beta_0,\eta_0, \sigma_0)$ and 
  $C_{n,3} = M(\wtbeta,\wteta, \wsigma)- M(\wtbeta,\wteta, \sigma_0)$. 
Note that the strong consistency of $\wsigma$ and the fact that $\wtbeta \in \itM_{p_1}^{(1)}$ and $ \wteta \in \itM_{p_2}^{(2)}$ entail that $|C_{n,1}|$ can be bounded by  $A_n$, so that $|C_{n,1}|\convpp 0$. Arguing as above when bounding $A_{n,3}$, we conclude that $ C_{n,3}\convpp 0$. Finally, using that $\|\wtbeta-\beta_0\|_{\infty} +\|\wteta-\eta_0\|_{\infty} \to 0$ entail that for any $(y,X.z)$,   $y-\langle X, \wtbeta \rangle +\wteta(z) \to y-\langle X, \beta_0 \rangle +\eta_0(z)$, together with the continuity and boundedness of $\rho$ and the bounded convergence theorem leads to $ C_{n,2} \to 0$.
Hence, we get that 
$$0\le M(\wbeta,\weta, \sigma_0)-M(\beta_0,\eta_0, \sigma_0) \convpp 0\,,$$
that is, $ M(\wbeta,\weta, \sigma_0)\convpp M(\beta_0,\eta_0, \sigma_0)$, concluding the proof of a). Note that the fact that $A_{n,3}\convpp 0$ entails also that  $ M(\wbeta,\weta, \wsigma)\convpp M(\beta_0,\eta_0, \sigma_0)$. \qed

 \vskip0.2in
\noi \textbf{Lemma \ref{sec:appen}.4.} \textsl{Let $\rho$ be  a bounded  function   satisfying   \ref{ass:rho} and \ref{ass:psi}  and such that  $M(\beta_0,\eta_0,\sigma_0)= b_{\rho}<1$. Let  $p_1=k_{n,\beta}$, $p_2=k_{n,\eta}$ and $(\wbeta,\weta)\in \itM_{p_1}^{(1)}\times \itM_{p_2}^{(2)}$ be such that $ M(\wbeta,\weta, \sigma_0)\convpp M(\beta_0,\eta_0, \sigma_0)$.   Assume that  $\esp\|X\|^2<\infty$   and that   \ref{ass:beta0-eta0} and \ref{ass:probaX*} hold with $c<1-b_{\rho}$. Then,  we have that, 
  there exists $L$ such that $  \prob( \cup_{m\in \natu}\cap_{n\ge m}\|\wbeta -\beta_0\|_{\itL_1}+\|\weta -\eta_0\|_{\itL_1}\le L)=1$.}
 
Note that  $M(\beta_0,\eta_0,\sigma_0)\le b<1$ whenever $\rho\le \rho_{0}$, so if $c<1-b$ then $c<1-b_{\rho}$ and the condition $c<1-b$ was also a requirement in Yohai (1987).
 
 \vskip0.2in
\noi \textsc{Proof.}  Recall that  $\itV_1^{(1)}$ is a compact set for the topology of the   norm $\|\cdot\|_{\infty}$, that is, as merged in $\itC([0,1])$.
Given $\delta>0$, define $K_\delta$ such that for any $K\ge K_\delta$,
\begin{equation}
\label{eq:XK} 
\prob(\|X\|\ge K)<\delta\,.
\end{equation}
Let $\itB=\{(\beta,\eta) : \|\beta\|_{\itL_1}+\|\eta\|_{\itL_1}=1\}\subset \itV_1^{(1)}\times \itV_1^{(1)} $ and fix   $\theta=(\beta,\eta) \in \itB$. Let $\phi_{\theta}>0$ be a continuity point of the distribution of $|\langle X, \beta \rangle +\eta (z)|$ such that
$$\prob \left(|\langle X, \beta \rangle +\eta (z)|< \phi_{\theta}\right)< c \,.$$
Then, if $\beta^{\star}, \eta^{\star} \in \itC([0,1])$ is such that $ \max\left(\|\beta^{\star}-\beta\|_{\infty}, \|\eta^{\star}-\eta\|_{\infty}\right)< \vartheta_{\theta}$, where $\vartheta_{\theta}= \phi_{\theta}/(2(K+1))$, we have that
\begin{multline*}
\prob \left(|\langle X, \beta^{\star} \rangle +\eta^{\star} (z)|   \ge  \frac{ \phi_{\theta}}2 \right)  \ge \prob \left(|\langle X, \beta \rangle +\eta (z)|\ge \phi_{\theta}\right) 
- \prob\left(\vartheta_{\theta} (\|X\|+1) \ge \frac{ \phi_{\theta}}2 \right) \, =\, A(\theta) \, .
\end{multline*}
Hence, noting that $A(\theta)>  1-c-\delta$, we conclude that
\begin{equation}
\label{eq:infimodelta}
\inf_{\max\left(\|\beta^{\star}-\beta\|_{\infty}, \|\eta^{\star}-\eta\|_{\infty}\right)< \vartheta_{\theta}}\prob \left(|\langle X, \beta^{\star} \rangle +\eta^{\star} (z)|  \ge  \frac{ \phi_{\theta}}2 \right)
\ge  A(\beta,\eta)>  1-c-\delta\,.
\end{equation}
Let us consider the covering  of $\itB$ given by $\{ \itB(\theta, \vartheta_{\theta})\}_{\theta\in\itB}$, where $\itB(\theta, \rho)$ stands for the open ball with center $\theta$ and radius $\rho$, that is, $\itB(\theta, \rho)=\{(f, g)\in \itC([0,1]): \max\left(\|f-\beta\|_{\infty},\right.$ \linebreak $\left.\|g-\eta\|_{\infty}\right)<\rho \}$. The fact that $\itV_1^{(1)}\times \itV_1^{(1)}$ is a compact set in $\itC([0,1])\times\itC([0,1])$ entails that $\itB$ is also compact, so there exist $(\beta_j, \eta_j)\in\itB$, $1\le j\le s$,  such that $\itB \subset \cup_{j=1}^s \itB(\theta_j,\vartheta_j )$ with $\vartheta_j=\vartheta_{\theta_j}$. Therefore, from \eqref{eq:infimodelta}, we obtain that
$$
\min_{1\le j\le s}\inf_{\max\left(\|\beta -\beta_j\|_{\infty}, \|\eta -\eta_j\|_{\infty}\right)< \vartheta_j}\prob \left(|\langle X, \beta  \rangle +\eta  (z)|> \frac{\phi_j}{2}\right)> 1-c-\delta\,.
$$
with $\phi_j=\phi_{\theta_j}$, meaning that for any $(\beta,\eta)\in \itB$, there exist $1\le j\le s$ such that 
\begin{equation}
\label{eq:infimoenC1}
 \prob \left(|\langle X, \beta \rangle +\eta  (z)|> \frac{\phi_j}{2}\right)> 1-c-\delta\,.
\end{equation}

Let $\itN$ be such that $\prob(\itN)=0$ and for each $\omega \notin \itN$,   $ M(\wbeta,\weta, \sigma_0)\to M(\beta_0,\eta_0, \sigma_0)=b_{\rho}$. Fix $\omega \notin \itN$ and let $\xi>0$ such that $b+\xi<1-c$. Then, there exists $n_0\in \natu$ such that for each $n\ge n_0$, $M(\wbeta_n,\weta_n, \sigma_0) \le b_{\rho}+\xi/2$, where to strength the dependence on $n$ we have denoted $(\wbeta,\weta)=(\wbeta_n,\weta_n)$.

We want to show that there exists $L>0$ such that, for $\omega \notin \itN$,   $\limsup_{n\to \infty}\|\wbeta_n-\beta_0\|_{\itL_1}+\|\weta_n-\eta_0\|_{\itL_1}\le L$. For that purpose it will be enough to show that there exist $L>0$  such that,
$$\inf_{\|\beta-\beta_0\|_{\itL_1 }+\|\eta-\eta_0\|_{\itL_1}>L} M(\beta,\eta, \sigma_0) \ge b_{\rho}+\xi\;.$$ 
Denote as $R(u)=\esp\rho\left( \epsilon  - u/\sigma_0\right)$.  First note that, the independence between the errors and covariates entails that
\begin{multline*}
M(\beta ,\eta, \sigma_0) \, = \, 
\esp\rho\left(\epsilon -\frac{ \langle X, \beta-\beta_0 \rangle +\eta (z)-\eta_0(z)}{\sigma_0}\right) 
\, = \, \esp R\left(\langle X, \beta-\beta_0\rangle +(\eta-\eta_0)(z) \right) \, .
\end{multline*}
Using that  $\lim_{|u|\to +\infty} R(u)=1$, we get that for any $\delta>0$, there exists  $u_0$ such that, for any $u$ such that $|u|\ge u_0$,  
\begin{equation}
\label{eq:epsM}
 R(u) >1-\delta\,.
\end{equation}
Choose $L> 2 \; u_0/ \min_{1\le j\le s}(\phi_j)$, where $\phi_j$ is given in \eqref{eq:infimoenC1} and   let $(\beta_k, \eta_k)\in \itL_1([0,1])\times \itL_1([0,1])$ be such that $\nu_k=\|\beta_k-\beta_0\|_{\itL_1 }+\|\eta_k-\eta_0\|_{\itL_1}>L$ and 
$$M(\beta_k,\eta_k, \sigma_0)\to \inf_{\|\beta-\beta_0\|_{\itL_1 }+\|\eta-\eta_0\|_{\itL_1}>L} M(\beta,\eta, \sigma_0)\,.$$ 
Denote as   $\wtbeta_k=(\beta_k-\beta_0)/\nu_k$ and $\wteta_k=(\eta_k-\eta_0)/\nu_k$, then $(\wtbeta_k,\wteta_k)\in \itB$, thus using \eqref{eq:infimoenC1}, we obtain  that there exists $1\le j=j(k)\le s$ such that 
$$
 \prob \left(|\langle X, \wtbeta_k \rangle +\wteta_k (z)|> \frac{\phi_j}{2}\right)> 1-c-\delta\,. 
$$
Using that  $\nu_k>L>2\, u_0/\phi_j$ and denoting as $u_k(X,z)=\nu_k (\langle X, \wtbeta_k\rangle +\wteta_k(z) )$, we obtain that  $|u_k(X,z)|>u_0$ whenever $|\langle X, \wtbeta_k\rangle +\wteta_k(z) |> {\phi_j}/2$, which together with \eqref{eq:epsM} leads to 
\begin{eqnarray*} 
M(\beta_k,\eta_k, \sigma_0)&=&  \esp R\left(  \langle X, \beta_k\rangle +\eta_k(z)  \right)=   \esp R\left(u_k(X,z) \right)\\
& \ge & \esp R\left(u_k(X,z)\right)\indica_{\left|\langle X, \wtbeta_k\rangle +\wteta_k(z)\right|>\frac{\phi_j}2}  \\
& > & (1-\delta)\, \prob\left(\left|\langle X, \wtbeta_k\rangle +\wteta_k(z)\right|>\frac{\phi_j}2\right) \\
&>& (1-c-\delta)\,(1-\delta)\,,
\end{eqnarray*}
where the last inequality follows from  \eqref{eq:infimoenC1}. Therefore,
$$\inf_{\|\beta-\beta_0\|_{\itL_1 }+\|\eta-\eta_0\|_{\itL_1}>L} M(\beta,\eta, \sigma_0)\ge (1-c-\delta)(1-\delta)\,.$$
The proof follows now easily noting that $\lim_{\delta\to 0} (1-c-\delta)(1-\delta)=1-c>b+\xi$, so we can choose $\delta$ and consequently $L$ such that   
$$\inf_{\|\beta-\beta_0\|_{\itL_1 }+\|\eta-\eta_0\|_{\itL_1}>L} M(\beta,\eta, \sigma_0)>b_{\rho}+\xi>M(\wbeta_n,\weta_n, \sigma_0)\,,$$
so $\|\wbeta_n-\beta_0\|_{\itL_1 }+\|\weta_n-\eta_0\|_{\itL_1}\le L$, concluding the proof. \qed

\vskip0.2in

\noi \textsc{Proof of Proposition \ref{sec:consis}.1.}
Recall that we have defined  $r(\beta,\eta) = y -\langle X, \beta \rangle - \eta(t)$ and we have assumed that $\sigma_0=S(\beta_0, \eta_0)$ where $S(\beta_0, \eta_0)$ is the solution of 
  \begin{equation*}
    \esp\rho_{0}\left(\frac{r(\beta_0,\eta_0)}{S(\beta_0, \eta_0)} \right) = b\,,
  \end{equation*}
  meaning that $\esp\rho_{0}\left(\epsilon \right) = b$. Besides,  the scale estimators $\wsigma=s_n(\wbeta_{\ini}, \weta_{\ini})$ satisfy  
  \begin{equation*}
   \frac{1}{n-(p_1+p_2)} \sum_{i=1}^n \rho_{0}
    \left( \frac{y_i-\langle X_i, \wbeta_{\ini} \rangle - 
        \weta_{\ini}(t_i)}{s_n(\wbeta_{\ini},\weta_{\ini})}\right) = b\,.
  \end{equation*}
 To avoid burden notation, we will briefly denote     $\wbeta$ and $\weta$ instead of $\wbeta_{\ini} $ and $\weta_{\ini}$.

We will show that for any  $\delta > 0$, with probability $1$ there exists $n_0\ge 1$ such that for $n\ge n_0$, we have  $|\wsigma-\sigma_0| \leq \delta$.

Using Lemma  \ref{sec:appen}.2 with $\rho=\rho_0$, we have that  
$$\sup_{\sigma >0\,, \beta \in \itM_{p_1}^{(1)}, \eta \in \itM_{p_2}^{(2)}} \left| \frac 1{n-p_1-p_2} \sum_{i=1}^n \left[\rho_0\left(\frac{y_i - \langle X_i, \beta\rangle -\eta(z_i)}{\sigma}\right)-M(\beta, \eta, \sigma)\right]\right| \convpp 0\,.$$
Then, there exists a null set $\itN_1$ such that, for any $\omega \notin \itN_1$, 
\begin{equation}
\label{eq:convunif1}
\sup_{\sigma >0, \beta \in \itM_{p_1}^{(1)}, \eta \in \itM_{p_2}^{(2)}} \left| \frac 1{n-p_1-p_2} \sum_{i=1}^n \left[\rho_0\left(\frac{y_i - \langle X_i, \beta\rangle -\eta(z_i)}{\sigma}\right)-M(\beta, \eta, \sigma)\right]\right| \to 0\,,
\end{equation}
holds. On the other hand, the strong law of large numbers  entails that
\begin{equation*}
    \frac{1}{n} \sum_{i=1}^n \rho_0\left(\frac{\sigma_0\, \epsilon_i}{\sigma_0+\delta}\right)
    \convpp \esp \rho\left(\frac{\sigma_0\, \epsilon}{\sigma_0+\delta}\right) < \esp \rho\left( {\epsilon} \right) = b\,,
  \end{equation*}
which together with the fact that $ (n-p_1-p_2)/n\to 1$ implies that 
\begin{equation*}
     \frac 1{n-p_1-p_2} \sum_{i=1}^n \rho_0\left(\frac{\sigma_0\, \epsilon_i}{\sigma_0+\delta}\right)
    \convpp \esp \rho_0\left(\frac{\sigma_0\, \epsilon}{\sigma_0+\delta}\right) \,.
  \end{equation*}
  Thus, there exists a null set $\itN_2$ such that, for any $\omega \notin \itN_2$, 
\begin{equation}
\label{eq:convepsilon}
A_n(\delta)= \frac 1{n-p_1-p_2} \sum_{i=1}^n \rho_0\left(\frac{\sigma_0\, \epsilon_i}{\sigma_0+\delta}\right)\to  \esp \rho_0\left(\frac{\sigma_0\, \epsilon}{\sigma_0+\delta}\right)\,. 
    \end{equation}
    Finally, taking into account that $\esp\left(\|X\| \right)<\infty$, from the strong law of large numbers and the fact that  $ (n-p_1-p_2)/n\to 1$, we get that there exists a null set $\itN_3$ such that, for any $\omega \notin \itN_3$,
    \begin{equation}
\label{eq:convX}
 \frac 1{n-p_1-p_2} \sum_{i=1}^n \|X_i\| \to    \esp\left(\|X\| \right)\,. 
    \end{equation}
Fix    $\omega \notin \cup_{i=1}^3\itN_i$.

We will begin by showing that, there exists $n_0$ such that  $\wsigma \leq \sigma_0 + \delta$, for $n\ge n_0$. 
  Using  \ref{ass:beta0-eta0} and \ref{ass:spacing} and the results in Schumaker (1981),  we get that there exists
  $\wtbeta \in \itM_{p_1}^{(1)}$ and $\wteta \in \itM_{p_2}^{(2)}$ such that
 \begin{equation}
 \label{eq:approximacion}
 \|\wtbeta-\beta_0\|_{\infty}=O(n^{-r\,\nu})\qquad \qquad
  \|\wteta-\eta_0\|_{\infty}=O(n^{-r\,\nu})\,.
  \end{equation}
  A Taylor's expansion of order one leads to 
  \begin{align*}
   \frac 1{n-p_1-p_2} \sum_{i=1}^n \rho_0 \left (\frac{y_i - \langle X_i, \wtbeta \rangle - 
      \wteta(t_i)}{\sigma_0+\delta}\right) &=  
      \frac 1{n-p_1-p_2} \sum_{i=1}^n\rho_0 \left ( \frac{\sigma_0\,\epsilon_i + \langle X_i, 
      \beta_0-\wtbeta \rangle + (\eta_0-\wteta)(t_i)}{\sigma_0+\delta}\right)\\
    &= \frac 1{n-p_1-p_2} \sum_{i=1}^n \rho_0 \left 
      (\frac{\sigma_0\, \epsilon_i}{\sigma_0+\delta}\right) + 
      R_n \\
      & =A_n(\delta)+R_n\,,
  \end{align*}
  where 
  $$R_n=\frac 1{n-p_1-p_2} \sum_{i=1}^n  \psi_0(\xi_i) \, \frac{\langle X_i, \beta_0-\wtbeta \rangle + 
      (\eta_0-\wteta)(t_i)}{\sigma_0+\delta}\,,$$
  $\psi_0=\rho_0^{\prime}$    and $\xi_i$ is an intermediate point. From \eqref{eq:convepsilon}, we get immediately that   
      $$A_n(\delta)\to  \esp \rho_0\left(\frac{\sigma_0\,\epsilon}{\sigma_0+\delta}\right) =b_1
    < \esp \rho_0\left( {\epsilon} \right) = b\,.$$
Besides, the bound
  \begin{equation*}
    |R_n| \leq \frac {n}{n-p_1-p_2}\;  \|\psi_0\|_{\infty} \frac{1}{\sigma_0+\delta} 
    \left(\|\eta_0-\wteta\|_{\infty}+\|\beta_0-\wtbeta\|_{\infty}   \frac 1{n} \sum_{i=1}^n \|X_i\|
       \right) \,,
  \end{equation*}
  together with \eqref{eq:convX} and \eqref{eq:approximacion} lead to $ |R_n|\to 0$. Hence, 
  $$\frac 1{n-p_1-p_2} \sum_{i=1}^n \rho_0 
      \left (\frac{y_i - \langle X_i, \wtbeta \rangle - 
      \wteta(t_i)}{\sigma_0+\delta}\right) \to  b_1 \,.$$
 Choose $\delta_1>0$ such that $b_1+\delta_1<b$, then there exists $n_0\in \natu$ such that for $n\ge n_0$,
  \begin{equation}
  \label{eq:cotab1}
    \frac 1{n-p_1-p_2} \sum_{i=1}^n \rho_0 
      \left (\frac{y_i - \langle X_i, \wtbeta \rangle - 
      \wteta(t_i)}{\sigma_0+\delta}\right) < b_1+\delta_1< b\,.
  \end{equation}
  Noting that
  $$ \frac 1{n-p_1-p_2}  
       \sum_{i=1}^n \rho_0 
    \left( \frac{y_i-\langle X_i, \wtbeta \rangle - 
        \wteta(t_i)}{s_n(\wtbeta,\wteta)}\right) =b \,,$$
from \eqref{eq:cotab1} and the fact that $\rho$ is non-decreasing we immediately obtain that   
  $s_n(\wtbeta,\wteta) < \sigma_0+\delta$. Using that $\wsigma=  \min_{\beta \in \itM_{p_1}^{(1)}, \eta \in \itM_{p_2}^{(2)} }  s_n(\beta , \eta )$ and the fact that $\wbeta \in \itM_{p_1}^{(1)}$ and $ \weta \in \itM_{p_2}^{(2)}$, we conclude that for $n\ge n_0$,  
  $\wsigma = s_n(\wbeta,\weta) \leq \sigma_0+\delta$. 
    
  It remains to show that there exists $n_1\in \natu$ such that for any $n\ge n_1$,   $\wsigma \geq \sigma_0 - \delta$. 
  
  The fact that $\rho_0$ is non--decreasing together with \ref{ass:densidad} implies that $M(\beta_0,\eta_0,\sigma_0-\delta) > M(\beta_0,\eta_0,\sigma_0) = b$ (see Lemma 3 in Salibi\'an--Barrera, 2006). Let $\delta_2>0$ be such that $M(\beta_0,\eta_0,\sigma_0-\delta) = b_2=b + \delta_2$. 
Using that \eqref{eq:convunif1} holds, we get that  there exists  $n_1\in \natu$ such that for any $n\ge n_1$,
$$\sup_{\sigma >0, \beta \in \itM_{p_1}^{(1)}, \eta \in \itM_{p_2}^{(2)}} \left| \frac 1{n-p_1-p_2} \sum_{i=1}^n \left[\rho_0\left(\frac{y_i - \langle X_i, \beta\rangle -\eta(z_i)}{\sigma}\right)-M(\beta, \eta, \sigma)\right]\right|< \frac{\delta_2}2\,.$$
Hence, 
$$\left| \frac 1{n-p_1-p_2} \sum_{i=1}^n  \rho_0\left(\frac{y_i - \langle X_i, \wbeta\rangle -\weta(z_i)}{\wsigma}\right)-\frac{n}{n-p_1-p_2}M(\wbeta, \weta, \wsigma)\right|< \frac{\delta_2}2 \,,$$
leading to
$$\frac{n}{n-p_1-p_2}M(\wbeta, \weta, \wsigma)< \frac 1{n-p_1-p_2} \sum_{i=1}^n  \rho_0\left(\frac{y_i - \langle X_i, \wbeta\rangle -\weta(z_i)}{\wsigma}\right)+  \frac{\delta_2}2=b+ \frac{\delta_2}2\,.$$
On the other hand, the fact that $(n-p_1-p_2)/n\to 1$ and $\rho$ is bounded implies that 
$$ \left|\frac{n}{n-p_1-p_2}M(\wbeta, \weta, \wsigma)-  M(\wbeta, \weta, \wsigma)\right|\le \left|\frac{n}{n-p_1-p_2}-1\right|\to 0 \,,$$
so    without loss of generality, we may assume that   for any $n\ge n_1$,
$$M(\wbeta, \weta, \wsigma)< \frac{n}{n-p_1-p_2}M(\wbeta, \weta, \wsigma)+\frac{\delta_2}2<  b+  \delta_2 \,.$$
Note that Lemma \ref{sec:appen}.1 entails that $M(\beta_0,\eta_0,\wsigma) \le  M(\wbeta,\weta,\wsigma)$, thus we get 
  \begin{equation*}
    M(\beta_0,\eta_0,\wsigma) < b +\delta_2   = M(\beta_0,\eta_0,\sigma_0-\delta)\,,
  \end{equation*}
 which implies that $\wsigma \geq \sigma_0 - \delta$ for $n\ge n_1$, concluding the proof. \qed

\vskip0.2in

\noi \textsc{Proof of Theorem \ref{sec:consis}.1.}
 For the sake of simplicity let $\theta=(\beta,\eta)$ and $\theta_0=(\beta_0,\eta_0)$. From Lemmas  \ref{sec:appen}.3 and  \ref{sec:appen}.4 with $\rho=\rho_1$, it will be enough to show that, for any $\epsilon>0$, 
$$\inf_{(\beta,\eta)\in \itA_\epsilon} M(\beta,\eta, \sigma_0) > M(\beta_0, \eta_0, \sigma_0)\,,$$
where $\itA_{\epsilon}=\{(\beta, \eta)\in \itL_1([0,1])\times \itL_1([0,1])  \|\beta-\beta_0\|_{\itL_1 }+\|\eta-\eta_0\|_{\itL_1}\le L\,, \, d(\theta, \theta_0) \ge \epsilon\}$ and $ d(\theta, \theta_0)=\|\beta -\beta_0\|_{\infty}+\|\eta -\eta_0\|_{\infty}$. 

As in the proof of Lemma \ref{sec:appen}.4, let $(\beta_k, \eta_k)\in \itA_{\epsilon}$ be such that  
$$M_k=M(\beta_k,\eta_k, \sigma_0)\to \inf_{(\beta,\eta)\in \itA_{\epsilon}} M(\beta,\eta, \sigma_0)\,,$$
 and denote $\nu_k=\|\beta_k-\beta_0\|_{\itL_1 }+\|\eta_k-\eta_0\|_{\itL_1}$. Then, from the fact that $\nu_k$ is bounded, using the Arzela--Ascoli Theorem, we obtain that there exists a subsequence $ {k_j}$ such that $f_j=\beta_{k_j}-\beta_0$ and $g_j=\eta_{k_j}-\eta_0$ converge uniformly to functions $f$ and $g$. Denote as $\wtbeta=f+\beta_0$ and $\wteta=g+\eta_0$ the uniform limit of  $\beta_{k_j}$ and $\eta_{k_j}$, respectively. Hence, we have that $\|\beta_{k_j}-\wtbeta\|_\infty+\|\eta_{k_j}-\wteta\|_\infty\to 0$, $\lim_j\|f_j\|_{\infty}=\|f\|_{\infty}$ and $\lim_j\|g_j\|_{\infty}=\|g\|_{\infty}$, so that $d(\wttheta, \theta_0)\ge \epsilon$ with $\wttheta=(\wtbeta , \wteta)$.   Using that $\rho_1$ is a bounded continuous function, from the Bounded Convergence Theorem, we obtain that $M_{k_j}\to M(\wtbeta , \wteta , \sigma_0)$ implying that  $\inf_{(\beta,\eta)\in \itA_\epsilon} M(\beta,\eta, \sigma_0)= M(\wtbeta , \wteta , \sigma_0)$. Lemma \ref{sec:appen}.1 together with the fact that $d(\wttheta, \theta_0)\ge \epsilon$, entail that $M(\wtbeta , \wteta , \sigma_0)>M(\beta_0, \eta_0, \sigma_0)$ concluding the proof. \qed

\subsection{Proof of Theorem \ref{sec:consis}.2}{\label{sec:dem-teo-32}}

Let us denote as  $\Theta=\itL_r([0,1])\times \itL_r([0,1])$ and as $\Theta_n = \itM_{p_1}^{(1)}\times  \itM_{p_2}^{(2)}\cap \{ \theta=(\beta,\eta)\in \Theta: \|\beta -\beta_0\|_{\infty}+\|\eta -\eta_0\|_{\infty}\le  \epsilon_0\}$, where $\epsilon_0$ is given in assumption \ref{ass:cotainf}.  Note that, except for a null probability set, $\wtheta=(\wbeta,\weta) \in \Theta_n$, for $n$ large enough.   From now on, for a   function $\varphi(y,X,z)$ we denote as $\|\varphi\|_2=\{\esp\left(\varphi^2(y,X,z)\right)\}^{1/2}$, that is, the $L_2(P)-$norm.

The following Lemma gives conditions under which \ref{ass:cotainf} holds.

\vskip0.2in
\noi \textbf{Lemma \ref{sec:dem-teo-32}.1.} \textsl{Let $\rho$ be a bounded function satisfying \ref{ass:rho} and \ref{ass:psi} and such that
$\rho^{\prime}= \psi$ is continuously differentiable with bounded derivative $\psi^{\prime}$ and $\esp \psi^{\prime}(\epsilon)>0$.
If   there exists $C>0$ such that $\prob(\|X\|\le C)=1$, then \ref{ass:cotainf} holds.}
 
 \vskip0.2in
\noi \textsc{Proof.} As in the proof of Theorem \ref{sec:consis}.1, denote  $\theta=(\beta,\eta)$ and $\theta_0=(\beta_0,\eta_0)$, then we have that
\begin{eqnarray*}
  M( \theta , \sigma)-M( \theta_0, \sigma) 
&=&   \esp\left[ \rho\left(\frac{\sigma_0\, \epsilon -\langle X, \beta-\beta_0\rangle+\eta(z )-\eta_0(z)}{\sigma}\right) -\rho\left(\frac{\sigma_0\,\epsilon}{\sigma}\right) \right]\\
  &=&  \esp\left[ \psi\left(\frac{\sigma_0\, \epsilon}{\sigma}\right) \left(\langle X, \beta-\beta_0\rangle + \eta(z )-\eta_0(z )\right)\right] \\
 & +& \frac 1{2}\;   \esp\left[   \psi^{\prime}\left(\frac{\sigma_0\epsilon+\xi}{\sigma}\right) \left(\langle X, \beta-\beta_0\rangle + \eta(z )-\eta_0(z )\right)^2 \right]  \\  
          &=& \frac 1{2}\;   \esp\left[    \psi^{\prime}\left(\frac{\sigma_0\epsilon+\xi}{\sigma}\right)  \left(\langle X, \beta-\beta_0\rangle + \eta(z )-\eta_0(z )\right)^2 \right]  \,,
\end{eqnarray*}
where $\xi$ is an intermediate point between $g(X,z)=\langle X, \beta-\beta_0\rangle+\eta(z )-\eta_0(z)$ and $0$. Note that
$|g(X,z) |\le \|X\| \|\beta-\beta_0\|_{\infty}  + \|\eta-\eta_0\|_{\infty}$, hence if $\|\beta-\beta_0\|_{\infty}  + \|\eta-\eta_0\|_{\infty}<\epsilon_0$, we get that $|\xi|\le (C+1) \epsilon_0$ with probability 1.

The fact that $\varphi=\esp \psi^{\prime}(\epsilon)>0$ and the continuity of $\psi^{\prime}$ entails that for $\delta$ small enough
$$ \inf_{\sigma>0, |\sigma-\sigma_0|<\delta, |a|<\delta} \esp \psi^{\prime}\left(\frac{\sigma_0\epsilon+a}{\sigma}\right)>\frac{\varphi}2>0\,,$$
Hence, if $\itV=\{\sigma>0: |\sigma-\sigma_0|<\delta \}$ and $\epsilon_0=\delta/(C+1)$, we have that
\begin{eqnarray*}
 M( \theta , \sigma)-M( \theta_0, \sigma) &=&  \frac 1{2}\;   \esp\left[    \psi^{\prime}\left(\frac{\sigma_0\epsilon+\xi}{\sigma}\right)  \left(\langle X, \beta-\beta_0\rangle + \eta(z )-\eta_0(z )\right)^2 \right] \\
 & > & \frac{\varphi}2 \esp\left[  \left(\langle X, \beta-\beta_0\rangle + \eta(z )-\eta_0(z )\right)^2 \right]= \frac{\varphi}2 \pi^2(\theta, \theta_0)\,,
\end{eqnarray*}
concluding the proof. \qed

Recall that $\pi^2( \theta, \theta_0)=\esp \left[\langle X, \beta-\beta_0\rangle +\eta(z)-\eta_0(z)\right]^2$ for any vectors of coefficients $\bb\in \real^{p_1}$ and $\ba\in \real^{p_2}$, $\beta_{\bb}(t)= \sum_{j=1}^{p_1} b_j \,   B_j^{(1)}(t)$ and
 $\eta_{\ba} (z)=\sum_{j=1}^{p_2} a_j \,   B_j^{(2)}(z)$. In order to prove Theorem \ref{sec:consis}.2, we will need the following Lemma. 

\vskip0.2in
\noi \textbf{Lemma \ref{sec:dem-teo-32}.2.} \textsl{Let $\rho$ be a bounded function satisfying \ref{ass:rho} and \ref{ass:psi}. Given $\bb_0\in \real^{p_1}$ and $\ba_0\in \real^{p_2}$, let $\wttheta_{0}=(\wtbeta_0, \wteta_0)\in\itM_{p_1}^{(1)}\times  \itM_{p_2}^{(2)}$ be such that $\wtbeta_0 = \beta_{\bb_0}$ and $\wteta_0 = \eta_{\ba_0} $. Define the class of functions
\begin{align*}
\itG_{n, c, \wttheta_{0}} & = \{f_{\theta,\sigma}=V_{  \theta,  \sigma}-V_{  \theta_{0}^{\star},  \sigma}:  \|\beta -\wtbeta_0\|_{\infty}+\|\eta -\wteta_0\|_{\infty} \leq c\,,     \theta\in \Theta_n\,, \,  \sigma\in \itV=[\sigma_1, \sigma_2]\}\,,
\end{align*} 
with $\sigma_1=\sigma_0/2$, $\sigma_2= (3/2)\, \sigma_0$,    $\theta_{0}^{\star}=(\beta_0^{\star}, \eta_{0}^\star)\in \Theta$ a fixed point   and
\begin{equation*}
V_{  \theta,  \sigma}=\rho\left(\frac{y- \langle X, \beta \rangle+ \eta(z)}{\sigma}\right) \,,
\label{eq:defV}
\end{equation*}
 for $ \theta=(\beta,\eta)$. Then, if $\esp \|X\|^2<\infty$, there exists   some constant  $A >0$ independent of $n$, $\wttheta_{0}$, $\theta_{0}^{\star}$ and $\epsilon$, such that
 $$N_{[\;]}(\epsilon, \itG_{n,c, \wttheta_{0}}, L_2(P))\le  3\,{\sigma_0} \left(\frac{ \max(1,c )  \, A }{\epsilon}\right)^{ p_1+p_2+1} \,.$$
}
 
 \noi \textsc{Proof.}  First note that, for any $\theta\in \Theta_n$, $\theta=(\beta_{\bb}, \eta_{\ba})$, we have that there exist a constant $D$ depending only on the degree $\ell$ of the considered splines such that
 \begin{equation}
 \label{eq:normainf}
 D\, \|\bb\|_{\infty}\le \| \beta_{\bb}\|_{\infty} \le \|\bb\|_{\infty}\qquad D\, \|\ba\|_{\infty}\le \| \eta_{\ba}\|_{\infty} \le \|\ba\|_{\infty}\,,
  \end{equation}
 where for a vector $\bc\in \real^s$, $\|\bc\|_{\infty}=\max_{1\le j\le s} |c_j|$ (see de Boor, 1973, Section 3).
 
 Then, we have that   $\itH_{c,\wtbeta_0}\subset \{  \sum_{j=1}^{p_1} b_j\, B_j^{(1)}(t)\,, \bb \in\itB_{\bb_0, p_1}(c_1)\}$ and $\itH_{c,\wteta_0}\subset  \{\sum_{j=1}^{p_2}   a_j\, B_j^{(2)}(z)\,, \ba \in\itB_{\ba_0, p_2}(c_1)\}$, where $c_1=c/D$, $\itB_{\bb, p_1}(\delta)  = \{\bb \in \real^{p_1}: \|\bb - \bb_0\|_{\infty} <\delta\}$, $\itB_{\ba_0, p_2}(\delta)  = \{\ba \in \real^{p_2}: \|\ba - \ba_0\|_{\infty} <\delta\}$,
\begin{align*}
\itH_{c,\wtbeta_0} & =\left\{\beta(t)=\sum_{j=1}^{p_1} b_j\, B_j^{(1)}(t)\,, \bb \in \real^{p_1}, \|\beta- \wtbeta_0\|_{\infty}\le c \right\}\\
\itH_{c,\wteta_0} & =\left\{\eta(z)=\sum_{j=1}^{p_2}   a_j\, B_j^{(2)}(z)\,, \ba \in \real^{p_2}, \|\eta- \wteta_0\|_{\infty} \le c \right\}\,.
\end{align*}
Recall that the ball $\itB_{\bb, p_1}(\delta)$ can be covered by at most $\{(4\delta+\epsilon)/\epsilon\}^{p_1}$ balls (with respect to the norm $\|\cdot\|_{\infty}$) of radius $\epsilon$, when $\epsilon<\delta$, while if $\epsilon>\delta$ the covering number  equals 1. Hence, using  the upper bounds given in \eqref{eq:normainf} and using that for any class of functions $\itH$, $ N_{[\;]}(\epsilon, \itH , L_{\infty})\le  N (\epsilon, \itH, L_{\infty})$, we obtain that
 \begin{align}
 \label{bracketHbeta}
 \log N_{[\;]}(\epsilon, \itH_{c,\wtbeta_0}, L_{\infty}) & \le    p_1 \log\left(5\,c_1/\epsilon\right)\,,\\
  \label{bracketHeta}
 \log N_{[\;]}(\epsilon, \itH_{c,\wteta_0}, L_{\infty}) & \le   p_2 \log\left(5\,c_1 /\epsilon\right)\,.
 \end{align}
 for $0<\epsilon< c/ D$. 
 Henceforth, using   \eqref{bracketHbeta} and \eqref{bracketHeta}, we get that,  for any $0<\epsilon< c/ D$, $\itH_{c,\wtbeta_0}$ can be covered by a finite number $ M_1(\epsilon)\le \left(5\,(c/D) /\epsilon\right)^{ p_1}$ of $\epsilon-$brackets   $\{[\beta_{j,L}, \beta_{j,U}]\,,   1\le j\le   M_1(\epsilon)\}$, while $\itH_{c,\wteta_0}$ can be covered by a finite number $ M_2(\epsilon)\le \left(5\,(c/D) /\epsilon\right)^{ p_2}$ of $\epsilon-$brackets   $\{[\eta_{j,L}, \eta_{j,U}]\,,   1\le j\le   M_2(\epsilon)\}$.
 
 On the other hand, the set $\itV=[\sigma_1,\sigma_2]=\{\sigma: |\sigma-\sigma_0|\le \sigma_0/2\}$ can be covered by $M_3(\epsilon)\le C_{\sigma_0} (1/\epsilon)$ balls of radius $\epsilon$ (when $\epsilon<\sigma_0/2$) and center $\sigma^{(s)}$,  $1\le s\le M_3(\epsilon)$, where $ C_{\sigma_0}=3 \sigma_0$.  
 
 Recall that  $\psi$ is bounded, so that, for $\sigma \in [\sigma_1,\sigma_2]$, 
 $$\left|\frac{\partial}{\partial u} \rho\left(\frac{y-u}{\sigma}\right)\right| \le \frac{\|\psi\|_{\infty}}{\sigma}\le 2\, \frac{\|\psi\|_{\infty}}{\sigma_0}\,.$$
Define $\epsilon_1=\epsilon/A_1$ where
$$A_1=\;  \frac{4}{\sigma_0} \,  \left(\|\psi \|_{\infty} \left( \esp \|X\|^2\right)^{1/2} \;+ \|\zeta\|_{\infty} +\|\psi\|_{\infty}\right) \,.$$
Given $f_{\theta,\sigma}\in \itG_{n,c, \wttheta_{0}}$, let $j$, $m$ and $s$ be such that $\beta$ belongs to the   $\epsilon_1-$bracket    $[\beta_{j,L}, \beta_{j,U}]$, that is, $\beta_{j,L}\le \beta\le \beta_{j,U}$ and $\|\beta_{j,U}- \beta_{j,L}\|_{\infty} <\epsilon_1$, $\eta$ belongs to the   $\epsilon_1-$bracket    $[\eta_{m,L}, \eta_{m,U}]$ and $|\sigma-\sigma^{(s)}|<\epsilon_1$. Denote as 
\begin{eqnarray*}
f_{j,m,s}(y,X,z)&=& \rho\left(\frac{y-\langle X,  \beta_{j,U}\rangle +\eta_{m,U}(z)}{\sigma^{(s)}}\right)-\rho\left(\frac{y- \langle X,  \beta_0^\star  \rangle+  \eta_0^\star(z)}{\sigma^{(s)}}\right) \,,\\
f_{j,m}(y,X,z)&=& \rho\left(\frac{y-\langle X,  \beta_{j,U} \rangle +\eta_{m,U}(z)}{\sigma}\right)-\rho\left(\frac{y- \langle X,  \beta_0^\star \rangle+  \eta_0^\star(z)}{\sigma}\right) \,.
\end{eqnarray*}
Using a Taylor's expansion of order one and the fact that $\zeta(u)= u\, \psi(u)$ is bounded,  we get that
\begin{eqnarray*}
|f_{\theta,\sigma}-f_{j,m,s}|& \le & |f_{\theta,\sigma}-f_{j,m}|+|f_{j,m}- f_{j,m,s}| \\
&\le &  \frac{2}{\sigma_0} \,\|\psi\|_{\infty} \left\{ \|X\|\;\left\|\beta -\beta_{j,U}\right\| +\left|\eta(z)-\eta_{m,U}(z)\right|\right\}+ 
2 \frac{ \|\zeta\|_{\infty}}{\sigma_0}\; |\sigma-\sigma^{(s)}|\\
&\le & \epsilon_1  \frac{2}{\sigma_0} \, \left\{\|\psi\|_{\infty}  \left(\|X\|+1\right)+  \|\zeta\|_{\infty} \right\}  \,,
\end{eqnarray*}
where  the last inequalities follow  from the fact that $\eta_{m,L}\le \eta(z) \le \eta_{m,U}(z)$, $\| \eta_{j,L}(z)-\eta_{j,U}(z) \|_{\infty}\le \epsilon_1 $,  
$0\le \beta_{j,U}(t)- \beta(t) \le \beta_{j,U}(t)- \beta_{j,L}(t)$, so that $\int_0^1 \left[\beta_{j,U}(t)- \beta(t)\right]^2 dt \le \int_0^1 \left[\beta_{j,U}(t)- \beta_{j,L}(t)\right]^2 dt \le \|\beta_{j,U}- \beta_{j,L}\|_{\infty}^2<\epsilon_1^2$ and $|\sigma-\sigma^{(s)}|<\epsilon_1$. Define the functions 
\begin{align*}
\varphi_{j,m,s}^{(U)}(y,X,z)&=f_{j,m,s}(y,X,z) +\epsilon_1  \frac{2}{\sigma_0} \, \left\{\|\psi\|_{\infty}  \left(\|X\|+1\right)+  \|\zeta\|_{\infty} \right\}    \,, \\
\varphi_{j,m,s}^{(L)}(y,\bx,tX,z)&=f_{j,m,s}(y,X,z) - \epsilon_1  \frac{2}{\sigma_0} \, \left\{\|\psi\|_{\infty}  \left(\|X\|+1\right)+  \|\zeta\|_{\infty} \right\}     \,.
\end{align*}
Then, we have that $\varphi_{j,m,s}^{(L)}\le f_{\theta,\sigma} \le \varphi_{j,m,s}^{(U)}$ and  taking into account that $\esp \|X\|^2 < \infty$, we obtain 
\begin{align*}
\|\varphi_{j,m,s}^{(U)}-\varphi_{j,m,s}^{(L)}\|_2
&\le         \epsilon_1 \;  \frac{4}{\sigma_0} \,  \left(\|\psi \|_{\infty} \left( \esp \|X\|^2\right)^{1/2} \;+ \|\zeta\|_{\infty} +\|\psi\|_{\infty}\right)  = \epsilon \,,
\end{align*}
which means that the total number of brackets of size $\epsilon$ needed to cover $\itG_{n,c, \widetilde{\theta}_{0}}$ is bounded by 
$$\prod_{i=1}^3 M_i(\epsilon_1)\le 3\,{\sigma_0} \left(\frac{5\,(c/D)}{\epsilon_1}\right)^{(p_1+p_2)}\, \left(\frac{1}{\epsilon_1}\right)\le
3\,{\sigma_0} \left(\frac{A\,\max(1,c )}{\epsilon}\right)^{p_1+p_2+1}\,,$$
with $A=    5\,A_1/D $, where we have used that $D\le 1$, concluding the proof of the first inequality. \qed
\vskip0.2in

\noi \textbf{Remark \ref{sec:dem-teo-32}.1.}  Note that if $\prob(\|X\|\le C)=1$, we further have that 
 $$N_{[\;]}(\epsilon, \itG_{n,c, \wttheta_{0}}, L_{\infty})\le  3\,{\sigma_0} \left(\frac{ \max(1,c )  \, A }{\epsilon}\right)^{ p_1+p_2+1} \,,$$
taking     $A=5\,A_1/D $ where $D$ is given in \eqref{eq:normainf} and
$$A_1=\;  \frac{4}{\sigma_0} \,  \left(\|\psi \|_{\infty}   C + \|\zeta\|_{\infty}  +\|\psi\|_{\infty}\right) \,.$$


The following Lemma is needed in the proof of Theorem \ref{sec:consis}.2. Its proof follows using similar arguments to those considered in the proof of  Theorem 3.2.5 of van der Vaart and Wellner (1996), we include it for completeness. In the statement of Lemma \ref{sec:dem-teo-32}.3, $\Theta= \itL_r([0,1])\times \itL_r([0,1])$, $\Theta^{(1)}_n$ corresponds to the set $\itM_{p_1}^{(1)}\times  \itM_{p_2}^{(2)}$, while $\Theta_n=\Theta^{(1)}_n\cap \{ \theta=(\beta,\eta)\in \Theta: \|\beta -\beta_0\|_{\infty}+\|\eta -\eta_0\|_{\infty}\le  \epsilon_0\}$ as defined above.

\vskip0.2in 
 \noi \textbf{Lemma \ref{sec:dem-teo-32}.3.} \textsl{Let $M_n$   be an stochastic process  indexed by $ \Theta^{(1)}_n\times (0,+\infty)$ and $M:\Theta\times (0,+\infty)\to \real$ where $\Theta^{(1)}_n\subset \Theta$. Let $\wsigma$ be an estimator of $\sigma_0$ such that $\prob(\wsigma \in \itV)\to 1$ where $\itV\subset (0,+\infty)$ and  denote as $\wtheta\in  \Theta^{(1)}_n$ the minimizer of  $M_n( \theta, \wsigma)$ over   $\Theta^{(1)}_n$, that is,  $M_n( \wtheta, \wsigma)  \le  M_n( \theta, \wsigma)$, for any $\theta \in  \Theta^{(1)}_n$. Let $\delta_n\ge 0 $ be a fixed sequence such that $\delta_n \to 0$ and fix $\upsilon>0$ with $0\le \delta_n< \upsilon$ for all $n$. Assume that $\prob(\wtheta \in \Theta_n)\to 1$ and 
$\pi(\wtheta,  \theta_{0,n})   \convprob  0 $,  where  $ \theta_{0,n}\in \Theta_n\subset \Theta^{(1)}_n$ is a fixed sequence. 
Furthermore, assume that there exists a function $\phi_n$ such that $\phi_n(\delta)/\delta$ is decreasing on $(\delta_n, \infty)$   and that  for any $\delta_n<\delta \le \upsilon$, we have
\begin{eqnarray}
\sup_{ \theta\in \Theta_{n,\delta}, \sigma \in \itV} M( \theta_{0,n}, \sigma)- M( \theta,\sigma) &\lesssim & -\delta^2 \,,
\label{aprobar1}\\
 \esp^{*} \sup_{ \theta\in \Theta_{n,\delta}, \sigma \in \itV}  \sqrt{n} \left |(M_n( \theta, \sigma)- M( \theta,\sigma))-(M_n( \theta_{0,n}, \sigma)- M( \theta_{0,n},\sigma)) \right | & \lesssim &\phi_n(\delta) \,,
\label{aprobar2}
\end{eqnarray}
where $\Theta_{n,\delta}=\{ \theta\in \Theta_n: \delta / 2  <  \pi( \theta, \theta_{0,n}) \leq \delta\}$, the symbol $\lesssim$ means \textit{less or   equal up to a universal constant}
 and  $\esp^{*}$ stands for the outer expectation. Then, if $\gamma_n$ is such that $\delta_n\,\gamma_n=O(1)$ and 
 $ \gamma_n^2 \phi_n\left(\gamma_n^{-1}\right)\le \sqrt{n}$,
 for every $n$, we have that $\gamma_n \pi(\wtheta, \theta_{0,n}) = O_{\prob}(1)$. }

\vskip0.2in
\noi   
\textsc{Proof.} Note that for each fixed $K$, 
$$\{ \theta\in \Theta_n: \gamma_n \, \pi(\theta,\theta_{0,n}) > 2^{K}  \}\subset \{ \theta\in \Theta_n: \gamma_n \, \pi(\theta,\theta_{0,n}) > 2^{K-1} \}=\dst \bigcup_{j\ge K} \itA_{j,n}\,,$$
where $\itA_{j,n}=\{ \theta\in \Theta_n:   2^{j-1}<\gamma_n \, \pi(\theta,\theta_{0,n}) \le 2^{j}\}$. Let $\itB_n=\{ \theta\in \Theta_n:\pi(\theta,  \theta_{0,n}) \le \upsilon/2 \}$, then we have 
\begin{align*}
\prob\left(\gamma_n \, \pi(\wtheta,\theta_{0,n}) > 2^{K}\right) & \le \prob\left(\wtheta \notin \Theta_n\right)+ \prob(\wsigma \notin \itV) +\prob\left(\pi(\wtheta,  \theta_{0,n}) >\frac{\upsilon}2\right) + \sum_{j\ge K} \prob\left( \wtheta\in   \itA_{j,n} \cap \itB_n, \wsigma \in \itV \right)\\ 
& \le \prob\left(\wtheta \notin \Theta_n\right)+ \prob(\wsigma \notin \itV) +\prob\left(\pi(\wtheta,  \theta_{0,n}) >\frac{\upsilon}2\right) 
 + \mathop{\sum_{j\ge K}}_{2^j\le \upsilon\; \gamma_n}\prob\left(  \wtheta\in  \itA_{j,n} ,   \wsigma \in \itV   \right)\,.
\end{align*}
For any $j\ge K$ such that  $2^j\le \upsilon \gamma_n$, denote $ \delta^{(j)}=  2^{j}/\gamma_n $. Using  \eqref{aprobar1}, we get that, for any $j\ge K$ such that  $2^j\le \upsilon \gamma_n$, $\theta\in \itA_{j,n}$ and   $\sigma \in \itV$, $  M( \theta_{0,n}, \sigma)- M( \theta,\sigma)  \le   - \, C\, (\delta^{(j)})^2 = - C\,  { 2^{2j}}/{\gamma_n^2}$, where $C$ is a universal constant independent of $j$, $n$. In particular, when $\wsigma \in \itV$ and  $\wtheta\in \itA_{j,n}$, we have
$$   M( \wtheta, \wsigma)- M( \theta_{0,n},\wsigma)  \ge   C\, \frac{ 2^{2j}}{\gamma_n^2}\,.$$
Besides, $M_n( \theta_{0,n}, \wsigma)- M_n( \wtheta, \wsigma)\ge 0$ since  $\wtheta$ minimizes  $ M_n( \theta, \wsigma)$ over $\Theta^{(1)}_n$ and $\theta_{0,n}\in  \Theta^{(1)}_n$.
Thus, if we denote $W_n(\theta, \sigma)= M_n( \theta, \sigma)- M( \theta,\sigma)$, we have that, whenever $\wtheta\in \itA_{j,n}$ and $\wsigma \in \itV$,  
\begin{align*}
 W_n( \theta_{0,n}, \wsigma)-W_n( \wtheta, \wsigma)  & =
  \left\{M_n( \theta_{0,n}, \wsigma)- M_n( \wtheta, \wsigma)\right\} - \left\{M( \theta_{0,n},\wsigma)- M( \wtheta,\wsigma)\right\}\\
  & \ge  M( \wtheta,\wsigma)- M( \theta_{0,n},\wsigma) \ge C \,  \frac{2^{2j}}{\gamma_n^2}\,,
\end{align*}
so 
$$ \sup_{\theta\in \itA_{j,n}, \sigma \in \itV}  \left|W_n( \theta_{0,n}, \sigma) -W_n( \theta, \sigma) \right|  \ge C \,  \frac{ 2^{2j}}{\gamma_n^2}$$
leading to
\begin{align*}
\prob\left(\gamma_n \, \pi(\wtheta,\theta_{0,n}) > 2^{K}\right) & \le \prob\left(\wtheta \notin \Theta_n\right)+ \prob(\wsigma \notin \itV) +\prob(\pi(\wtheta,  \theta_{0,n}) >\upsilon/2 )  \\ 
&+ \mathop{\sum_{j\ge K}}_{2^j\le \upsilon \gamma_n}\prob\left(  \sup_{\theta\in \itA_{j,n}, \sigma \in \itV}  \left|W_n( \theta_{0,n}, \sigma) -W_n( \theta, \sigma)\right|  \ge C \,  \frac{ 2^{2j}}{\gamma_n^2}  \right)\\
&\le \prob\left(\wtheta \notin \Theta_n\right)+ \prob(\wsigma \notin \itV) +\prob(\pi(\wtheta,  \theta_{0,n}) >\upsilon/2 )  \\ 
&+ C^{-1} \sum_{j\ge K}  \frac{\gamma_n^2}{2^{2j}}\;   \esp^{*} \left\{\sup_{\theta\in \itA_{j,n}, \sigma \in \itV}  \left|W_n( \theta_{0,n}, \sigma) -W_n( \theta, \sigma)\right|\right\}\\
&\le \prob\left(\wtheta \notin \Theta_n\right)+ \prob(\wsigma \notin \itV) +\prob(\pi(\wtheta,  \theta_{0,n}) >\upsilon/2 )  \\ 
&+ C^{\star} \sum_{j\ge K}  \phi_n\left(\frac{ 2^{j}}{\gamma_n}  \right)\, \frac{\gamma_n^2}{2^{2j}\; \sqrt{n}}
\end{align*}
where the last inequality is a consequence of Markov's inequality and \eqref{aprobar2} and $C^{\star}$ is a universal constant.

The fact that $\phi_n(\delta)/\delta$ is decreasing entails that 
$$ \phi_n\left(\frac{ 2^{j}}{\gamma_n}  \right)\le   2^{j} \phi_n\left(\frac{1}{\gamma_n}  \right)$$
which together with the assumption that $ \gamma_n^2 \phi_n\left(\gamma_n^{-1}\right)\le \sqrt{n}$, implies that 
$$\sum_{j\ge K}  \phi_n\left(\frac{ 2^{j}}{\gamma_n}  \right)\, \frac{\gamma_n^2}{2^{2j}\; \sqrt{n}} \le \sum_{j\ge K}     \frac{1}{2^{j}}= \frac{1}{2^{K-1}}\,,$$
which converges to 0 when $K\to \infty$. Thus, using that $\prob(\wsigma \in \itV)\to 1$,  $\prob(\wtheta \in \Theta_n)\to 1$ and 
$\pi(\wtheta,  \theta_{0,n})   \convprob  0 $, we obtain that for any $\epsilon>0$, there exists $n_0$ and $K$ such that for $n\ge n_0$,  
$\prob\left(\gamma_n \, \pi(\wtheta,\theta_{0,n}) > 2^{K}\right) <4\epsilon$, concluding the proof. \qed

\vskip0.2in
 
\noi  \textsc{Proof of Theorem \ref{sec:consis}.2.} 
 As in the proof of Lemma \ref{sec:appen}.3, let $\wtbeta \in \itM_{p_1}^{(1)}$ and $ \wteta \in \itM_{p_2}^{(2)}$, such that
 $\|\wtbeta-\beta_0\|_{\infty}=O(n^{-r\,\nu})$, $ \|\wteta-\eta_0\|_{\infty}=O(n^{-r\,\nu})$ and denote $ \theta_{0,n} = (\wtbeta,\wteta)$. Furthermore, denote as $\bb_{0,n}\in \real^{p_1}$ and  $\ba_{0,n}\in \real^{p_2}$ the vectors such that  $\wtbeta(t)= \bb_{0,n}\trasp\bB^{(1)}(t)$ and $\wteta(z)=\ba_{0,n}\trasp \bB^{(2)}(z)$ where  $\bB^{(1)}(t)=\left( B_1^{(1)} , \dots, B_{p_1}^{(1)}(t)\right)\trasp$ and  $\bB^{(2)}(z)= \left(B_1^{(2)}(z), \dots, B_{p_2}^{(2)}(z)\right)\trasp$.  

 In order to get the convergence rate  of our estimator $\wtheta = (\wbeta,\weta)$ we will apply Lemma \ref{sec:dem-teo-32}.3. Let $\delta_n=A\left(\|\beta_0-\wtbeta\|_{\infty}+\|\eta_0-\wteta\|_{\infty}\right)$ , where $A=4\,\sqrt{(C_0\, \left(\esp  \| X\|^2+1\right)  +A_0)/C_0}$ with $A_0= 4\, \|\psi_1^{\prime}\|_{\infty}  \left(\esp  \| X\|^2+1\right)$, $\psi_1=\rho_1^{\prime}$  and $C_0$ is given in \ref{ass:cotainf}. The consistency of $\wsigma$ entails that $\prob(\wsigma \in \itV)\to 1$, while from Theorem \ref{sec:consis}.1 we obtain that  $\prob(\wtheta \in \Theta_n)\to 1$. Furthermore,   the condition $M_n( \wtheta, \wsigma)  \le  M_n( \theta, \wsigma)$  for any $\theta \in \Theta_n^{(1)}$ is   trivially fulfilled.

 The triangular inequality implies that $\pi ( \theta, \theta_{0,n})\le     \pi ( \theta, \theta_0)+   \pi ( \theta_{0,n}, \theta_0)  $, therefore
 \begin{equation}
  \pi^2( \theta, \theta_{0,n})  \le     2 \left\{\pi^2( \theta, \theta_0) +    \left(\esp  \| X\|^2+1\right) \,\left(\|\wtbeta-\beta_0\|_{\infty}^2+ \|\wteta -\eta_0 \|_{\infty}^2\right)\right\} \,.
\label{eq:dnbetan} 
\end{equation} 
Using that $\pi^2(\wtheta, \theta_0)\convpp 0$,  $\|\wtbeta-\beta_0\|_{\infty}=O(n^{-r\,\nu})$, $ \|\wteta-\eta_0\|_{\infty}=O(n^{-r\,\nu})$ and \eqref{eq:dnbetan}  we obtain  that $\pi(\wtheta,  \theta_{0,n})   \convprob  0 $ as required in  Lemma \ref{sec:dem-teo-32}.3.

 It remains to show that  there exists a function $\phi_n$ such that $\phi_n(\delta)/\delta^{\nu_1}$ is decreasing on $(\delta_n, \infty)$ for some $\nu_1<2$ and that  for any $\delta>\delta_n$, we have
\begin{eqnarray}
\sup_{ \theta\in \Theta_{n,\delta}, \sigma \in \itV} M( \theta_{0,n}, \sigma)- M( \theta,\sigma) &\lesssim & -\delta^2 \,,
\label{aprobar1n}\\
 G_n  & \lesssim &\phi_n(\delta) \,,
\label{aprobar2n}
\end{eqnarray}
where $\Theta_{n,\delta}= \{ \theta\in \Theta_n: \delta / 2  <  \pi( \theta, \theta_{0,n}) \leq \delta\}$ and 
$$G_n =\esp^{*}\sup_{ \theta\in \Theta_{n,\delta}, \sigma \in \itV} \sqrt{n} \left |(M_n( \theta, \sigma)- M( \theta,\sigma))-(M_n( \theta_{0,n}, \sigma)- M( \theta_{0,n},\sigma)) \right |\,. $$ 
Assumption \ref{ass:cotainf} implies that for any $ \theta\in \Theta_n$, $M( \theta, \sigma)-M( \theta_0, \sigma)\ge C_0\,\pi^2( \theta, \theta_0)$. Besides,  the fact that the errors have a symmetric distribution and are independent of the covariates entail  that
 $$\esp\left[ \psi_1\left(\frac{y-\langle X, \beta_0\rangle+\eta_0(z )}{\sigma}\right) \left(\langle X, \wtbeta-\beta_0\rangle + \wteta(z )-\eta_0(z )\right)\right]=0 \,,$$ 
 so 
\begin{eqnarray*}
  M( \theta_{0,n}, \sigma)-M( \theta_0, \sigma)  
    &=& \frac 1{2}\;   \esp\left[    \psi_1^{\prime}\left(\frac{\xi}{\sigma}\right)  \left(\langle X, \wtbeta-\beta_0\rangle + \wteta(z )-\eta_0(z )\right)^2 \right]  \\
  &\leq  &\frac 1{2}\;  \|\psi_1^{\prime}\|_{\infty}\esp \left(\langle X, \wtbeta-\beta_0\rangle + \wteta(z )-\eta_0(z )\right)^2 \\
  &\le & \frac 1{2}\;  \|\psi_1^{\prime}\|_{\infty} 4\, \left(\esp  \langle X, \wtbeta-\beta_0\rangle^2+ \esp \left(  \wteta(z )-\eta_0(z )\right)^2\right)\\ & \le & 2\, \|\psi_1^{\prime}\|_{\infty}  \, \left(\esp  \| X\|^2\, \|\wtbeta-\beta_0\|_{\infty}^2+ \|\wteta -\eta_0 \|_{\infty}^2\right)\\
  &\le & A_0  \left[\|\wtbeta-\beta_0\|_{\infty}^2+ \|\wteta-\eta_0\|_{\infty}^2\right] = O(n^{-2\,r\nu} )\,,
\end{eqnarray*}
where $A_0= 4\, \|\psi_1^{\prime}\|_{\infty} (\left(\esp  \| X\|^2+1\right)$ and $ \xi $ is an  intermediate values between $y-\langle X, \wtbeta\rangle+\wteta(z )$ and $y-\langle X, \beta_0\rangle+\eta_0(z )$.
Thus, using \eqref{eq:dnbetan} and that  $\delta / 2  <  \pi( \theta, \theta_{0,n}) $ we obtain that
\begin{eqnarray*}
M( \theta, \sigma)-  M( \theta_{0,n}, \sigma)& =&\left\{M( \theta, \sigma)-  M( \theta_{0}, \sigma)\right\}- \left\{M( \theta_{0,n}, \sigma)-M( \theta_0, \sigma) \right\}\\
& \ge  & C_0\,\pi^2( \theta, \theta_0)-  A_0\, \left[\|\wtbeta-\beta_0\|_{\infty}^2+ \|\wteta-\eta_0\|_{\infty}^2\right]\\
& \ge & \frac{C_0}2 \pi^2( \theta, \theta_{0,n}) - \left(C_0 \left(\esp  \| X\|^2+1\right)+A_0\right) \,\left(\|\wtbeta-\beta_0\|_{\infty}^2+ \|\wteta -\eta_0 \|_{\infty}^2\right)  \\
 &\ge&   \frac{C_0}2 \pi^2( \theta, \theta_{0,n}) - \left(C_0 \left(\esp  \| X\|^2+1\right)+A_0\right) \left(\|\beta_0-\wtbeta\|_{\infty}+\|\eta_0-\wteta\|_{\infty}\right)^2\\ 
&\ge & \frac{C_0}8 \delta^2 - \frac{1}{A^2} \left(C_0 \left(\esp  \| X\|^2+1\right)+A_0\right) \delta_n^2 =\frac{C_0}8 \delta^2- \frac{C_0}{16} \delta_n^2\ge \frac{C_0}{16} \delta^2\,,
 \end{eqnarray*}
where the last inequality follows from the fact that $\delta >\delta_n$, concluding the proof of \eqref{aprobar1n}.

We have now to  find $\phi_n(\delta)$ such that $\phi_n(\delta)/\delta$ is decreasing in $\delta$ and \eqref{aprobar2n} holds.  
Define the class of functions
$$\itF_{n,\delta} =  \{V_{  \theta,  \sigma}-V_{  \theta_{0,n},  \sigma}:   \theta\in \Theta_{n,\delta}\,, \, \sigma\in \itV\}\subset  \{V_{  \theta,  \sigma}-V_{  \theta_{0,n},  \sigma}:   \theta\in \Theta_{n}\,, \, \sigma\in \itV\}\,,$$ 
with 
$$V_{  \theta,  \sigma}=\rho_1\left(\frac{y- \langle X, \beta \rangle+ \eta(z)}{\sigma}\right) \,,$$
 for $ \theta=(\beta,\eta)$. The inequality \eqref{aprobar2n} involves an empirical process indexed by $\itF_{n,\delta}$, since
$$G_n \le \esp^{*} \sup_{f\in \itF_{n,\delta}} \sqrt{n} |(P_n-P) f|\,.$$
For any $f\in \itF_{n,\delta} $ we have that $\|f\|_{\infty} \le A_1 = 2 \|\rho_1\|_{\infty}=2 $. Furthermore, if $A_2= 2\, \|\psi_1\|_{\infty}/\sigma_0  $ using that for any $ \sigma  \in \itV$, we have
\begin{align*}
|V_{  \theta, \sigma}-V_{  \theta_{0,n}, \sigma}| & = \left|\rho_1\left(\frac{y- \langle X, \beta \rangle+ \eta(z)}{\sigma}\right)-\rho_1\left(\frac{y- \langle X, \wtbeta \rangle+ \wteta(z)}{\sigma}\right)\right| \\
& \le 2\, \|\psi_1\|_{\infty}   \left|\frac{ \langle X, \beta-\wtbeta \rangle+ \eta(z)- \wteta(z)}{\sigma_0}\right|\,,
\end{align*}
and the fact that $\pi( \theta, \theta_{0,n})\le \delta$,  we get that
$$P f^2\le \frac{4\, \|\psi_1\|_{\infty}^2}{\sigma_0^2} \esp\left(  \left[\langle X, \beta-\wtbeta \rangle+ \eta(z)- \wteta(z)\right]^2\right)= A_2^2\,  \pi^2( \theta, \theta_{0,n})\le  A_2^2\, \delta^2\,.$$
  Lemma 3.4.2 van der Vaart and Wellner (1996) leads to
$$\esp^{*} \sup_{f\in \itF_{n,\delta}} \sqrt{n} |(P_n-P) f|\le J_{[\;]}\left( A_2 \delta,\itF_{n,\delta}, L_2(P)\right) \left ( 1+ A_1 \frac{J_{[\;]}(A_2 \,\delta,\itF_{n,\delta}, L_2(P))}{A_2^2 \delta^2 \; \sqrt{n}}   \right ) \,,$$ 
where $J_{[\;]}(\delta, \itF, L_2(P)) =\int_0^\delta \sqrt{1+ \log N_{[\;]}(\epsilon, \itF, L_2(P)) } d\epsilon$ is the bracketing integral of the class $\itF$.

Recall that $\|\wtbeta-\beta_0\|_{\infty}+ \|\wteta-\eta_0\|_{\infty}< \epsilon_0$, so that, for any $\theta=(\beta,\eta)\in \Theta_n$, we have    $\|\wtbeta-\beta\|_{\infty}+ \|\wteta-\eta\|_{\infty}< 2\,\epsilon_0$. 
Hence,     $\itF_{n,\delta}\subset \itG_{n,c,\wttheta_0}$ with $c= 2\,\epsilon_0$, $\wttheta_{0}=\theta_{0}^{\star}=\theta_{0,n}$ and   the bound given in Lemma \ref{sec:appen}.6 leads to
$$N_{[\;]}\left( \epsilon,\itF_{n,\delta}, L_2(P)\right)\le B_1  \left(\frac{B_2 }{\epsilon}\right)^{p_1+p_2+1}\,,$$
for some positive constants $B_1$ and $B_2$ independent of $n$, $\theta_{0,n}$ and $\epsilon$. 
Therefore,
\begin{align*}
J_{[\;]}\left( A_2 \delta,\itF_{n,\delta}, L_2(P)\right) &\le  \int_{0}^{A_2\,\delta}  \sqrt{1+\log\left(B_1  \left(\frac{ B_2 }{\epsilon}\right)^{p_1+p_2+1}\right)} d\epsilon\\
& \le \int_{0}^{A_2\,\delta}  \sqrt{1+\log(B_1)+ (p_1+p_2+1)\log\left(\frac{B_2 }{\epsilon}\right)} d\epsilon\\
& \le 2\, \max\left(1,\sqrt{\log(B_1)}\right)(p_1+p_2+1)^{1/2}\int_{0}^{A_2\,\delta}  \sqrt{1+ \log\left(\frac{B_2 }{\epsilon}\right)} d\epsilon \\
& = 2\, B_2\, \max\left(1,\sqrt{\log(B_1)}\right)(p_1+p_2+1)^{1/2}  \int_{0}^{\frac{A_2}{B_2}\;\delta}  \sqrt{1+ \log\left(\frac{1}{\epsilon}\right)} d\epsilon  \,.
\end{align*}
Note that  $\int_{0}^{\delta} \sqrt{1+\log(1/\epsilon)}\, d\epsilon=O(\delta  \sqrt{\log(1/\delta)})$ as $\delta\to 0$, hence there exists $\delta_0>0$ and a constant $C>0$ such that for any $\delta<\delta_0$,  $\int_{0}^{\delta} \sqrt{1+\log(1/\epsilon)}\, d\epsilon\le C \, \delta \,\sqrt{\log(1/\delta)}$. This implies that, for $\delta<\delta_0\, B_2/A_2$,
$$J_{[\;]}( A_2 \delta,\itF_{n,\delta}, L_2(P)) \lesssim \delta\, \sqrt{\log\left(\frac{1}{\delta}\right)}   \sqrt{p_1+p_2+1}\,.$$
If we denote $q_n = p_1 + p_2+1$, we obtain that for some constant $A_3$ independent of $n$ and $\delta$,
$$  G_n \leq A_3\,\left[\delta \, q_n^{1/2}  \sqrt{\log\left(\frac{1}{\delta}\right)}    + \frac{ q_n  }{ \sqrt{n}}\; \log\left(\frac{1}{\delta}\right)\right]\,.  $$
Choosing
$$\phi_n(\delta)=\delta \, q_n^{1/2}  \sqrt{\log\left(\frac{1}{\delta}\right)}     + \frac{ q_n  }{ \sqrt{n}} \; \log\left(\frac{1}{\delta}\right)\,,$$
we have that $\phi_n(\delta)/\delta$ is decreasing in $\delta$, concluding the proof of \eqref{aprobar2n}.

Note that, since  $\gamma_n= O( n^{ r\nu})$ and $\delta_n=A\left\{\|\beta_0-\wtbeta\|_{\infty}+ \|\eta_0-\wteta\|_{\infty}\right\}=O(n^{-r\nu})$,  we have that $\delta_n\;\gamma_n =O(1)$ as required in Lemma  \ref{sec:dem-teo-32}.3. To apply Lemma   \ref{sec:dem-teo-32}.3, we have to prove that $\gamma_n^2\phi_n \left(1/{\gamma_n}\right)\lesssim \sqrt{n}$, since  $\phi_n(c\delta)\le c \,\phi_n(\delta)$, for $c>1$. 
 Note that 
 $$\gamma_n^2\phi_n \left(\frac{1}{\gamma_n}\right)=\gamma_n  q_n^{1/2} \, \sqrt{\log(\gamma_n)} + 
\gamma_n^2\, \log(\gamma_n)\; \frac{ q_n }{\sqrt{n}} =\sqrt{n}\; a_n(1+a_n) \,\,,$$ 
where $a_n=\gamma_n \, \sqrt{\log(\gamma_n)}\;  q_n^{1/2}/\sqrt{n}$.
Hence, to derive that  $\gamma_n^2\phi_n \left(1/{\gamma_n}\right)\lesssim \sqrt{n}$, it is enough to show that $a_n=O(1)$, which follows easily since  $q_n=O(n^{\nu})$ and $\gamma_n \sqrt{\log(\gamma_n)} = O(n^{(1-\nu)/2})$, concluding the proof.

 Hence, from Lemma   \ref{sec:dem-teo-32}.3,  we get that $\gamma_n^2 \pi^2( \theta_{0,n},\wtheta) = O_{\prob}(1)$.
On the other hand, $\pi( \theta_{0,n}, \theta_0)\le  \|\wtbeta-\beta_0\|_{\infty}(\esp \|X\|^2)^{1/2}+  \|\wteta-\eta_0\|_{\infty}=O(n^{-r\nu})$ together with the fact that $\gamma_n= O( n^{ r\nu})$ entail that $\gamma_n \, \pi( \theta_{0,n}, \theta_0)= O(1)$. Thus, from the triangular inequality we immediately get that   $\gamma_n^2 \pi^2( \theta_{0},\wtheta) = O_{\prob}(1)$, concluding the proof.  \qed


\end{document}